\documentclass[12pt,letter]{article}

\usepackage{subfigure}
\usepackage{graphicx, epsfig, color}
\usepackage{amsmath,amssymb}
\usepackage{slashed}
\usepackage[compat=1.1.0]{tikz-feynman}
\def\to{\rightarrow}
\def\sr2{\sqrt{2}}

\def\bi{\begin{itemize}}
\def\ei{\end{itemize}}

\def\tf{\tilde f}

\def\tchi{\tilde\chi}

\def\tu{\tilde u}
\def\sps1ap{SPS1a$^\prime$}
\def\c1p{C1$^\prime$}

\def\tb{\tilde b}
\def\tf{\tilde f}
\def\td{\tilde d}

\def\tst{\tilde t}
\def\ttau{\tilde \tau}

\def\tg{\tilde g}
\def\tnu{\tilde\nu}
\def\tell{\tilde\ell}

\def\tw{\widetilde W}
\def\tz{\widetilde Z}
\def\alt{\lesssim}
\def\agt{\gtrsim}
\def\be{\begin{equation}}  
\def\ee{\end{equation}}  
\def\bea{\begin{eqnarray}}  
\def\eea{\end{eqnarray}}  
\def\beas{\begin{eqnarray*}}  
\def\eeas{\end{eqnarray*}}

\begin{document}
\begin{titlepage}
\begin{flushright}
OU-HEP-211030
\end{flushright}

\vspace{0.5cm}
\begin{center}
{\Large \bf The cosmological moduli problem and naturalness
}\\ 
\vspace{1.2cm} \renewcommand{\thefootnote}{\fnsymbol{footnote}}
{\large Kyu Jung Bae$^{1}$\footnote[1]{Email: kyujung.bae@knu.ac.kr},
Howard Baer$^{2,3}$\footnote[1]{Email: baer@nhn.ou.edu },
Vernon Barger$^{3}$\footnote[2]{Email: barger@pheno.wisc.edu },
and Robert Wiley Deal$^{2,3}$\footnote[3]{Email: rwileydeal@ou.edu}
}\\ 
\vspace{1.2cm} \renewcommand{\thefootnote}{\arabic{footnote}}
{\it 
$^1$Dep't of Physics,
Kyungpook National University, Daegu 41566, Korea \\
}
{\it 
$^2$Homer L. Dodge Dep't of Physics and Astronomy,
University of Oklahoma, Norman, OK 73019, USA \\
}
{\it 
$^3$Dep't of Physics,
University of Wisconsin, Madison, WI 53706, USA \\
}
\end{center}

\vspace{0.5cm}
\begin{abstract}
\noindent 
Nowadays, the cosmological moduli problem (CMP) comes in three parts: 
1. potential violation of Big-Bang nucleosynthesis (BBN) constraints from late decaying moduli fields, 
2. the moduli-induced gravitino problem wherein gravitinos are overproduced and their
decays violate BBN or dark matter overproduction bounds and
3. the moduli-induced lightest SUSY particle (LSP) overproduction problem. 
Also, the CMP may be regarded as either a problem or else a solution to scenarios
with dark matter over- or under-production.
We examine the cosmological moduli problem and its connection to 
electroweak naturalness. 
We calculate the various two-body decay widths of a light modulus field 
into MSSM particles and gravitinos within general supersymmetric models.
We include both phase space and mixing effects.
We examine cases without and with helicity suppression of modulus decays
to gravitinos (cases 1 \& 2) and/or gauginos (cases A \& B).
For case B1, we evaluate regions of gravitino mass $m_{3/2}$ vs. modulus mass $m_\phi$ parameter space constrained by BBN, by overproduction of gravitinos and by 
overproduction of neutralino dark matter, along with connections to naturalness. 
For this case, essentially all of parameter space is excluded unless $m_\phi\agt 2.5\times 10^3$ TeV with $m_\phi<2m_{3/2}$. 
For a potentially most propitious case B2 with $\phi$ decay to Higgs and matter turned off, then modulus branching fractions to SUSY and to gravitinos become highly suppressed at large $m_\phi$. But since the modulus number density increases faster than 
the branching fractions decrease, there is still gross overproduction of neutralino
dark matter.
We also show that in this scenario the thermally produced gravitino problem is fixed 
by huge entropy dilution, but non-thermal gravitino production from moduli decay remains
a huge problem unless it is kinematically suppressed with $m_\phi < 2m_{3/2}$.
In a pedagogical appendix, we present detailed calculations of modulus field two-body
decay widths.
\vspace*{0.8cm}

%\noindent PACS numbers: 12.60.Jv,14.80.Va,14.80.Ly

\end{abstract}

\end{titlepage}

\tableofcontents

\section{Introduction}
\label{sec:intro}

Supersymmetric (SUSY) models of particle physics have long been the dominant paradigm for
Beyond the Standard Model physics~\cite{Baer:2006rs,Baer:2020kwz} due to their clean solution to the gauge hierarchy
problem and their capacity to explain the dark matter in the universe. They are also supported
by a variety of virtual effects, including 1. gauge coupling unification~\cite{Amaldi:1991cn}, 2. top-Yukawa induced
radiative breaking of electroweak symmetry~\cite{Ibanez:1982fr}, 3. agreement between theory and experiment on the 
mass of the newly discovered Higgs boson~\cite{Slavich:2020zjv} and 
4. precision electroweak measurements~\cite{Heinemeyer:2013dia} which actually
favor a (heavy) TeV-scale SUSY spectrum over the Standard Model (SM). 
In contrast to these success stories, SUSY models have been under seige lately due to 
1. lack of superpartners at LHC with mass below 
early naturalness estimates (the SUSY naturalness or little hierarchy problem) and 
2. lack of a clear WIMP signal at direct or indirect dark matter detection experiments. The first of these problems
has been dispatched by realizing that conventional naturalness measures overestimate finetuning in supersymmetric\footnote{The log derivative measure was typically applied assuming multiple independent soft SUSY breaking parameters in the low energy 
effective field theory (EFT) which leads to 
overestimates of finetuning~\cite{Baer:2013gva,Mustafayev:2014lqa,Baer:2014ica} compared
to more complete theories wherein the soft parameters are correlated. 
The high scale measure $\Delta_{HS}=\delta m_h^2/m_h^2$ abandons some {\it dependent}
contributions which lead to cancellations when fully included~\cite{Baer:2013gva,Mustafayev:2014lqa,Baer:2014ica}.} models.
Applying the more conservative, parameter-independent $\Delta_{EW}$ measure~\cite{Baer:2012up,Baer:2012cf} 
allows for gluinos and squarks well beyond current LHC mass limits while only the 
several neutral and charged higgsinos need lie close to the electroweak scale: $m(higgsinos)\sim m_{weak}\sim m_{W,Z,h}$.
Thus, in natural SUSY, we expect the usual dark matter candidate, the lightest neutralino, to be dominantly higgsino-like. 

The second problem, the lack of a clear WIMP signal, was perhaps already presaged 
by the difficulty of thermally-produced SUSY WIMPs to make up the measured 
dark matter relic density~\cite{Baer:2010wm}.
Binos tend to thermally overproduce dark matter due to their low annihilation cross sections
in the early universe unless special mechanisms such as co-annihilation~\cite{Griest:1990kh,Ellis:1998kh} or resonance
annihilation~\cite{Baer:2000jj,Ellis:2001msa} or tuning/tempering~\cite{Arkani-Hamed:2006wnf,Baer:2006te} are invoked. 
Wino dark matter, as expected in anomaly-mediation models,
thermally underproduce the measured abundance for $m(wino)\alt 3$ TeV while higgsino-like
WIMPs underproduce DM for $m(higgsino)\alt 1$ TeV (well beyond expected bounds from naturalness which require $m(higgsino)\alt 350$ GeV). 
One way around the underproduction issue is to invoke
the axion solution to the strong CP problem so that both axions and WIMPs make up the dark matter~\cite{Baer:2011hx,Bae:2013bva}.
The relic density calculation then requires solving eight coupled Boltzmann equations which include
contributions from axions, axinos, saxions, neutralinos, and gravitinos~\cite{Baer:2011uz,Bae:2014rfa}.

An alternative solution to the WIMP underproduction problem has been to invoke 
non-thermal DM production via one or more hypothesized moduli fields-- gravitationally coupled scalar fields with a classically flat potential. Moduli fields are highly motivated from 
compactified string theory where they determine the size and shape of the 6-7 compact dimensions of the required 10-11 dimensional spacetime. 
The properties of the moduli fields are critical for
predictivity in string theory since their vacuum expectation values (vevs) determine the otherwise free parameters of the low energy EFT such as gauge and Yukawa couplings 
and soft SUSY breaking terms. 
A primary concern of string phenomenologists is to understand moduli stabilization: how
perturbative or non-perturbative effects can stabilize the moduli so that their vevs can be determined. 
Two popular schemes for moduli stabilization in II-B string theory include the KKLT~\cite{Kachru:2003aw} approach, where
complex structure moduli are stabilized by flux to gain ultra-high (decoupled) masses whilst 
the K\"ahler moduli are assumed stabilized by non-perturbative effects and may gain much lower  masses. KKLT is characterized by a mass hierarchy $m_T\gg m_{3/2}\gg m_{soft}$
where $m_T$ is the presumed mass of the lightest K\"ahler modulus~\cite{Choi:2005ge}. 
The second scheme, the large volume scenario (LVS)~\cite{Balasubramanian:2005zx,Blumenhagen:2009gk}, 
stabilizes K\"ahler moduli via a combination of perturbative and non-perturbative contributions to the scalar potential which arise from compactification on a ``swiss cheese'' type of CY manifold where one modulus sets the overall size of the cheese
(large volume) and a second modulus describes the size of four-cycles (holes in the cheese). The volume modulus is expected as the lightest of the K\"ahler moduli and could 
have mass well below the soft SUSY breaking scale leading to the CMP.

Light moduli fields $\phi_i$ with TeV-scale masses ought to have a big impact on WIMP 
production rates in the early universe. They are expected to obtain Planckian field strengths
$\phi_0\sim m_P$ during the inflationary epoch, but then start oscillating at temperatures
$T_{osc}$ when $3H(T_{osc})\sim m_{\phi}(T_{osc})$:
\be
T_{osc}\simeq
\begin{cases}
    \left( 10/\pi^2g_*(T_{osc})\right)^{1/4}\sqrt{m_P m_{\phi}}
    &
    (T_{osc} \leq T_R)
    \\
    \left( 10 g_*(T_R)/\pi^2g_*^2(T_{osc})\right)^{1/8}
    \left(
        T_R^2 m_P m_{\phi}
    \right)^{1/4}
    &
    (T_{osc} > T_R),
\end{cases}
\label{eq:Tosc}
\ee
where $m_P\equiv M_{Pl}/\sqrt{8\pi}$ is the reduced Planck mass.
Since the oscillating moduli fields have the equation of state of matter, they will after some time come to dominate over 
the radiation produced from inflationary reheating.
Due to the extremely large masses the light moduli may possess, these oscillations may actually begin during the reheating phase.
In this case, expansion quickly dilutes their energy density relative to the radiation energy density, which at this phase is still being sourced by inflaton decay.
The moduli will then decay at a temperature $T_D$
\be
T_D\simeq \sqrt{\Gamma_\phi m_P}/(\pi^2 g_*/90)^{1/4} 
\label{eq:TD}
\ee
into radiation (SM particles), dark matter and possibly dark radiation (neutrinos and axions).
For a large enough modulus branching fraction into SUSY particles, then a vast overabundance of LSPs are produced from the modulus field decay 
so that they {\it re-annihilate} during the modulus decay period. 
Then the DM abundance is typically elevated beyond the thermally-produced
value, but not as high as two-WIMPs-per modulus quanta. 
The yield variable $Y_\chi^{\phi}\equiv n_{\chi}/s$ for re-annihilating LSPs 
can be simply estimated~\cite{Giudice:2000ex,Choi:2008zq} and is found to be
\be
Y_{\chi}^{\phi}\sim H(T_D)/\langle\sigma v\rangle s(T_D) 
\ee
leading to a non-thermally produced (NTP) relic abundance
\be
\Omega_{\chi}^{NTP}h^2\sim \Omega_{\chi}^{TP} (T_{fo}/T_D) .
\ee
For $T_D>T_{fo}$, the neutralino abundance will take its thermally-produced (TP) 
value while for $T_D<T_{fo}$ the neutralino abundance is enhanced over the TP expectation by a factor $T_{fo}/T_D$. 
Of course, $T_D$ should also be higher than the temperature $T_{BBN}$
where Big Bang Nucleosynthesis (BBN) begins in order to not upset the successful 
prediction of light element abundances via the standard BBN calculation. 
This latter constraint typically requires rather heavy moduli field masses 
$m_{\phi}\agt 30$ TeV since the gravitationally-coupled moduli decay late and 
$\Gamma_\phi\sim c/4\pi (m_{\phi}^3/m_P^2$ ) (where $c$ is a model-dependent constant of
order $\sim 1$ which is evaluated in this paper).
Thus, by adjusting $m_{\phi}$ to high enough values, one can gain modulus field decay
before the onset of BBN. This is the traditional solution to the BBN portion of
the CMP.

From the above discussion, it should be apparent that the presence of late-decaying moduli 
fields in the early universe can be seen as both a {\it problem} and a {\it solution} 
for dark matter cosmology. 
It is problematic in that late decaying moduli can conflict with BBN
(the original cosmological moduli problem) and/or overproduce dark matter 
(the moduli-induced dark matter problem). 
Also, if moduli can decay to gravitinos, then the gravitino overproduction
can violate BBN or overproduce dark matter (this is the moduli-induced gravitino problem).
In the case of thermally underproduced dark matter, then the presence of moduli can be 
invoked as a {\it solution} in that non-thermal dark matter production can bring the underproduced relic density into accord with measurements~\cite{Moroi:1999zb}.

\subsection{Brief review of previous work and plan for this work}

We break this subsection up into papers concerned mainly with cosmological moduli
as a problem and then those which regard it as a solution to other problems, 
mainly dark matter over- and under-production. Then we preview our perspective.

\subsubsection{Cosmological moduli as a problem}

The first paper to raise an alarm as to the cosmological moduli problem (CMP) was 
Coughlin {\it et al.}~\cite{Coughlan:1983ci} who considered already in 1983 a single hidden sector 
gravitationally coupled scalar field in the supergravity-breaking 
Polonyi superpotential and how its decay 
would produce excess entropy which could disrupt baryogenesis. Dine {\it et al.}
proposed multi-field hidden sectors that could allow the potentially disruptive scalar fields
to settle to much lower field values~\cite{Dine:1983ys}. 
Banks, Kaplan, and Nelson~\cite{Banks:1993en} noted in 1993 that hidden sector
models required $m_\phi\sim m_{3/2}\agt 30$ TeV to solve the CMP. 
This may conflict with {\it naturalness} and so make hidden sector (HS) models less palatable than the alternative of dynamical SUSY breaking (DSB). 
Also in 1993, de Carlos {\it et al.}~\cite{deCarlos:1993wie} placed the CMP on a firm string foundation, emphasizing the generic problem in string theory of 
light dilaton and moduli fields with $m_\phi\simeq m_{3/2}$ 
in place of the SUGRA-based Polonyi fields. 
They emphasize $m_\phi\agt 10$ TeV to avoid the CMP. 
The tension between the required high modulus mass and the 
natural scale of SUSY breaking is noted. 
In Ref. ~\cite{Randall:1994fr}, Randall and Thomas proposed a second stage of
weak scale inflation due to the same moduli flat directions as begat the CMP, 
with only a few $e$-foldings as a means to dilute the modulus field strength 
(see also Ref. ~\cite{Banks:1995dt}).
Dine, Randall and Thomas\cite{Dine:1995uk} (DRT) pointed out the presence of Hubble-induced 
SUSY breaking masses for scalar fields during inflation and proposed some symmetry 
relating the minima of inflationary and post-inflationary potentials to set the modulus
field strength to tiny values in the early universe. 
Dvali~\cite{Dvali:1995mj} proposed also that large contributions to moduli masses
help set the modulus condensate to small values as a possible solution to the CMP. 
Lyth and Stewart proposed in 1995~\cite{Lyth:1995ka} that the presence of flatons $\phi_f$-- 
scalar fields with weak scale masses but $\langle \phi_f\rangle\sim 10^{14}$ GeV-- could initiate a
period of thermal inflation which would act to dilute  the moduli fields and thus alleviate the CMP.
And in 1998, Hashimoto {\it et al.}~\cite{Hashimoto:1998mu}-- concerned with the apparent hierarchy $m_\phi\gg m_{3/2}$--
examined modulus decay to gravitinos $\phi\to\psi_\mu\psi_\mu$ which would create a {\it moduli-induced
gravitino problem} wherein overproduction of gravitinos followed by their late decays could disrupt
BBN. They concluded $m_\phi\agt 100$ TeV would be needed, thus furthering the little hierarchy between the lightest modulus mass and the natural scale for soft SUSY breaking.
In 2004, Kohri {\it et al.}~\cite{Kohri:2004qu,Kohri:2005ru} considered production and dilution of gravitinos by moduli decay 
including $\phi\to\psi_\mu\tilde{\phi}$, decay to gravitino plus modulino, 
along with more complete BBN bounds, for modulino mass $m_{\tilde{\phi}}\sim m_{3/2}$. Lightest SUSY particle (LSP) overproduction was also considered.
Constraints were plotted in the $m_\phi$ vs. $m_{3/2}$ plane.
They concluded that $m_\phi\agt 10$ TeV is required.
In 2006, Nakamura and Yamaguchi~\cite{Nakamura:2006uc} and also with Asaka~\cite{Asaka:2006bv}, and
Endo, Hamaguchi, and Takahachi~\cite{Endo:2006zj} calculated improved modulus decay rates, especially to gravitinos, and compared with BBN and dark matter overproduction constraints, 
concluding that moduli with masses $m_{\phi}\agt 10^2-10^3$ TeV would be required.
These sorts of disparities in moduli vs. gravitino mass scales appeared to be realized
in KKLT~\cite{Kachru:2003aw} models of moduli stabilization via fluxes and non-perturbative
effects where $m_\phi\sim\log (m_P/m_{3/2})m_{3/2}\sim (\log (m_P/m_{3/2})^2 m_{soft}$ seemed to emerge~\cite{Choi:2005ge}.
Thus, nonthermal dark matter from mirage mediation was examined by Nagai and Nakayama 
in 2007~\cite{Nagai:2007ud}.
The idea of softening the CMP via a late decaying saxion field was promoted
by Endo and Takahashi~\cite{Endo:2006ix}. 
Dine {\it et al.}~\cite{Dine:2006ii} further examined modulus decay to gravitinos 
to check if branching fractions could be suppressed due to helicity effects.
In 2014, Blinov {\it et al.}~\cite{Blinov:2014nla} re-examined the
CMP problem, especially with regard to overproduction of LSPs. 
Extensions of the MSSM to include additional hidden sector dark matter 
were invoked in an attempt to avoid the CMP. 

\subsubsection{Cosmological moduli as a solution}

Along with the above papers, an alternative thread developed using the cosmological
production of light moduli as a solution to the dark matter underproduction problem,
especially in SUSY theories with wino-like or higgsino-like LSPs.

This was begun in 1998, where Moroi and Randall~\cite{Moroi:1999zb} (MR) 
invoked nonthermal LSP production 
via moduli decay to bolster the underabundance of wino dark matter as expected
in theories of anomaly-mediated SUSY breaking (AMSB). 
In AMSB~\cite{Randall:1998uk}, soft SUSY breaking terms are suppressed from $m_{3/2}$ by a loop factor so $m_\phi$ and $m_{3/2}$ can both be $\sim 10-100$ TeV: 
in this case, decay modes of the modulus to $\psi_\mu\psi_\mu$ 
would be kinematically closed and the modulus could be heavy enough to evade 
BBN bounds. But also its decay can provide non-thermally the missing DM abundance. 
MR listed a variety of non-renormalizable operators via which the light modulus could decay to MSSM particles (used in this work).
Somewhat later, Pallis (2004)\cite{Pallis:2004yy} and Gelmini and Gondolo (2006)\cite{Gelmini:2006pw,Gelmini:2006pq} 
provided a rather general analysis by which the measured cold dark matter (CDM) abundance 
could be obtained non-thermally via $\phi$ decay for {\it almost any} 
thermally-produced over- or under-abundance. 
The analysis depended on just two parameters:
$b/m_\phi$ and $T_D$, the reheat temperature due to $\phi$ decay. 
(Here, $b$ is the number of neutralinos produced on average for each $\phi$ decay.) 
If $b=0$, then some value of $T_D$ can be used to diminish the thermal abundance via entropy dilution; 
for $b\ne 0$, then neutralino production from $\phi$ decay, usually followed by
$\chi$ re-annihilation, can augment almost any thermally-produced under-abundance. 
The analysis assumed $\phi$ decay to gravitinos was not available.
Acharya {\it et al.}~\cite{Acharya:2006ia} found in the case of $M$-theory 
compactified on a manifold of $G_2$-holonomy 
($G_2MSSM$~\cite{Acharya:2008zi}) a SUSY spectrum with 
$m(scalars)\sim m_{3/2}\sim 30-100$ TeV whilst gauginos gained masses from 
anomaly mediation, thus with a wino-like LSP. 
They explored non-thermal dark matter production and the CMP in a string-motivated framework~\cite{Acharya:2008bk} including calculations of moduli decay widths. 
This analysis also motivated their ``non-thermal WIMP miracle~\cite{Acharya:2009zt}'' scenario along with arguments providing a rationale that the
lightest modulus, which dominates non-thermal dark matter production, 
should have mass comparable to the gravitino mass in supergravity EFTs~\cite{Acharya:2010af} (see also ~\cite{Denef:2004cf}). 
Non-thermal dark matter production in the LVS moduli-stabilization scheme was 
examined in Ref. ~\cite{Allahverdi:2013noa}, Allahverdi {\it et al.}. 
Thermal effects on the production of dark matter in an early matter dominated
universe were considered by Drees and Hajkarim\cite{Drees:2017iod}
where they also mapped
out viable regions of model parameter space with regards to dark matter
production rates.\cite{Drees:2018dsj}.
A review of non-thermal dark matter from cosmological moduli decay
was given in Ref. ~\cite{Kane:2015jia}, Kane, Sinha, and Watson. 
Meanwhile, it was realized that data from indirect dark matter searches (IDD) 
could rule out wino-only dark matter scenarios~\cite{Cohen:2013ama,Fan:2013faa,Baer:2016ucr}. 
Higgsino-only dark matter was found to be ruled out up to 350 GeV
(the naturalness limit) in Ref. ~\cite{Baer:2018rhs} and below 550 GeV using 
AMS limits in Ref. ~\cite{Han:2019vxi}.
Non-thermally produced light higgsino dark matter was also considered in a 
stringy context in Aparicio {\it et al.}~\cite{Aparicio:2016qqb}. 
The seeming exclusion of wino dark matter led some authors to consider
the presence of hidden sector, inert dark matter which can arise from MSSM LSP (late) decays~\cite{Acharya:2016fge,Acharya:2017kfi} via portal mixing. 

\subsubsection{From one to many light moduli}

In this paper, we mainly analyze the case of a single light modulus field
and consider the remaining light moduli to be integrated out.
It is usually assumed that if the moduli masses are sufficiently spaced out, then
mainly the lightest modulus is most important since its latest decay will dilute the
effects of heavier moduli.
This may not be realistic since in string theory there can be potentially tens-to-hundreds
of lighter moduli, many with comparable masses.
In Ref. \cite{Dienes:2015bka}, the case of multiple light moduli is considered
starting with a toy model containing two light moduli which experience dynamical mass
generation during a cosmological phase transition.
The mass generation over a finite time interval coupled with mixing between the fields
leads to new effects which can increase or decrease the field energy densities by
orders of magnitude. 
A shortened summary is given in Ref. \cite{Dienes:2016mte}. 
In Ref. \cite{Acharya:2019pas}, the cosmological consequences of two light
moduli are also investigated in the context of dark matter and dark radiation production rates. 

\subsection{Work done here and brief conclusion}

From the preceeding discussion, we see that the presence of moduli fields in the early universe
can lead to alarming problems-- disruption of successful BBN by late-time entropy production, overproduction of gravitinos and neutralinos, or forays into electroweak finetuned 
parameter space via the requirement of heavy moduli and the associated soft breaking scale-- or they can be regarded as an important solution to problems of 
over- or under-production of SUSY dark matter. 
In this work we revisit the cosmological impact of late decaying moduli fields
in the context of {\it natural} supersymmetric models: 
those that naturally give rise to a weak scale $m_{weak}\sim m_{W,Z,h}\sim 100$ GeV, 
{\it i.e.} without unnatural or implausible finetunings.
These models are also motivated by the string landscape picture~\cite{Susskind:2003kw} which seems to 
favor soft breaking terms as large as possible~\cite{Susskind:2004uv,Douglas:2004qg,Arkani-Hamed:2005zuc} subject to the 
anthropic condition that the pocket-universe value of the weak scale $m_{weak}^{PU}$ is not displaced by more than a factor of a few from the value of the 
weak scale as measured in our universe~\cite{Agrawal:1998xa}. 
The string landscape picture leads to a mini-split supersymmetric spectra 
with rather light higgsinos $\mu\sim 100-350$ GeV but with TeV-scale gauginos
and top squarks and 
even heavier (first/second generation) scalars, and consequently heavier gravitinos
with $m_{3/2}$ in the 10-40 TeV range. We might also expect $m_\phi\sim m_{3/2}$ if the moduli 
receive their dominant mass contributions from soft SUSY breaking. In such a case, then
supersymmetric decay modes (along with SM decay modes) are likely always open for both
moduli and gravitinos, and we expect a severe CMP.

Motivated by these considerations, we evaluate the moduli two-body decay modes into MSSM particles and gravitinos
including both phase space and mixing effects. We evaluate regions of SUSY model 
parameter space in the $m_\phi$ vs. $m_{3/2}$ plane that are subject to severe constraints.
We find almost all parameter space to be excluded, so that our work may be included in the
litany of papers where the light moduli give rise more to problems than solutions.
At the end, we list a variety of ways out of the CMP as presented in the literature and
offer some critical perspective. We conclude that while some solutions have been proposed, there
is room for more work on this important issue. Perhaps most urgent is a deeper 
understanding of moduli properties such as stabilization and their mass scales.

\section{Two-body decays of modulus into MSSM particles and gravitinos}
\label{sec:decay}

From the above discussion, it is apparent that the modulus decay temperature 
$T_D\simeq \sqrt{\Gamma_\phi m_P}$ plays a central role
in determining the outcome of the CMP. This is because $T_D$ determines both whether
the light modulus field avoids BBN bounds (requiring $T_D>T_{BBN}$) and also
the amount of non-thermally produced LSP dark matter via reannihilation
through the ratio $T_{fo}/T_D$.
Thus, to determine $T_D$, the modulus field decay width $\Gamma_\phi$ must be
calculated. In many works, $\Gamma_\phi$ is simply estimated to be
$\Gamma_\phi\sim c/4\pi (m_\phi^3/m_P^2)$ on dimensional grounds, with
$c$ being a (undetermined) model-dependent numerical constant of order unity.
In Appendix \ref{appendix}, we list our assumed modulus-MSSM operators and present a 
rather pedagogical treatment for the calculation of couplings and mixings and 
partial widths for all modulus decays including phase space effects.
(These seem to be lacking in the general literature with the exception of Ref. ~\cite{Acharya:2008bk}). 
Depending on the details of the interaction, some decay modes can receive {\it chirality suppression}. 
We therefore summarize the different case scenarios of leading contributions in Table \ref{tab:cases}.
In this Section, we present some general discussion followed by numerics based upon a natural SUSY benchmark point.

\begin{table}[h!]
    \centering
    \begin{tabular}{c | c | c |}
    & Unsuppressed gravitinos & Suppressed gravitinos \\
    \hline
    Unsuppressed gauginos & Case \textbf{A1} & Case \textbf{A2} \\
    \hline 
    Suppressed gauginos & Case \textbf{B1} & Case \textbf{B2} \\
    \hline
    \end{tabular}
    \caption{Summary of case scenarios on leading decay modes depending on whether or not they receive chirality suppression.
    }
    \label{tab:cases}
\end{table}
    
\subsection{Modulus decay to gravitinos}

The modulus field $\phi$ is broken into real and imaginary parts $\phi =\frac{1}{\sqrt{2}}
(\phi_R+i\phi_I)$ where we are concerned with $\phi_R$ as $\phi_I$ 
may take on the role of an ALP (axion-like particle).
For modulus decay to gravitino ($\psi_\mu$) pairs 
\be
\phi_R\to\psi_\mu\psi_\mu
\ee
we use the formulae computed in Ref.~\cite{Nakamura:2006uc} (augmented by 
phase space factors as listed in Appendix \ref{app:gravitino}).
These widths are actually model dependent and depend on the
form of the K\"ahler function $G=K+\log|W|^2$ where $K$ is the K\"ahler potential
and $W$ is the superpotential. For case {\bf 1}, we have unsuppressed modulus decay to gravitinos so $\Gamma (\phi\to\psi_\mu\psi_\mu )\sim m_\phi^3/m_P^2$ while for
case {\bf 2} we have helicity-suppressed decay to gravitinos so that instead
$\Gamma (\phi\to\psi_\mu\psi_\mu )\sim m_\phi m_\psi^2/m_P^2$.

\subsection{Modulus decay to gauge bosons and gauginos}

An operator suggested by MR~\cite{Moroi:1999zb} is 
\be
{\cal L}\ni \int d^2\theta \frac{\lambda_G}{m_P}\phi W^\alpha W_\alpha
\label{eq:phiWW}
\ee
in two component notation, and where $G=1-3$ for the various SM gauge groups.
The interaction Lagrangian and resultant decay widths are listed in 
Appendix \ref{app:GF}.
For the $SU(3)_C$ gauge group, this leads to 
\be
\phi_R\to gg,\ \tg\tg
\ee 
while for the mixed $SU(2)_L$ and $U(1)_Y$ groups we have
\bea
\phi_R&\to & W^+W^-,\ Z^0Z^0,\ \gamma\gamma,\ \ {\rm and}\ \ \gamma Z^0 \\
&\to & \tchi^0_i\tchi^0_j,\ \ \tchi^+_k\tchi^-_l
\eea
where the neutralino indices $i,j=1-4$ and the chargino indices $k,l=1,2$.
These decay widths into -ino pairs all proceed as $\Gamma (\phi\to\lambda\lambda )\sim
m_\lambda^2m_\phi/m_P^2$ and so are helicity suppressed (case {\bf B}, assuming 
$F_\phi\sim 0$).
(Here, $\lambda$ denotes a generic gaugino state.)
It is worth noting that impact of the $\phi_I$ fields is highly 
model-dependent. 
While their couplings can be determined as in Eq.~\eqref{eq:phiWW}, 
their mass values depend strongly on how the modulus acquire its mass. 
In this respect we neglect the impact of $\phi_I$ by assuming its 
initial field value is tiny.

There are additional possible modulus-gaugino interactions written down in Ref.~~\cite{Asaka:2006bv,Nakamura:2006uc,Dine:2006ii} which are proportional to the vev $F_\phi$. 
In this case, the modulus decay to gauginos loses its helicity-suppression factor
so that instead the widths go as $\Gamma (\phi\to\lambda\lambda )\sim m_\phi^3/m_P^2$
(case {\bf A}).
Here, we maintain ignorance as to whether the lightest modulus is also the modulus
leading to SUSY breaking, and so here we assume $F_\phi\sim 0$.

\subsection{Modulus decay to Higgs pairs}

The modulus field can also decay to Higgs bosons via the Giudice-Masiero\cite{Giudice:1988yz} 
K\"ahler potential operator
\be
{\cal L}\ni\frac{\lambda_H}{m_P}\int d^4\theta (\phi H_U^*H_d^* +h.c.)
\ee
leading to the following decays (as calculated in Appendix \ref{app:HF}):
\bea
\phi_R &\to & hh,\ HH,\ hH,\ AA,\ \ {\rm and}\ \ H^+H^- .
\eea
(This operator may be suppressed by whatever symmetry is invoked to solve the
SUSY $\mu$ problem; for a review, see Ref. \cite{Bae:2019dgg}.)
The SUSY Higgs bosons $H, A\ {\rm and}\ H^\pm$ will further decay to both SM and SUSY
particles, with the exact branching fractions being model-dependent.
These widths all go as $\Gamma (\phi\to Higgs\ pairs)\sim m_\phi^3/m_P^2$.

In addition, there exist modulus-Higgs mixing effects due to the vevs of 
Higgs fields.
These result in more complex decay modes whose partial widths are 
normally $\Gamma\sim m_W^8/(m_{\phi}^5m_P^2)$.
They are sizable where $m_{\phi}\sim m_W$, but this region is clearly 
excluded by the BBN bound, so we neglect these mixing effects 
in the following discussion.

\subsection{Modulus decay to matter and sfermions}

We will also consider the MR operator
\be
{\cal L}\ni \int d^4\theta \frac{\lambda_Q}{m_P}(\phi Q^\dagger Q+h.c.)
\ee
where the superfields $Q$ stand for the various matter chiral superfields of the MSSM.
This operator leads to modulus decays, as calculated in Appendix \ref{app:MF},
\be
\phi_R\to u_i\bar{u}_i,\ d_i\bar{d}_i,\ \ell_i\bar{\ell}_i,\ \nu_i\bar{\nu}_i
\ee
where the decay widths to matter $f$ are all proportional to the matter
mass-squared $m_f^2$. Here, the index $i$ runs over the $i=1-3$ generations.
These widths go as $\Gamma (\phi\to SM\ fermion\ pairs)\sim m_f^2m_\phi/m_P^2$
so that they are suppressed by a factor of $m_f^2$.

There are also decays of the modulus to the MSSM sfermions $\tilde{f}_i$.
In this case, mixing effects are included. 
We find for generations $i=1-2$ that 
\be
\phi_R\to \tu_{Li}\tu_{Li}^*,\ \tu_{Ri}\tu_{Ri}^*,\ \td_{Li}\td_{Li}^*,\ 
\td_{Ri}\td_{Ri}^*,\ \tell_{Li}\tell_{Li}^*,\ \tell_{Ri}\tell_{Ri}^*,\ \tnu_{Li}\tnu_{Li}^* .
\ee
For third generation squarks and sleptons, we have
\be
\phi_R\to \tst_1\tst_1^*,\ \tst_2\tst_2^*,\ \tst_1\tst_2^*+c.c.,\ 
\tb_1\tb_1^*,\ \tb_2\tb_2^*,\ \tb_1\tb_2^*+c.c.,\ \ttau_1\ttau_1^*,\ 
\ttau_2\ttau_2^*,\ \ttau_1\ttau_2^*+c.c.,\ \tnu_{L3}\tnu_{L3}^*
\ee
These widths all go as $\Gamma (\phi\to sfermion\ pairs)\sim m_{\tf}^4/m_\phi m_P^2$
and so actually die off as $m_\phi$ gets large.

\subsection{A natural SUSY benchmark point}

To illustrate the modulus decay widths and branching fraction to SUSY particles, we
adopt a {\it natural} SUSY~\cite{Baer:2012cf} benchmark point from the three-extra-parameter-non-universal Higgs model (NUHM3)~\cite{Baer:2005bu}. 
We generate the sparticle and Higgs mass spectra using Isajet 7.88~\cite{Paige:2003mg}.
The NUHM3 model parameter space
is given by 
\be
m_0(1,2),\ m_0(3),\ m_{1/2},\ A_0,\ \tan\beta,\ m_{H_u},\ m_{H_d}
\ee
where the Higgs mass soft terms $m_{H_u}\ne m_{H_d}\ne m_0$. 
Using the EW minimization conditions, it is convenient to trade the high scale soft terms $m_{H_u},\ m_{H_d}$ for the weak scale parameters $\mu$ and $m_A$.
We assume the gravitino mass $m_{3/2}=m_0(1,2)$.
The sparticle and Higgs masses from the benchmark point are listed in Table \ref{tab:bm}
along with several observables as calculated by Isajet. The point has naturalness
measure $\Delta_{EW}=20$ so that it is indeed a natural SUSY benchmark point.
%
%BM point : \TABLE{
\begin{table}[h!]
\centering
\begin{tabular}{lc}
\hline
parameter & NUHM3 \\
\hline
$m_0(3)$      & 5 TeV \\
$m_0(1,2)$    & 10 TeV \\
$m_{1/2}$      & 1.2 TeV \\
$A_0$      & -8 TeV \\
$\tan\beta$    & 10  \\
\hline
$\mu$          & 200 GeV  \\
$m_A$          & 2 TeV \\
\hline
$m_{\tilde{g}}$   & 2927.4 GeV \\
$m_{\tilde{u}_L}$ & 10209.4 GeV \\
$m_{\tilde{u}_R}$ & 10288.5 GeV \\
$m_{\tilde{e}_R}$ & 9912.9 GeV \\
$m_{\tilde{t}_1}$&  1251.0 GeV \\
$m_{\tilde{t}_2}$&  3655.6 GeV \\
$m_{\tilde{b}_1}$ & 3697.1 GeV \\
$m_{\tilde{b}_2}$ & 5104.5 GeV \\
$m_{\tilde{\tau}_1}$ & 4729.8 GeV \\
$m_{\tilde{\tau}_2}$ & 5061.5 GeV \\
$m_{\tilde{\nu}_{\tau}}$ & 5030.0 GeV \\
$m_{\tilde{\chi}_1^\pm}$ & 209.1 GeV \\
$m_{\tilde{\chi}_2^\pm}$ & 1042.8 GeV \\
$m_{\tilde{\chi}_1^0}$ & 197.7 GeV \\ 
$m_{\tilde{\chi}_2^0}$ & 208.0 GeV \\ 
$m_{\tilde{\chi}_3^0}$ & 547.1 GeV \\ 
$m_{\tilde{\chi}_4^0}$ & 1052.6 GeV \\ 
$m_h$       & 125.3 GeV \\ 
\hline
$\Omega_{\tilde{z}_1}^{std}h^2$ & 0.011 \\
$BF(b\to s\gamma)\times 10^4$ & $3.0$ \\
$BF(B_s\to \mu^+\mu^-)\times 10^9$ & $3.8$ \\
$\sigma^{SI}(\tilde{\chi}_1^0, p)$ (pb) & $1.7\times 10^{-9}$ \\
$\sigma^{SD}(\tilde{\chi}_1^0, p)$ (pb)  & $3.6\times 10^{-5}$ \\
$\langle\sigma v\rangle |_{v\to 0}$  (cm$^3$/sec)  & $2.0\times 10^{-25}$ \\
$\Delta_{\rm EW}$ & 20 \\
\hline
\end{tabular}
\caption{Input parameters (TeV) and masses (GeV)
for a SUSY benchmark point from the NUHM3 model
with $m_t=173.2$ GeV and $m_0(1,2)=m_{3/2}=10$ TeV 
using Isajet 7.88~~\cite{Paige:2003mg}.
}
\label{tab:bm}
\end{table}

\subsection{Modulus decay widths and branching fractions}

Here we present the modulus decay widths and branching fractions into 
the possible final states shown in the preceding subsections. 
We first discuss the suppressed gaugino cases (case {\bf B}) 
assuming $F_{\phi}=0$.
Afterwards, we provide a brief discussion of the unsuppressed gaugino cases 
(case {\bf A}).

\subsubsection{Case {\bf B1}:}
\label{sssec:B1}

\begin{figure}[tbp]
\includegraphics[height=0.5\textheight]{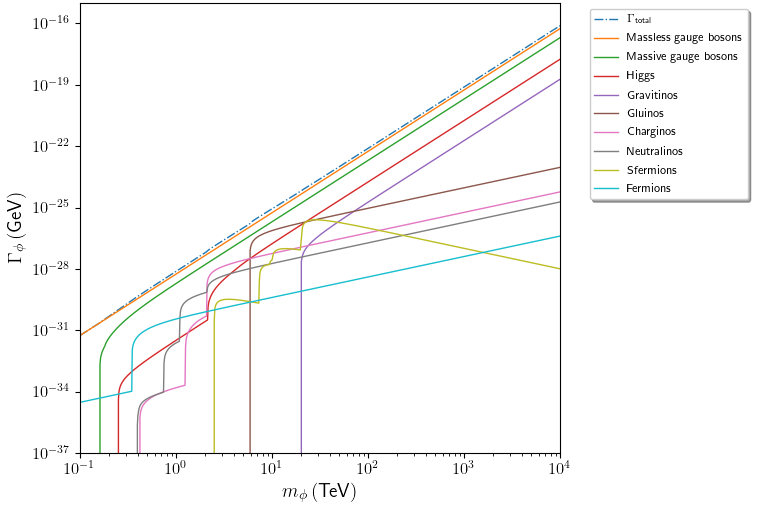}
\caption{Modulus decay widths into various MSSM particles and gravitinos
versus $m_{\phi}$ for the natural SUSY benchmark point with $m_{\psi_\mu}=10$ TeV.
We take all $\lambda_i$ couplings equal to one. The plot is for case {\bf B1}:
suppressed decays to gauginos but unsuppressed decays to gravitinos.
\label{fig:widthsB1}}
\end{figure}

In Fig. \ref{fig:widthsB1}, we show the modulus field $\phi$ partial widths 
versus modulus mass $m_\phi$ from 0.1 to $10^4$ TeV assuming the natural
SUSY benchmark point in Table \ref{tab:bm} with unsuppressed decay to gravitinos 
but suppressed decay to gauginos (case {\bf B1}). Here, we assume all $\lambda_i$ 
couplings equal to unity. For $m_\phi\sim 0.1$ TeV,
only decays to fermions ($u\bar{u},\ d\bar{d}$, {\it etc.}, blue curve)  and massless gauge bosons
({\it e.g.} $gg$ and $\gamma\gamma$, orange curve) are open and 
$\Gamma_\phi\sim 10^{-31}$ GeV,
corresponding to $T_D\sim 1$ keV, so that the modulus field would decay well after
BBN, which occurs beginning at $T_{BBN}\sim 3-5$ MeV. As $m_\phi$ increases, 
decays to massive vector bosons ($WW$, $ZZ$, green curve) turn on at 
$m_\phi\sim 160$ GeV, followed by  
decays into Higgs pairs, which turn on at $m_\phi\sim 250$ GeV (red curve). 
This is followed by $\phi\to t\bar{t}$ pairs (kink in blue curve) at
$m_\phi\sim 350$ GeV. As $m_\phi$ increases even further, decays to neutralinos (grey curve) and charginos (magenta curve) turn on around $m_\phi\simeq 2\mu\sim 400$ GeV.
These curves show several kinks as the various electroweakino (EWino) pair 
decay thresholds are passed. The decay to vector boson pairs increases as $m_\phi^3$
as expected, but decays to gauginos are helicity suppressed and increase instead 
as $m_\phi^1$. This asymptotic behavior also holds for $\phi\to t\bar{t}$ as well.
Decays to sfermions (yellow curve) turn on around 
$m_\phi\sim 2.5$ TeV ($\tst_1\tst_1^*$ mode) but then decrease with increasing $m_\phi$ as
$\Gamma (\phi\to \tf\tf^*)\propto 1/m_\phi$, and so become hardly significant for large
values of $m_{\phi}$. 
At $m_\phi\sim 6$ TeV, $\phi\to \tg\tg$ also turns on (brown curve), which temporarily 
makes a large contribution to the total width. 
Lastly, we see the decay to gravitinos $\phi\to\psi_\mu\psi_\mu$ 
turns on at 20 TeV (purple curve) with the partial width increasing as $m_\phi^3$.
This is an important threshold since gravitino production via modulus decay
is negligible for $m_\phi\alt 2m_{3/2}$ and so in this regime there is no
modulus-induced gravitino problem. Asymptotically, for high $m_\phi$, then
the branching fraction BF($\phi\to\psi_\mu\psi_\mu )\sim 10^{-3}$.

\begin{figure}[tbp]
\includegraphics[height=0.5\textheight]{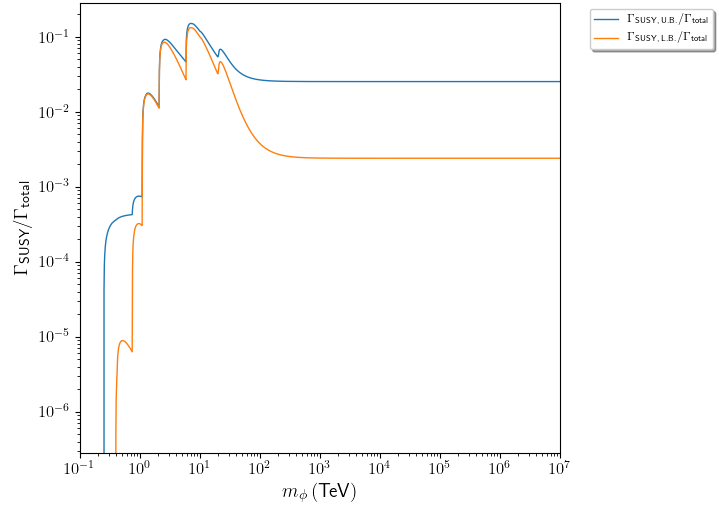}
\caption{Modulus branching fraction to various SUSY particles including gravitinos
versus $m_{\phi}$ for the natural SUSY benchmark point with $\lambda_i=1$ 
and $m_{3/2}=10$ TeV for case {\bf B1}.  
\label{fig:BFB1}}
\end{figure}

The overall modulus branching fraction to SUSY particles is an important
quantity, as emphasized by Gelmini and Gondolo~\cite{Gelmini:2006pw}: 
its value contributes to the modulus-induced LSP problem. 
In Fig. \ref{fig:BFB1}, we show the
modulus branching fraction into SUSY particles vs. $m_{\phi}$ for the same BM point 
in Table \ref{tab:bm}. The branching into MSSM sparticles and gravitinos is
easy to compute, but the $\phi$ decay into Higgs pairs can also contribute
depending on the heavy Higgs boson ($H, \ A$ and $H^\pm$) 
branching fractions into MSSM particles. These BFs are model dependent, so we
show an upper and a lower bound on the $BF(\phi\to SUSY )$ assuming
{\it no} Higgs decay to SUSY (lower orange curve) or the upper bound with 
100\% Higgs decay to SUSY (blue curve). The BF can reach as high as
$\sim 0.2$ for $m_\phi\sim 10$ TeV but asymptotically can range between
$0.002-0.02$ for $m_\phi\gg 10^2$ TeV.

\subsubsection{Case {\bf B2}, $\lambda_H=\lambda_Q=0$:}
\label{sssec:B2}

\begin{figure}[tbp]
\includegraphics[height=0.5\textheight]{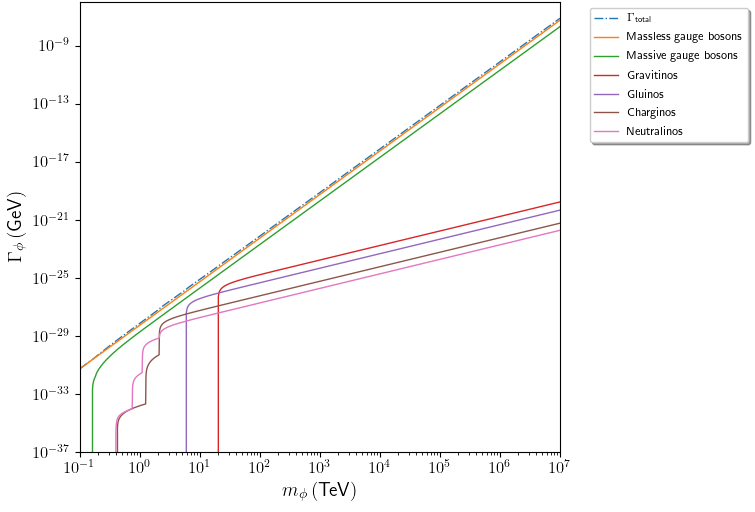}
\caption{Modulus decay widths into various MSSM particles and gravitinos
versus $m_{\phi}$ for a natural SUSY benchmark point with $m_{3/2}=10$ TeV.
We take all $\lambda_{1,2,3}=1$ but all $\lambda_H$ and $\lambda_Q =0$.
This is for case {\bf B2}: suppressed decays to gravitinos and gauginos.
\label{fig:widthsB2}}
\end{figure}

As a second example, we show in Fig. \ref{fig:widthsB2} various modulus partial
widths for the case {\bf B2} with chirality suppressed decays to both
gauginos and gravitinos. We also assume $\lambda_H=0$ which might occur if the
Higgs fields carry PQ charge but the $\phi$ field doesn't so that the 
modulus-Higgs operator is disallowed via the same symmetry that forbids the $\mu$ parameter.
The modulus-matter operator may similarly be forbidden so we take
$\lambda_Q=0$. This case would then have highly suppressed decays to SUSY
particles but with allowed decays to gauge bosons. From the plot, we see that for
$m_\phi\sim 1-20$ TeV, then modulus decay to SUSY particles is comparable to decay widths
to SM particles. But as $m_\phi$ increases, then the helicity suppressed decays to
gauginos and gravitinos sets in and these widths become increasingly suppressed
compared to decays to SM particles (mainly gluon pairs).  

\begin{figure}[tbp]
\includegraphics[height=0.5\textheight]{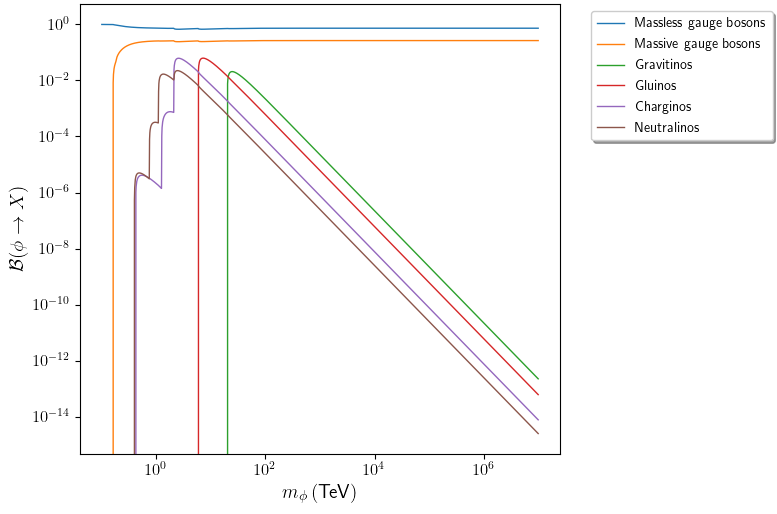}
\caption{Modulus branching fraction to various SUSY particles including gravitinos
versus $m_{\phi}$ for a natural SUSY benchmark point with with $m_{3/2}=10$ TeV.
We take all $\lambda_{1,2,3}=1$ but all $\lambda_H$ and $\lambda_Q =0$.
This is for case {\bf B2}: suppressed decays to gravitinos and gauginos.
\label{fig:BFB2}}
\end{figure}

The situation is shown more clearly in Fig. \ref{fig:BFB2} where we show instead the modulus field branching fraction into gauge bosons (blue and orange curves)
along with modulus branching fractions into gauginos (red, purple, and brown curves)
and the branching fraction into gravitinos (green curve). As $m_\phi$ increases, the
helicity suppression sets in and the branching into SUSY particles falls off sharply
compared to the branching into SM particles. This case is engineered to try to avoid
overproduction of neutralino dark matter and overproduction of gravitinos while possibly diluting any relics produced at temperatures $T>T_D$. We will see shortly why this approach unfortunately fails to solve the CMP.
Further plots can be shown for cases {\bf A1} and {\bf A2} with unsuppressed
moduli decays to gauginos. These cases yield always more neutralino dark matter
overproduction while the total width will be roughly the same, so in the interest of simplicity, we refrain from showing these plots.

\section{Constraints on $m_{\phi}$ and $m_{3/2}$ from BBN,
dark matter, and gravitino overproduction: fun with temperatures}
\label{sec:B1}

\subsection{Constraints on $m_\phi$}

The evolution of the early universe in the presence of light moduli fields 
depends on several important temperature values.
\bi
\item $T_R$, the reheat temperature induced by inflaton decay at the end of inflation,
\item $T_{osc}$, the temperature at which the lightest modulus field begins to oscillate,
\item $T_e$, the temperature at which modulus and radiation energy densities become equal,
\item $T_{fo}$, the temperature at which thermally-produced WIMPs freeze out,
\item $T_D$, the temperature of radiation at the time of modulus decay,
\item $T_{3/2}$, the temperature of radiation at the time of gravitino decay and
\item $T_{BBN}\sim 3-5$ MeV, the temperature of radiation at the onset of BBN
  \footnote{For additional perspective on BBN bounds, see {\it e.g.}
    Ref's \cite{Kawasaki:1999na,Kawasaki:2000en,Hasegawa:2019jsa}.}.
\ei

A rough picture of the evolution of the early universe can then be depicted in
terms of these temperatures and related quantities such as entropy-dilution ratio 
$r\equiv S_f/S_i$ (the ratio of entropy before and after modulus decay)
and non-thermal dark matter production due to modulus decay. 
Here, we will first assume that the reheat temperature $T_R$ is the maximal
temperature scale after inflation resulting in a radiation dominated universe
at temperatures $T$ at or just below $T_R$. 
Here, we note that such high $T_R$ values are at the upper bound from
Buchmuller {\it et al.} (BHLR)~\cite{Buchmuller:2004xr,Buchmuller:2004tz} 
where it is shown that thermal effects from $T_R\agt T_{BHLR}\sim 10^{12}$ GeV
would destabilize the dilaton potential and cause a runaway dilaton field.
We also assume the simplest scenario for reheating in the cases where 
that factors into the calculations (see e.g. ~\cite{Kolb:1990vq}).

\begin{figure}[tbp]
\includegraphics[height=0.5\textheight]{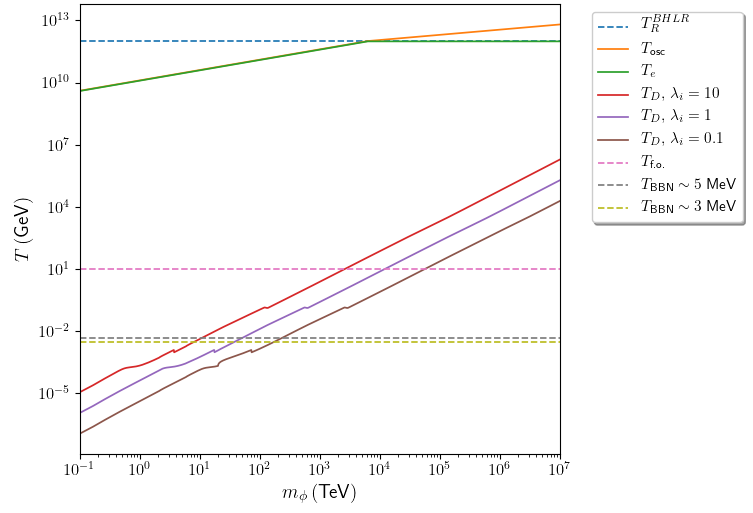}
\caption{Modulus decay temperatures $T_D$ for the natural SUSY benchmark point
for $\lambda_i =\ 0.1,\ 1$, and 10 with $m_{\psi_\mu}=10$ TeV.
We also show the onset of BBN lines at $T_{BBN}\sim 3-5$ MeV, and the
neutralino freeze-out temperature $T_{fo}$ along with the modulus
oscillation temperature $T_{osc}$ and the modulus-radiation
equality temperature $T_e$ assuming $\phi_0=\sqrt{2/3} m_P$.
\label{fig:temps}}
\end{figure}

At temperatures $T$ with $T > T_{osc}$, the modulus field is
effectively frozen in its value due to the preponderance of the friction term 
in its equation of motion: 
\be
\ddot\phi +3H(T)\dot{\phi}+\frac{dV}{d\phi}=0  .
\ee
The scalar potential can be approximated as a simple harmonic oscillator (SHO) form 
$V\simeq m_{\phi}^2\phi^2/2$ for small $\phi$ values. 
As the universe expands, the Hubble constant $H=\sqrt{\rho_T/3m_P^2}$ 
(where $\rho_T$ is the total energy density and $m_P$ is the reduced Planck mass) decreases. 
When $3H(T)\sim m$, the modulus field begins to oscillate at 
the oscillation temperature $T_{osc}$ given in Eq. \eqref{eq:Tosc}.
Specifically, if $m_\phi \lesssim 6000$ TeV (for our assumed $T_R=10^{12}$~GeV), oscillations commence during the assumed initial radiation dominance 
with $\rho_T\sim \rho_{rad}=\pi^2 g_*T^4/30$ so that 
\be 
H(T) \simeq \frac{ \pi T^2 }{m_P} \sqrt{g_*(T) / 90}.
\ee
If, however $m_\phi \gtrsim 6000$ TeV (again, for our assumed $T_R$), oscillations will begin during reheating.
As $\rho_{rad}$ scales roughly as $(1/R)^{3/2}$ during this era ~\cite{Kolb:1990vq} where $R$ is the scale factor of the universe, we can estimate $H(T)$ by comparing $\rho_{rad}(T > T_R)$ to $\rho_{rad} (T=T_R)$.
Noting that the reheating period is effectively matter dominated (i.e. dominated by coherent oscillations of the inflaton), we have then 
\be
H(T) \simeq H(T_R) \frac{g_*(T) T^4}{g_*(T_R) T_R^4}.
\ee
The values of $T_{osc}$ along with an assumed value of $T_R=10^{12}$ GeV are shown in 
Fig. \ref{fig:temps} versus $m_\phi$ (orange curve). 

At this point, the oscillating $\phi$ field
behaves with an equation of state of matter and so  the
energy density of the $\phi$ field diminishes as 
$\rho_\phi=\frac{1}{2}m_\phi^2\phi_0^2(R_{osc}/R)^3$ due to the expansion of the
universe, while $\rho_{rad}$ diminishes as $(1/R)^4$ (once the reheating period has ended).
Here, $R_{osc}$ is the scale factor at $T=T_{osc}$.

The temperature of radiation/modulus energy density equality, determined by 
requiring $\rho_R=\rho_\phi$, is found to be
\be
T_e\equiv 
\begin{cases}
    \left(15 / \pi^2 g_*(T_e) \right)^{1/4} \sqrt{m_\phi \phi_0}
    & 
    (T_{osc} < T_{e} < T_R)
    \\
    (3/2 m_P^2)\phi_0^2\sqrt{m_Pm_\phi}
    \left( 10/\pi^2 g_*(T_e) \right)^{1/4}
    &
    (T_e < T_{osc} < T_R)
    \\
    (3/2 m_P^2)\phi_0^2 T_R
    &
    (T_e < T_R < T_{osc} )
\end{cases}
\label{eq:Te}
\ee
At $T=T_e$, the $\phi$ field begins to dominate the energy density of the universe (depicted by the green curve in Fig.~\ref{fig:temps}). 
Before proceeding, a few comments are warranted on Eq.~\eqref{eq:Te}.
If oscillations begin after reheating ($T_{osc} < T_R$), one would clearly expect to encounter the first case ($T_{osc} < T_e < T_R$) assuming that $\phi_0 \sim m_P$.
Since $T_e > T_{osc}$, indeed the modulus has actually begun to dominate the energy density shortly before it begins to oscillate - i.e. its energy density is so large that it will come to dominate over radiation while still being frozen by Hubble friction.
However, if some mechanism can lower the expected value of $\phi_0$, then it would be possible to encounter the second case ($T_e < T_{osc} < T_R$).
In the event that oscillations begin during reheating ($T_{osc} > T_R$), note that it is actually not possible to achieve a radiation dominated universe after reheating unless $\phi_0 \lesssim \sqrt{2/3} m_P$ - the modulus energy density is so large that reheating should lead directly to a period of early matter domination!
We leave the case of initial matter domination for future work (which requires a more careful coupled Boltzmann approach to modeling the reheating period where the presence of the oscillating modulus contributes to $H$).
However, we do not expect the main results to differ significantly from those we find here, and where appropriate adopt $\phi_0 \sim \sqrt{2/3} m_P$ for consistency with initial radiation domination.
Assuming again the case where $T_{osc} > T_R$ and $\phi_0 \lesssim \sqrt{2/3} m_P$, the modulus will actually dominate over radiation during the reheating period.
The modulus then falls below radiation due to the relative scalings.
This briefly leads to a radiation dominated universe once reheating concludes, and the modulus quickly becomes dominant once more as radiation is no longer sourced by inflaton decay.

In any of the above cases, the modulus field then dominates the energy density of the universe until it 
decays at the decay temperature $T_D$ given by Eq.~\eqref{eq:TD} which is approximately $T_D\simeq \sqrt{m_P\Gamma_\phi}$ and 
which obviously depends on the modulus decay width discussed in the previous 
Section and calculated in the Appendix. 
We show three values of $T_D$ in Fig. \ref{fig:temps} corresponding to setting all 
$\lambda_i$ values to 0.1 (brown curve), 
$\lambda_i=1$ (purple curve), and $\lambda_i=10$ (red curve).
If $T_D<T_{BBN}$, then the modulus field decays after the onset of BBN, leading to destruction of the light element abundances. 
For the purple $T_D$ curve with all $\lambda_i=1$, 
then we can read off from Fig. \ref{fig:temps} that $m_\phi$ is required to be
$m_\phi\agt 40$ TeV (using the $T_{BBN}=5$ MeV value, dashed grey curve).
This is the traditional solution to the old CMP. We also show in Fig. \ref{fig:temps}
the thermally-produced neutralino freeze-out temperature $T_{fo}\sim m_\chi/20$ 
as the magenta dashed line. For $T_D>T_{fo}$, then the relic density of
neutralino dark matter will just be its thermally-produced value $\Omega_\chi^{TP}h^2$
which for our natural SUSY benchmark point is $\Omega_\chi^{TP}h^2\sim 0.01$, about a 
factor 10 below the measured dark matter abundance.
It is worth noting however that this approximation is not necessarily valid for neutralinos produced by the decay of non-thermally produced gravitinos, which may decay at a lower temperature - we will discuss this case in the next section.

For the case where $T_{BBN}<T_D<T_{fo}$, then a further concern for the CMP 
is the direct overproduction of dark matter $\chi$ from moduli cascade decays. 
The initial modulus energy density is 
$\rho_\phi=m_\phi^2\phi_0^2/2$ and the modulus number density is roughly 
$n_\phi\sim \rho_\phi/m_\phi$. Assuming each modulus particle decays 
with branching fraction $B(\phi\to\chi)$, and accounting for the expansion
of the universe between $T_{osc}$ and $T_D$, we find
\be
n_\chi^D\sim B(\phi\to\chi )m_\phi \phi_0^2\left(\frac{g_*(T_D)T_D^3}{g_*(T_{osc})T_{osc}^3}\right),
\label{eq:nD}
\ee 
{\it i.e.} naively, the neutralinos initially inherit the modulus field number density, 
subject to branching ratio and expansion effects.

\begin{figure}[tbp]
\includegraphics[height=0.5\textheight]{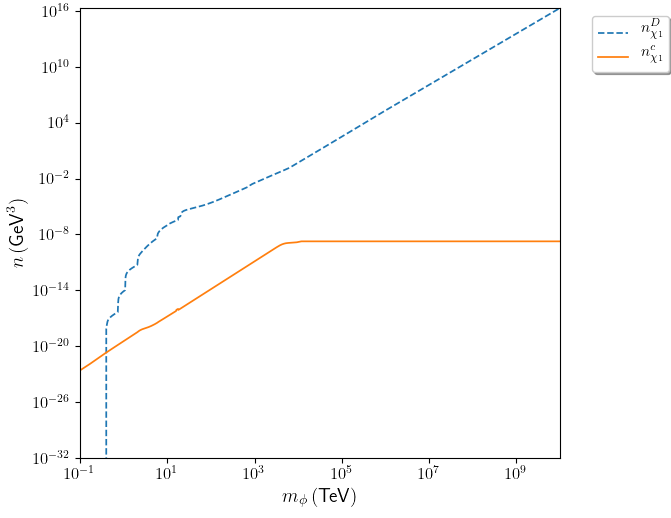}
\caption{Neutralino number densities $n_\chi^c$ and $n_\chi^D$ 
as a function of modulus mass $m_\phi$ for the natural SUSY benchmark point with 
case {\bf B1} for $\lambda_i =1$, $m_{3/2} = m_\phi$, and assuming the no Higgs decay to SUSY. 
Case {\bf B2} is nearly identical, as $n_\chi^D$ is mostly produced by the gauginos if the gravitino channel is kinematically closed.
\label{fig:nD}}
\end{figure}

However, this accounting can be greatly modified if the modulus field 
decays at temperature $T_D<T_{fo}$ and the number density 
$n_\chi$ exceeds the critical density $n_\chi^c$ above which 
neutralino {\it reannihilation effects} may be important~\cite{Giudice:2000ex,Choi:2008zq,Baer:2011hx}. 
The Boltzmann equation for the neutralino number density is given by
\be
\frac{dn_\chi}{dt}+3Hn_\chi=-\langle\sigma v\rangle n_{\chi}^2
\ee
and if $\langle \sigma v\rangle n_\chi(T_D)>H(T_D)$, then neutralinos
will reannihilate after modulus decay. The Boltzmann equation, rewritten
in terms of the yield variable $Y_\chi\equiv n_\chi/s$, where $s$ is the
entropy density ($s=\frac{2\pi^2}{45}g_{*S}T^3$ for radiation) is given by
\be
\frac{dY_\chi}{dt}=-\langle\sigma v\rangle Y_\chi^2 s
\ee
which, assuming $\langle\sigma v\rangle$ is dominated by the constant term, 
can be easily integrated to find $Y_\chi^{reann}\simeq H(T_D)/\langle\sigma v\rangle s(T_D)$ or
\be
n_\chi^c \simeq H/\langle\sigma v\rangle |_{T=T_D}
\ee
so that 
\be
\Omega_\chi^{reann}h^2\simeq \Omega_\chi^{TP}h^2(T_{fo}/T_D),
\ee
{\it i.e.} the reannihilation abundance is enhanced from its TP value
by a factor $T_{fo}/T_D$. The final neutralino number density is then given by
\be
n_\chi\sim \min\left\{ n_\chi^c,n_\chi^D\right\}
\label{eq:nc}
\ee
with $\Omega_\chi h^2= m_\chi n_\chi/\rho_c$
where $\rho_c$ is the critical closure density and $h$ is the scaled Hubble constant.
An example of the neutralino critical number density $n_\chi^c$ and the neutralino
number density from moduli decay $n_\chi^D$ is shown in Fig. \ref{fig:nD}
versus $m_\phi$ for the natural SUSY benchmark point with case {\bf B1}
with helicity-suppressed decay to gauginos. We take $m_\phi\sim m_{3/2}$
so that gravitinos do not enter the plot. 
Actually in this case, Fig. \ref{fig:nD} also describes case {\bf B2} as well since all other sparticle channels are highly subdominant compared to the gauginos.
The critical density 
increases with $m_\phi$ until $T_D>T_{fo}$ whence the $n_\chi^c$ assumes
its thermally produced value. The value of $n_\chi^D$ is almost always far
larger than $n_\chi^c$ due to the huge assumed modulus field strength
$\phi_0\sim m_P$. Thus, we would expect from this plot that the 
neutralino relic density would take its non-thermally-produced reannihilation value
over all allowed values of $m_\phi$.
Provided reannihilation effects are important, the reannihilation
enhancement factor $T_{fo}/T_D$ can be read off from Fig. \ref{fig:temps}.
Since $\phi_0$ is so large, almost always $n_\chi^c<n_\chi^D$ so that the
neutralino relic abundance takes the reannihilation value for $T_D<T_{fo}$.

\begin{figure}[tbp]
\includegraphics[height=0.5\textheight]{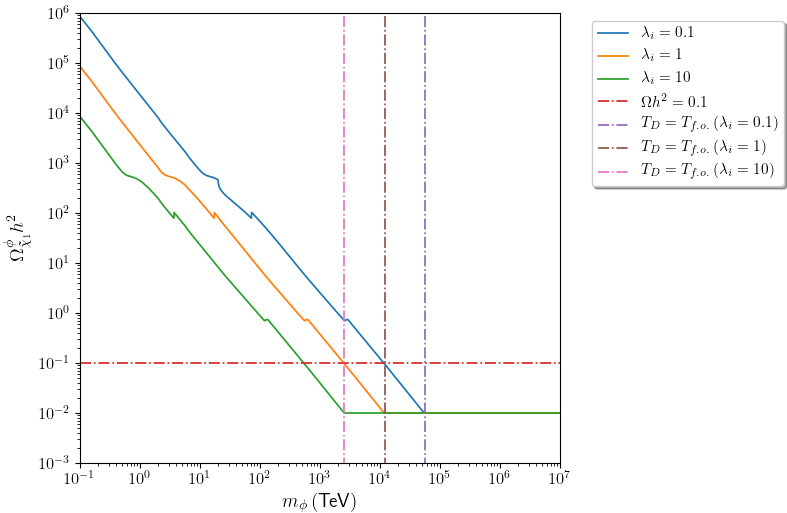}
\caption{Neutralino relic density as a function of modulus mass $m_\phi$
for the natural SUSY benchmark point with case {\bf B1}
for $\lambda_i =\ 0.1,\ 1$, and 10 with $m_{3/2} = m_\phi$ TeV.
As with Fig. \ref{fig:nD}, this plot also describes the case {\bf B2} well if the gravitino channel is kinematically closed.
\label{fig:Oh2}}
\end{figure}

In Fig. \ref{fig:Oh2}, we show the resulting neutralino relic abundance from
both thermal and non-thermal production versus $m_\phi$ for the natural SUSY BM point
and case {\bf B1},
and for the three different values of $\lambda_i=0.1,\ 1$, and 10. 
The orange dot-dashed line shows the measured DM abundance at $\Omega_{CDM} h^2=0.12$. 
We see that for the orange curve with $\lambda_i=1$ and $m_\phi\sim 0.1-2.5\times 10^3$ TeV, 
then neutralinos are greatly overproduced: the {\it moduli-induced LSP
overproduction problem}. Depending on the $\lambda_i$ values, the measured
DM abundance can be achieved, but only for $m_\phi\sim 500-10^4$ TeV.
If the light moduli receive masses from gravity-mediated SUSY breaking, then 
one expects $m_\phi\sim m_{3/2}\sim m_{soft}$ and so to avoid neutralino 
overproduction, one runs into a {\it naturalness} problem in that
sparticles are expected in the $500-10^4$ TeV range. Such high sparticle
masses would lead to a severe Little Hierarchy Problem (LHP), 
wherein one would be hard-pressed to understand why the weak scale is just 
$m_{weak}\sim m_{W,Z,h}\sim 100$ GeV whilst sparticles that contribute to the
weak scale lie in the $500-10^4$ TeV range.

\subsection{Moduli-induced gravitino problem}

\subsubsection{Thermally-produced gravitinos: not a problem}

The usual gravitino problem in supersymmetric theories has to do
with overproduction of thermally produced gravitinos along with their
subsequent late decays which can lead to 1. violation of BBN limits~\cite{Jedamzik:2006xz,Kawasaki:2008qe,Kawasaki:2017bqm}
and 2. overproduction of neutralino dark matter. 
These two constraints depend on the temperature $T_{3/2}$ at which the
gravitinos decay, and on their putative thermally-produced relic abundance, 
had they not decayed: $\Omega_{3/2}^{TP}h^2$. 
The gravitino decay temperature is 
\be
T_{3/2}\simeq \sqrt{\Gamma_{3/2}m_P}/(\pi^2g_*/90)^{1/4}
\label{eq:T32}
\ee
and so we need the gravitino decay widths. 
These have been calculated in Kohri {\it et al.}~\cite{Kohri:2005wn} and 
programmed into the coupled Boltzmann computer
code~\cite{Baer:2011uz,Bae:2014rfa} which we use. We also use the thermally-produced
gravitino production rates as calculated by Pradler and Steffen~\cite{Pradler:2006qh} (see also Ref's \cite{Rychkov:2007uq,Eberl:2020fml}), 
which depends linearly on $T_R$.
Typically, $T_R\alt 2\times 10^9$ GeV is required to avoid gravitino overproduction 
followed by decay to the stable LSP thus leading to dark matter overproduction 
and possible conflict with BBN bounds\cite{Kawasaki:2008qe,Kawasaki:2017bqm}. 
For the case with a decaying light modulus, then entropy is injected during modulus decay
with a dilution factor 
\be
r=S_f/S_0\simeq 4m_\phi Y_\phi/2T_D=T_e/T_D .
\ee
For moduli fields with $\phi_0\sim m_P$, then the entropy dilution is enormous
since $T_e\sim \sqrt{m_\phi \phi_0}$, 
and all thermally-produced gravitinos will be diluted to 
below bounds from BBN and DM overproduction.
If $\phi_0 \lesssim \sqrt{2/3} m_P$, entropy dilution can still be enormous for larger values of $\phi_0$ since $T_e \sim (\phi_0 / m_P)^2 \sqrt{m_\phi m_P}$ if $T_{osc} < T_R$ and $T_e \sim (\phi_0 / m_P)^2 T_R$ if $T_{osc} > T_R$. If, however, $\phi_0$ is lowered far below the Planck scale, the entropy dilution falls rapidly.

\begin{figure}[tbp]
\includegraphics[height=0.5\textheight]{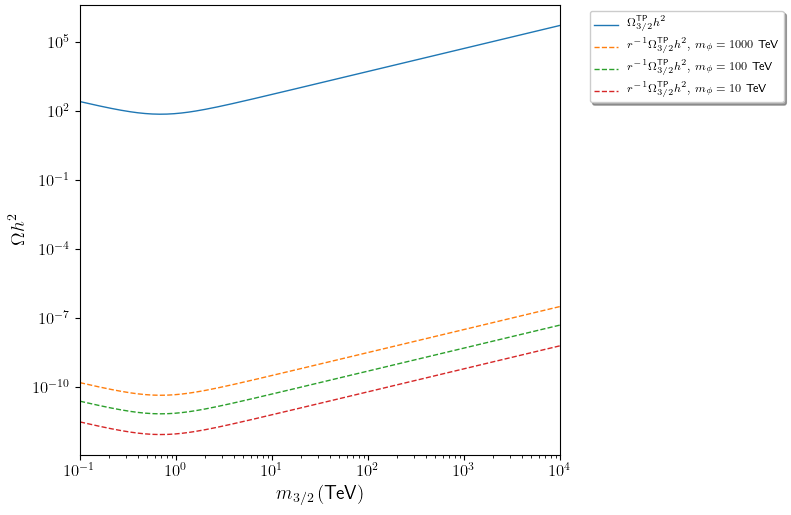}
\caption{Putative thermally-produced gravitino abundance vs. $m_{3/2}$ 
for $T_R=10^{12}$ GeV. We also show the relic would-be gravitino
abundance after modulus decay, which is multiplied by the light modulus
field entropy dilution factor $r=T_e/T_D$. The enormous entropy release
from light modulus decay dilutes the putative thermally-produced gravitino 
abundance to tiny levels. We show cases for $m_\phi=10,\ 100$, and 
1000 TeV.}
\label{fig:Oh2TP_32}
\end{figure}

In Fig. \ref{fig:Oh2TP_32}, we show the would-be thermally-produced
gravitino abundance (had the gravitino not decayed) $\Omega_{3/2}^{TP}h^2$
vs. gravitino mass $m_{3/2}$ for a $T_R=10^{12}$ GeV value, right at the 
BHLR bound.
At such high $T_R$ values, there is an enormous production of gravitinos in the early universe (blue curve). 
However, there is also enormous entropy production from light modulus decay.
We show the diluted gravitino abundance by the three dashed curves for 
three values of $m_\phi=10,\ 100$, and 1000 TeV. We see that after entropy dilution, 
the thermally-produced gravitino abundance has dropped to tiny levels. Thus, 
the late-decaying modulus field has eliminated (just) the thermally produced
gravitino problem. However, the moduli-induced non-thermally produced
gravitino problem is yet unresolved.

\subsubsection{Non-thermally produced gravitinos}

\begin{figure}[tbp]
\includegraphics[height=0.5\textheight]{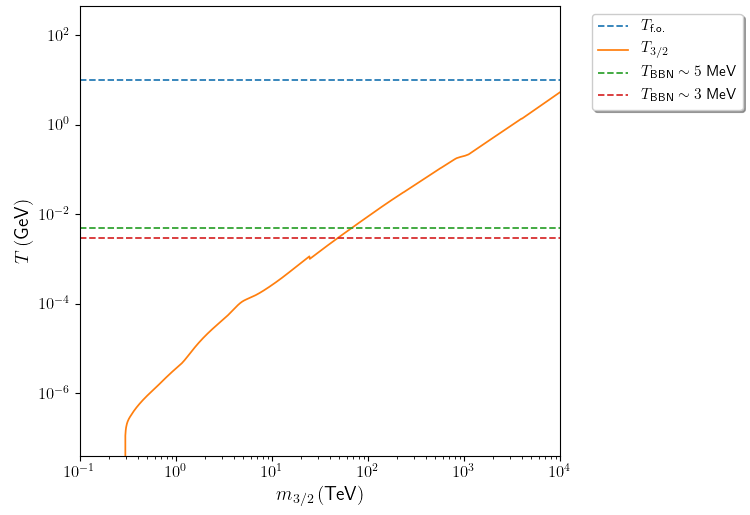}
\caption{Gravitino decay temperature $T_{3/2}$ vs. $m_{3/2}$
assuming our MSSM benchmark point spectrum. We also show 
$T_{BBN}=3$ and 5 MeV along with the neutralino freeze-out temperature 
$T_{fo}$.
}
\label{fig:T32}
\end{figure}

\begin{figure}[tbp]
    \includegraphics[height=0.5\textheight]{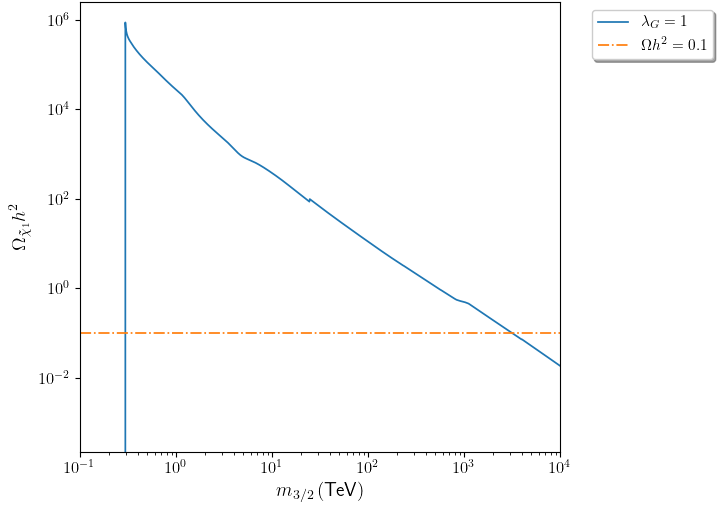}
    \caption{Neutralino relic density as a function of non-thermally produced gravitino mass $m_{3/2}$ for the natural SUSY benchmark point with case {\bf B1}.
    In this case, the produced neutralino number density $n_{\chi}^D$ is always well above the critical number density $n_{\chi}^c$ if the channel is kinematically open and hence the relic density receives an enhancement factor over its freeze-out value.
}
\label{fig:oh232B1}
\end{figure}

Like the modulus field, there are intricate constraints from BBN on late-decaying 
{\it non-thermally-produced} gravitinos in the early universe. 
To examine these, we plot in Fig. \ref{fig:T32} the gravitino decay temperature
$T_{3/2}$ vs. $m_{3/2}$ assuming an MSSM spectrum from our natural SUSY benchmark point.
From the plot, we see that the gravitino decays at temperatures below
$T_{BBN}$ for $m_{3/2}\alt 50$ TeV. For higher $T_{3/2}$ values, then
the gravitino typically is safe from BBN bounds, but not from neutralino
overproduction. For $T_{3/2}\agt T_{BBN}$, then LSPs will be produced
at a huge rate from gravitino decay. The neutralino reannihilation
relic abundance after gravitino decay is expected to be 
$\Omega_{\chi}^{TP}h^2\times (T_{fo}/T_{3/2})$ (using a similar treatment
as for moduli decays to LSPs). As can be seen in Fig. \ref{fig:oh232B1}, the non-thermally-produced
neutralino abundance from gravitino decay is enhanced by several orders of magnitude
(depending on $m_{3/2}$) beyond its thermally-produced value. This is the
moduli-induced, gravitino-induced LSP overproduction problem.

\subsection{Overview of $m_{3/2}$ vs. $m_\phi$ parameter space: case {\bf B1}}

\begin{figure}[tbp]
\includegraphics[height=0.6\textheight]{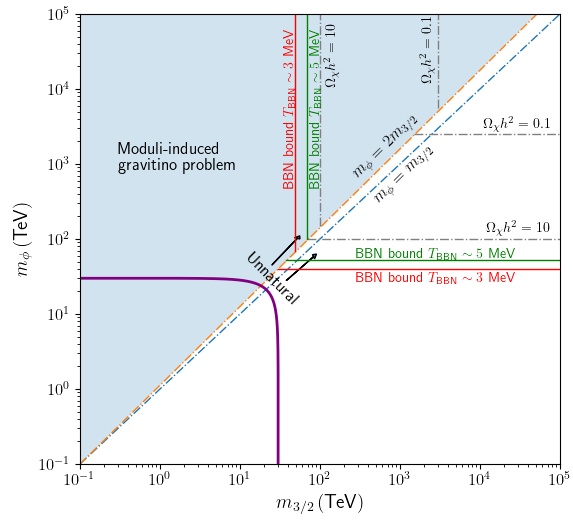}
\caption{Overview of constraints in the $m_{3/2}$ vs. $m_\phi$ plane
for the Natural SUSY benchmark point in case {\bf B1}: helcity-suppressed
decay to gauginos with all $\lambda_i=1$.}
\label{fig:planelog}
\end{figure}

In Fig. \ref{fig:planelog}, we display the $m_{3/2}$ vs. $m_\phi$ plane
assuming the natural SUSY spectra (in this case, uncorrelated with $m_{3/2}$
or $m_\phi$). The region below or to the left of the $T_{BBN}$ 
lines would be excluded by  $T_D$ or $T_{3/2}\alt T_{BBN}$, for two choices of 
$T_{BBN}=3$ and 5 MeV. 
The region to the lower-left of the purple line could be regarded as natural if 
$m_{soft}\sim m_{3/2}\sim m_\phi$ as expected in models where the modulus 
$\phi$ receives a dominant mass contribution from soft SUSY breaking.
Then models with increasing values of $m_\phi$ and $m_{3/2}$ would
become increasingly unnatural. 
The diagonal $m_\phi =m_{3/2}$ line would also be favored by dominant
lightest modulus mass from gravity-mediated soft SUSY breaking~\cite{Acharya:2010af}.
The region to the left of the $m_\phi =2m_{3/2}$ curve we would expect to have
a severe moduli-induced gravitino problem. We also show curves of 
$\Omega_\chi^{NTP}h^2=10$ and 0.1 
(grey dot-dashed lines at constant $m_\phi$ and constant $m_{3/2}$).
Thus, all of the plane shown has a moduli-induced LSP problem 
except the upper-right region where very large values of $m_\phi$
and $m_{3/2}$ are required. 
The fact that almost all of the $m_{3/2}$ vs. $m_\phi$ plane as shown is {\it excluded}
shows the severity of the CMP in the light of naturalness.

\subsection{Case {\bf B2}: why the CMP is still not solved}

\begin{figure}[tbp]
\includegraphics[height=0.5\textheight]{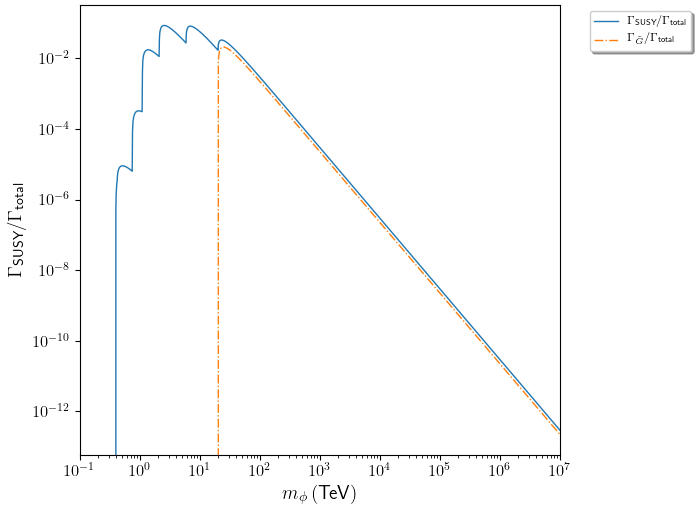}
\caption{Modulus branching fraction into SUSY particles 
versus $m_{\phi}$ for a natural SUSY benchmark point NS1 with $m_{\psi_\mu}=10$ TeV
for case {\bf B2}. We take $\lambda_G=1$ but $\lambda_H=\lambda_Q=0$.
\label{fig:BFSUSYB2}}
\end{figure}

\begin{figure}[tbp]
    \includegraphics[height=0.5\textheight]{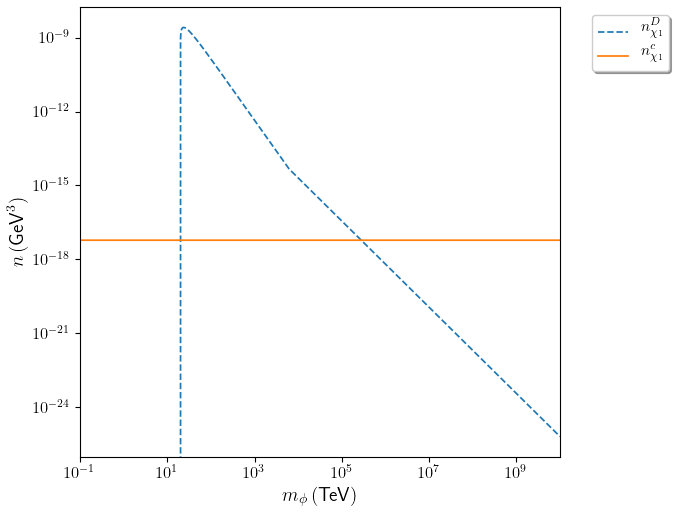}
    \caption{Neutralino number density produced from gravitino decay in case {\bf B2}
    versus $m_{\phi}$ for a natural SUSY benchmark point NS1 with $m_{\psi_\mu}=10$ TeV. 
    We take $\lambda_G=1$ but $\lambda_H=\lambda_Q=0$.
    Note that one needs $m_{\phi} \gtrsim 10^5$ TeV before chirality suppression reduces the produced number density below the critical value (evaluated at $T_{3/2}$).
\label{fig:nDB2}}
\end{figure}

\begin{figure}[tbp]
\includegraphics[height=0.5\textheight]{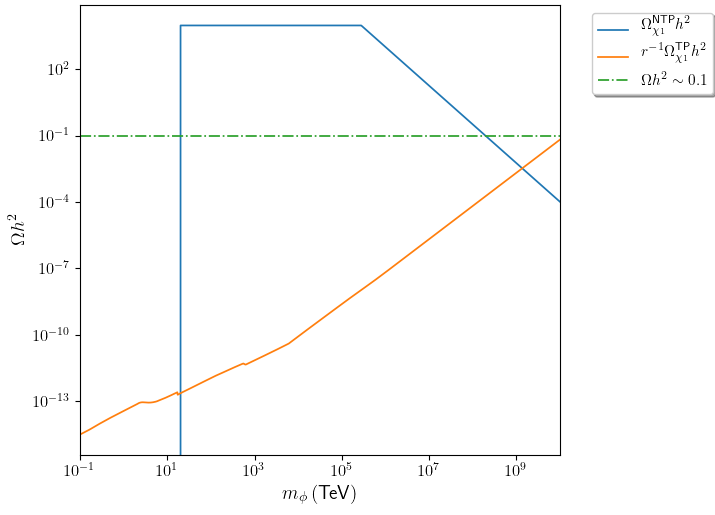}
\caption{Neutralino relic density from case {\bf B2}
versus $m_{\phi}$ for a natural SUSY benchmark point NS1 with $m_{\psi_\mu}=10$ TeV. 
We take $\lambda_G=1$ but $\lambda_H=\lambda_Q=0$.
\label{fig:Oh2B2}}
\end{figure}

In case {\bf B2}, taking $\lambda_H = \lambda_Q = 0$ ensures a highly suppressed branching fraction to SUSY final states, as seen in Fig. \ref{fig:BFSUSYB2}. 
While one might initially expect this case to allow much more parameter space due to this extreme suppression, we find that this is generally not the case.
As seen in Fig. \ref{fig:nDB2},
the produced number density of the neutralinos from the decay of the NTP gravitinos begins roughly nine orders of magnitude above the critical value for $m_{3/2} = 10$ TeV.
In fact, Fig. \ref{fig:Oh2B2} shows that the helicity suppression does not reduce the neutralino relic density to the measured $\Omega h^2 \sim 0.1$ until $m_\phi \sim 10^8$ TeV.
This corresponds to a suppression of $BF( \phi \rightarrow SUSY ) \sim 10^{-14}$!
Although the helicity suppression can technically allow for a natural gravitino mass without overproduction of non-thermal neutralinos, one then has to understand why $m_\phi$ is roughly seven orders of magnitude above $m_{3/2}$.
In addition, entropy dilution due to modulus decay becomes significantly less effective with such a large $m_\phi$,
and hence thermal relics can again become problematic without some other mechanism to deal with them.
We would finally like to point out that inclusion of Higgs or matter superfield decays will make the issue of non-thermal overproduction worse due to the increase in $BF(\phi \rightarrow SUSY)$ while leaving the diminishing entropy dilution $r$ relatively unchanged.
Indeed, it is challenging to reconcile naturalness with the CMP even in this best case scenario.

\label{sec:B2}


\subsection{Moduli-induced baryogenesis problem}

A further cosmological problem arising from weak scale moduli in the
early universe pertains to baryogenesis (for some recent reviews, see {\it e.g.}
  Refs. \cite{Dine:2003ax,Enqvist:2003gh,Bae:2015efa}).
We have seen that a late-decaying modulus field can inject an enormous
amount of entropy into the early universe at temperatures $T_D\sim 5-5000$ MeV.
Such entropy dilution of any relics present at the time of decay is
problematic for many baryogenesis scenarios which occur at higher
temperatures. For instance, EW baryogenesis\cite{Morrissey:2012db}
(perhaps no longer viable in a SUSY context\cite{Quiros:1999tx}) is expected to
occur at or around $T\sim T_{weak}\agt 10^2-10^4$ GeV and so any baryon asymmetry
produced via this mechanism would be wiped out.
Similarly, thermal leptogenesis\cite{Buchmuller:2005eh}
via heavy neutrino production\cite{Fukugita:1986hr} requires $T_R\agt 10^9$ GeV, with conversion to a baryon asymmetry via sphaleron
effects at temperatures around $T_{sphaleron}\agt 10$ TeV\cite{Kuzmin:1985mm};
thus, this sort of asymmetry would also likely be wiped out
by radiation production from late-time modulus decay. A third mechanism--
Affleck-Dine (AD) baryogenesis\cite{Affleck:1984fy,Dine:1995kz} via coherent
production of a baryon or lepton number carrying scalar field along a classical flat direction--
can be highly efficient at producing the baryon asymmetry so that
entropy dilution from late-time modulus decay can actually bring the
baryon asymmetry into accord with its measured value\cite{Dine:2003ax}.
In addition, a range of new mechanisms have been proposed which make use of
the properties of the modulus decay to generate both a baryon asymmetry
and the dark matter abundance. These models purport then to explain why the
seemingly unrelated abundances of baryons and dark matter are
comparable in the present epoch: $\Omega_{baryons}/\Omega_{DM}\sim 1/5$.
Starting with Thomas in 1995\cite{Thomas:1995ze}, some relevant works include
\cite{Kitano:2008tk,Allahverdi:2010im,Allahverdi:2010rh,Ishiwata:2013waa,Dhuria:2015xua,Kane:2019nes}.

\section{Some paths towards solving the CMP}
\label{sec:paths}

In this Section, we discuss some of the proposed solutions to the CMP
as illustrated in Fig. \ref{fig:planelog}.

\begin{figure}[tbh]
\begin{center}
\includegraphics[height=0.5\textheight]{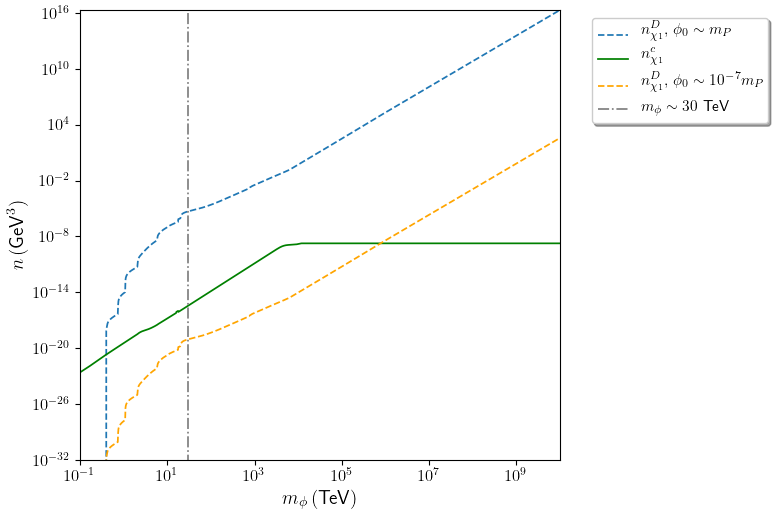}
\caption{LSP number density $n_{\chi}^D$ from modulus decay along with
critical number density $n_{\chi}^c$ which is expected to arise from
LSP reannihilation after modulus decay in the early universe.
\label{fig:nchi}}
\end{center}
\end{figure}

\bi
\item The first solution is to conjecture that the dominant contribution to
moduli masses is not from soft SUSY breaking, but rather from other
possibly non-perturbative effects. Then the lightest modulus may have masses
far beyond the weak scale whilst MSSM sparticles that contribute to the 
weak scale and are associated with naturalness have far smaller values. 
This issue is entangled with the issue of moduli stabilization~\cite{Bose:2013fqa}, 
and so far
can be rather speculative due to lack of experimental guidance and lack of
knowledge of the details of the 6-7 dimensional compactified spaces
whose properties are determined by the various moduli.
\item In Ref. ~\cite{Randall:1994fr} a period of low (weak) scale
inflation is invoked in order to dilute all relics after the period of dark matter
and gravitino production. Similarly, a second period of late thermal 
inflation is invoked in Ref.~~\cite{Lyth:1995ka} which could dilute all relics.
\item In DRT (Ref. ~\cite{Dine:1995uk}), Hubble-induced soft masses are invoked 
in the early universe but some symmetry could cause the modulus field to settle 
to the same minimum as in the low energy theory. 
Then $\phi_0$ could be small and the energy release from modulus decay could be miniscule.
\item In Ref. ~\cite{Endo:2006ix}, a late-decaying saxion field is invoked to dilute
all relics. In this case, the saxion would be but one element of an axion superfield, 
so one would expect additional axion dark matter and additional LSPs produced from
axino decay in the early universe. Plus, one would have to arrange for the saxion to 
not decay into SUSY particles or dark radiation~\cite{Bae:2013qr} so that its role is solely as a source of entropy dilution of all relics at the saxion decay temperature.
\item In Ref. ~\cite{Blinov:2014nla}, several specific hidden sectors are suggested to solve the moduli-induced LSP overproduction. In a hidden $U(1)_x$ extension which is
spontaneously broken by a pair of hidden chiral multiplets, 
then a hidden sector gaugino $\chi_1^x$ is the LSP and the MSSM LSP decays to it via
portal interactions. Typical values of $m_{\chi_1^x}\sim 1-5$ GeV. The $\chi_1^x\chi_1^x$
annihilation cross section must be large enough to generate 
$\Omega_{\chi_1^x}h^2\sim 0.12$ but then the $\chi_1^x$ can annihilate strongly enough
into $\gamma$s that IDD bounds come into play. To obey IDD bounds, the $\chi_1^x$
typically then forms only a small fraction of the total DM abundance.
Perhaps axions can make up the remainder~\cite{Baer:2011hx}. 
Blinov {\it et al.} also 
consider a hidden $U(1)_X$ with asymmetric dark matter 
and a non-Abelian hidden gauge group $SU(N)_X$ with similar conclusions: the addition 
of extra light hidden sector states might help evade the CMP but perhaps at the cost
of implausible parameter and/or model choices.
\item In Ref. ~\cite{Acharya:2006ia,Acharya:2007rc}, Acharya {\it et al.} explain how moduli are stabilized and how exponentially suppressed scales emerge in 
11-d M-theory compactified on a manifold of $G_2$-holonomy (which leads to the
MSSM plus moduli plus axions as the 4-d EFT). They present arguments that the lightest
modulus field should have mass nearby to the gravitino mass $m_{3/2}$ 
(which sets the mass scale for scalar fields in the theory)~\cite{Acharya:2010af}
(see also Ref. ~\cite{Denef:2004cf}). Since $m_\phi\sim m_{3/2}$, then the lightest 
modulus decay mode to gravitinos is closed and there is no moduli-induced gravitino problem. To avoid BBN constraints, then $m_\phi\agt 30$ TeV and so also 
$m_{3/2}\agt 30$ TeV. Since gaugino masses could be suppressed either dynamically or via symmetries compared to scalar masses, they can be much lighter: $m(gauginos)\sim 1$ TeV. 
The authors initially expected AMSB-like masses for gauginos with a wino as LSP
which is thermally underproduced by typically two orders of magnitude from the 
measured value. Then the wino abundance can be non-thermally enhanced via 
modulus decay to near its measured value for $m_\phi\sim 50-100$ TeV.
This basic scenario of wino DM produced from moduli decay was first introduced 
by Moroi and Randall~\cite{Moroi:1999zb}.
Since then, the wino as a DM candidate seems ruled out due to constraints from
indirect DM detection~\cite{Cohen:2013ama,Fan:2013faa,Baer:2016ucr}. 
Acharya {\it et al.}~\cite{Acharya:2016fge} have since then explored the possibility 
of an inert hidden sector DM candidate which would still be allowed. 
\item Recently a landscape solution to the CMP has been proposed~\cite{Baer:2021zbj}. In that paper (which overlaps considerably with this one), the huge 
non-thermally produced dark matter abundance from modulus decay is noted as yielding
a huge dark-matter-to-baryon ratio for which there may be anthropic limits in the multiverse: {\it e.g.} structure might appear as virialized dark matter clouds with little baryonic content~\cite{Linde:1987bx,Wilczek:2004cr,Tegmark:2005dy,Freivogel:2008qc,Bousso:2013rda}. Then if the values of $\phi_0$ are spread uniformly in different pocket universes (PU) within the multiverse, only those PUs with $\phi_0\alt 10^{-7}m_P$
would lead to livable universes. The huge suppression of $\phi_0$ would lead to
$n_\chi^D< n_\chi^c$ so that the neutralinos would inherit the suppressed modulus field
number density which could bring the dark matter abundance into accord with measured
values. The case is illustrated in Fig. \ref{fig:nchi} which we reproduce here for 
the convenience of the reader.
\ei

\begin{figure}[tbh]
\begin{center}
\includegraphics[height=0.5\textheight]{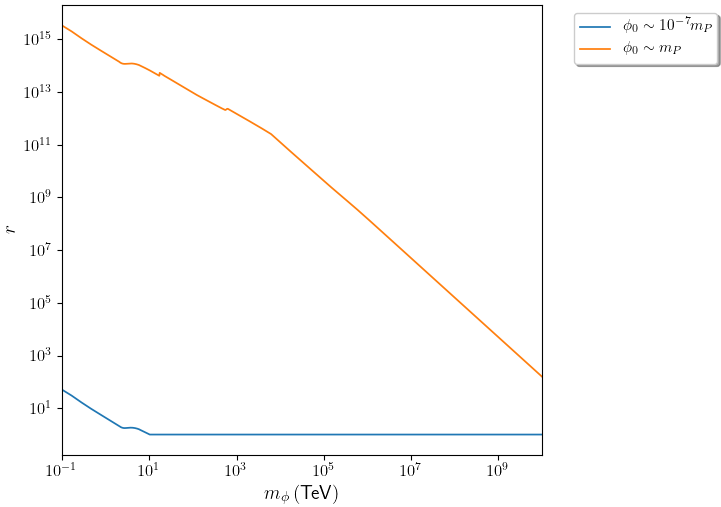}
\caption{Entropy dilution factor $r$ versus $m_{\phi}$ for the cases
  of $\phi_0=m_P$ and for $10^{-7}m_P$ assuming our
  natural SUSY BM point.
\label{fig:r}}
\end{center}
\end{figure}

The anthropic solution to the CMP also presents a solution to the
moduli-induced baryogenesis problem. Namely, with $\phi_0\sim 10^{-7}m_P$,
then $T_e$ is highly reduced and entropy dilution of all relics $r=T_e/T_D$
can be reduced to $\sim 1$, {\it i.e.} no entropy dilution:
see Fig. \ref{fig:r}.
Then the several baryogenesis mechanisms that require $T_R\agt m_{weak}$
become once again viable. On the other hand, there may now be no moduli-dilution
of thermally-produced gravitinos. In this case, $T_R\alt 10^9$ GeV may be
required to avoid the thermally-produced gravitino problem\cite{Bae:2015efa}.

\section{Summary and conclusions}
\label{sec:conclude}

In this paper we have investigated the CMP with regard to 1. limits imposed 
by late decaying moduli which disrupt the successful predictions of BBN, 2. 
overproduction of gravitinos from moduli which lead to BBN violations or DM overproduction
(the moduli-induced gravitino problem) and 3. non-thermal overproduction of 
neutralino dark matter via moduli decays (the moduli-induced LSP problem).
We also confronted the CMP with expectations from SUSY naturalness, as 
exhibited by SUSY models with low $\Delta_{EW}\alt 30$.
While the first of these can be solved by taking $m_\phi\agt 30$ TeV, the second is
more serious but can be solved in cases where $m_\phi\alt 2 m_{3/2}$ so that 
moduli decay to gravitino pairs is kinematically not open. The more serious is the third of these: dark matter overproduction from moduli decays. 
All these issues require computing the putative moduli decay widths into MSSM particles
and gravitino pairs. We perform this task including all phase space and mixing effects
in the Appendix to this paper.
One can then solve the moduli-induce dark matter problem by taking $m_\phi\agt 2.5\times 10^{3}$ TeV for the case of our natural SUSY benchmark model. However, if the 
lightest modulus mass $m_{\phi}$ gains mass dominantly from SUSY breaking so that
$m_\phi\sim m_{3/2}$, then we would also expect sparticles up around the
$10^3$ TeV range. This would require huge finetuning to understand why then the 
weak scale exists at scale $m_{weak}\sim m_{W,Z,h}\sim 100$ GeV.
We review other potential solutions to the CMP including the recent anthropic solution
that could arise from the string landscape in the context of an eternally inflating 
multiverse\cite{Baer:2021zbj}. 
We expect to address issues associated with dark radiation from moduli decay to axion-like-particles (ALPs) in a future work, as well as implementing our calculations into a coupled Boltzmann equation calculation which could include the effects of 
axions, saxions and axinos in addition to MSSM particles and gravitinos.

\section*{Acknowledgments}

We thank L. Randall for suggesting this project more than ten years ago.
This material is based upon work supported by the U.S. Department of Energy, 
Office of Science, Office of High Energy Physics under 
Award Number DE-SC-0009956 and U.S. Department of Energy (DoE) Grant DE-SC-0017647.
%The computing for this project was performed at the OU Supercomputing Center 
%for Education \& Research (OSCER) at the University of Oklahoma (OU).
The work of KJB was supported by the National Research Foundation of Korea (NRF) grant funded by the Korean government (NRF-2020R1C1C1012452).

\appendix

\section{Modulus decay widths}
\label{appendix}

In this Appendix, we shall use the notation $\tw_i$ and $\tz_j$ to denote
the $i=1-2$ chargino and $j=1-4$ neutralino mass eigenstates of the MSSM. 
This alternative notation is intended to remind the reader that all 
mixing notation in this Appendix is given in accord with the text 
{\it Weak Scale Supersymmetry: From Superfields to Scattering Events}\cite{Baer:2006rs}.

\subsection{Decay into gauge fields}
\label{app:GF}

Here, we adopt the Moroi-Randall (MR) operator \cite{Moroi:1999zb}
\begin{align*}
    \mathcal{L}_G 
    &= 
    \int d^2 \theta \, \frac{\lambda_G}{M_*} \phi W^\alpha W_\alpha 
    +
    \text{h.c.}
\end{align*}
(which leads to helicity-suppressed modulus decay to gauginos (case {\bf B})).
For the case of helicity-unsuppressed decay to gauginos, which occurs when fields
in the gauge kinetic function obtain a SUSY breaking vev\cite{Nakamura:2006uc}
(case {\bf A}), then we will replace factors of $m_{\phi} m_{\lambda}^2\to m_{\phi}^3$
in all decay width formulae for modulus to gauginos pairs (where $\lambda$ stands for
generic gaugino fields).

Expanding to the relevant terms (ignoring for now the modulino and setting $F_\phi = 0$ as we expect SUSY breaking to primarily come from other sources), we have from \cite{Wess:1992cp} Eq. 6.13 (converting from mostly-plus to mostly-minus Minkowski metric):
\begin{alignat*}{2}
    \mathcal{L}_G
    &\supset 
    \frac{\lambda_G}{M_*}
    &&\Bigg[
        \phi
        \left(
            2i \lambda \sigma^{m} \partial_m \overline{\lambda} 
            +
            \frac{1}{2} v_{mn} v^{mn} 
            -
            D^2
            +
            \frac{i}{4} \epsilon^{mnlk} v_{mn} v_{lk} 
        \right)
        \\ &
        &&+
        \phi^\dagger
        \left(
            -2i \partial_m \lambda \sigma^{m} \overline{\lambda} 
            + 
            \frac{1}{2} v_{mn} v^{mn} 
            -
            D^2
            -
            \frac{i}{4} \epsilon^{mnlk} v_{mn} v_{lk}
        \right)
    \Bigg]
    \\
    &=
    \frac{\lambda_G}{\sqrt{2} M_*}
    &&\Bigg[
        \phi_R
        \left(
            2i \lambda \sigma^{m} \partial_m \overline{\lambda} 
            -
            2i \partial_m \lambda \sigma^{m} \overline{\lambda} 
            +
            v_{mn} v^{mn} 
            -
            2 D^2
        \right)
        \\ &
        &&+
        i
        \phi_I
        \left(
            2i \lambda \sigma^{m} \partial_m \overline{\lambda} 
            +
            2i \partial_m \lambda \sigma^{m} \overline{\lambda} 
            +
            \frac{i}{2} \epsilon^{mnlk} v_{mn} v_{lk} 
        \right)
    \Bigg]
    \\
    &=
    \frac{\lambda_G}{\sqrt{2} M_*}
    &&\Bigg[
        \phi_R
        \left(
            2i \left(
                \lambda \sigma^{m} \partial_m \overline{\lambda} 
                +
                \overline{\lambda} \overline{\sigma}^{m} \partial_m \lambda           
            \right)
            +
            v_{mn} v^{mn} 
            -
            2 D^2
        \right)
        \\ &
        &&+
        i
        \phi_I
        \left(
            2i \left(
                \lambda \sigma^{m} \partial_m \overline{\lambda} 
                -
                \overline{\lambda} \overline{\sigma}^{m} \partial_m \lambda 
            \right)
            +
            \frac{i}{2} \epsilon^{mnlk} v_{mn} v_{lk} 
        \right)
    \Bigg]
\end{alignat*}
Since we are in the Weyl basis, we can combine the spinors and the Pauli matrices into the Weyl representation of the Clifford algebra:
\begin{alignat*}{2}
    \mathcal{L}_G
    &\supset
    \frac{\lambda_G}{\sqrt{2} M_*}
    &&\Bigg[
        \phi_R
        \left(
            2i 
            \begin{pmatrix}
                \lambda && \overline{\lambda}
            \end{pmatrix}
            \slashed{\partial} 
            \begin{pmatrix}
                \lambda \\
                \overline{\lambda}
            \end{pmatrix}
            +
            v_{mn} v^{mn} 
            -
            2 D^2
        \right)
        +
        i
        \phi_I
        \left(
            2i 
            \begin{pmatrix}
                \lambda && \overline{\lambda}
            \end{pmatrix}
            \gamma_5
            \slashed{\partial} 
            \begin{pmatrix}
                \lambda \\
                \overline{\lambda}
            \end{pmatrix}
            +
            \frac{i}{2} \epsilon^{mnlk} v_{mn} v_{lk} 
        \right)
    \Bigg]
\end{alignat*}
The term coupled to $\phi_I$ is then a surface term and will only arise in instanton corrections.
We are only left with $\phi_R$ directly contributing to interactions (as expected).
Changing to the Majorana representation and switching to the notation of \cite{Baer:2006rs}, we have 
\begin{alignat*}{2}
    \mathcal{L}_G
    &\supset
    \frac{4\lambda_G}{\sqrt{2} M_*}
    \phi_R
    &&\left(
        \frac{i}{2}
        \overline{\lambda}_A \slashed{D}_{AC} \lambda_C 
        -
        \frac{1}{4} F_{\mu \nu A} F_{A}^{\mu \nu} 
        +
        \frac{1}{2} D_A^2
    \right)
\end{alignat*}
where we have also made explicit the generalization to non-Abelian gauge groups.
We now need to rid ourselves of the auxiliary field $D_A$, 
for which we must first include the other relevant terms in the Lagrangian:
\begin{alignat*}{2}
    \mathcal{L}_G
    &\supset
    \frac{4\lambda_G}{\sqrt{2} M_*}
    \phi_R
    &&\left(
        \frac{i}{2}
        \overline{\lambda}_A \slashed{D}_{AC} \lambda_C 
        -
        \frac{1}{4} F_{\mu \nu A} F_{A}^{\mu \nu} 
        +
        \frac{1}{2} D_A^2
    \right)
    + 
    \frac{1}{2} D_A^2 
    -
    g \mathcal{S}_i^\dagger (t_A D_A) \mathcal{S}_i
    -
    \xi_A D_A
\end{alignat*}
and so the equation of motion for $D_A$ becomes 
\begin{align*}
    D_A \left( 
        1 + \frac{4 \lambda_G}{\sqrt{2} M_*} \phi_R
    \right)
    &\simeq
    D_A
    =
    g \mathcal{S}_i^\dagger t_A \mathcal{S}_i 
    +
    \xi_A
\end{align*}
where we neglect the contribution between $D_A$ and $\phi_R$ as it is Planck suppressed in comparison to the leading order term of the equation of motion.
Note that the D-term induces an interaction between $\phi_R$ and the matter scalars.
We ignore for now these interactions, as they will be 4-body decays and hence highly suppressed.
There is the possible exception of a 2-body decay in the presence of a Fayet-Iliopoulos term, but we ignore that contribution as well.

Therefore, for the gauge and gaugino fields, we have the relevant Lagrangian terms 
\begin{align}\label{modulusGaugeGauginoLagrangian}
    \mathcal{L}_G
    &\supset
    \frac{4\lambda_G}{\sqrt{2} M_*}
    \phi_R
    \left(
        \frac{i}{2}
        \overline{\lambda}_A \slashed{D}_{AC} \lambda_C 
        -
        \frac{1}{4} F_{\mu \nu A} F_{A}^{\mu \nu} 
    \right)
\end{align}

\subsection{Modulus decay into electroweakinos}

Looking at the first term of Eq. \ref{modulusGaugeGauginoLagrangian}, we can insert the $SU(2)_L$ and $U(1)_Y$ components:
\begin{alignat}{2}
    \mathcal{L}_{G}
    &\supset
    \frac{4\lambda_{SU(2)}}{\sqrt{2} M_*}
    &&\frac{i}{2}
    \phi_R
    \overline{\lambda}_A \slashed{D}_{AC} \lambda_C 
    +
    \frac{4\lambda_{U(1)}}{\sqrt{2} M_*}
    \frac{i}{2}
    \phi_R
    \overline{\lambda}_0 \slashed{\partial} \lambda_0
    \\
    &=
    \frac{4 \lambda_{SU(2)}}{\sqrt{2} M_*} && \frac{i}{2} \phi_R
    \left(
        \overline{\lambda}_1 \slashed{\partial} \lambda_1
        +
        \overline{\lambda}_2 \slashed{\partial} \lambda_2
        +
        \overline{\lambda}_3 \slashed{\partial} \lambda_3
    \right)
    +
    \frac{4\lambda_{U(1)}}{\sqrt{2} M_*}
    \frac{i}{2}
    \phi_R
    \overline{\lambda}_0 \slashed{\partial} \lambda_0  \nonumber
    \\ &
    &&- 
    \frac{4 \lambda_{SU(2)}}{\sqrt{2} M_*} \frac{ig}{2} \phi_R
    \left(
        - 
        \overline{\lambda}_1 \slashed{W}_3 \lambda_2 
        +
        \overline{\lambda}_2 \slashed{W}_3 \lambda_1
        +
        \overline{\lambda}_1 \slashed{W}_2 \lambda_3
        -
        \overline{\lambda}_2 \slashed{W}_1 \lambda_3
        -
        \overline{\lambda}_3 \slashed{W}_2 \lambda_1
        +
        \overline{\lambda}_3 \slashed{W}_1 \lambda_2
    \right) \nonumber
    \\
    &=
    \frac{4 \lambda_{G}}{\sqrt{2} M_*} && \frac{i}{2} \phi_R
    \left(
        2
        \overline{\lambda} \slashed{\partial} \lambda
        +
        \overline{\lambda}_3 \slashed{\partial} \lambda_3
        +
        \overline{\lambda}_0 \slashed{\partial} \lambda_0
    \right)
    +
    \frac{4 \lambda_{SU(2)}}{\sqrt{2} M_*} g \phi_R
    \left(
        \overline{\lambda} \slashed{W}_3 \lambda
        -
        \overline{\lambda}_3 \slashed{W}^+ \lambda
        -
        \overline{\lambda} \slashed{W}^- \lambda_3
    \right) \label{modulusGauginoComponents}
\end{alignat}
where we used the shorthand $\lambda_G$, with $G=SU(2)$ for 
$\lambda$ and $\lambda_3$ interactions, and $U(1)$ for $\lambda_0$ interactions.

This then gives us a 3-point interaction between a gaugino pair and the modulus, and a 4-point interaction between the $SU(2)_L$ gauge fields, a gaugino pair, and the modulus.
We want to move to the mass eigenstate basis with the help of the following identities:
\begin{align*}
    \lambda_3
    =
    \sum_i 
    v_3^{(i)} (i \gamma_5)^{\theta_i} \widetilde{Z}_i,
    \qquad
    \overline{\lambda}_3
    =
    \sum_i 
    v_3^{(i)} \overline{\widetilde{Z}}_i (i \gamma_5)^{\theta_i},
    \qquad
    \lambda_0
    &=
    \sum_i 
    v_4^{(i)} (i \gamma_5)^{\theta_i} \widetilde{Z}_i,
    \qquad
    &&
    \overline{\lambda}_0
    =
    \sum_i 
    v_4^{(i)} \overline{\widetilde{Z}}_i (i \gamma_5)^{\theta_i}
\end{align*}
and
\begin{alignat*}{2}
    \lambda
    &=
    \theta_{x} \cos \gamma_L (\gamma_5)^{\theta_{\widetilde{W}_2}} P_L \widetilde{W}_2
    +
    \sin \gamma_L (\gamma_5)^{\theta_{\widetilde{W}_1}} P_L \widetilde{W}_1
    &&+
    \theta_{y} \cos \gamma_R (\gamma_5)^{\theta_{\widetilde{W}_2}} P_R \widetilde{W}_2 
    +
    \sin \gamma_R (\gamma_5)^{\theta_{\widetilde{W}_1}} P_R \widetilde{W}_1
    \\
    \overline{\lambda}
    &=
    \theta_{x} \cos \gamma_L \overline{\widetilde{W}}_2 (-\gamma_5)^{\theta_{\widetilde{W}_2}} P_R
    +
    \sin \gamma_L \overline{\widetilde{W}}_1 (-\gamma_5)^{\theta_{\widetilde{W}_1}} P_R
    &&+
    \theta_{y} \cos \gamma_R \overline{\widetilde{W}}_2 (-\gamma_5)^{\theta_{\widetilde{W}_2}} P_L
    +
    \sin \gamma_R \overline{\widetilde{W}}_1 (-\gamma_5)^{\theta_{\widetilde{W}_1}} P_L
\end{alignat*}

\subsubsection{Relevant matrix elements and phase space formulae}
Once we are in the mass eigenstate basis, there are a few general forms for the interaction.
It is convenient to work out the Feynman rules, matrix elements, and phase space factors to derive general decay widths and insert the specific couplings and particles later.
All of these formulae are worked out in the rest frame of the modulus.

We start with the 3-point interaction formulae.
The first general interaction is of the form:
\begin{align}
    \mathcal{L}_I
    &=
    g
    \phi_R 
    \left(
        \overline{\chi} \gamma^\mu ( g_V - g_A \gamma_5 ) \partial_\mu \psi
    \right)
\end{align}
which corresponds to the diagram:
\newline
\begin{center}
    \begin{tikzpicture}
        \begin{feynman}
            \vertex (a) {\(\phi\)} ;
            \vertex [right=of a] (b) ;
            \vertex [below right=of b] (f1) {\( \overline{\chi} \)};
            \vertex [above right=of b] (f2) {\( \psi \)};
            \diagram* {
                (a) -- [scalar, momentum'={\(q\)}] (b),
                (b) -- [anti fermion, momentum'={\(k'\)}] (f1), 
                (b) -- [fermion] (f2),
            };
        \end{feynman}
    \end{tikzpicture}
\end{center}

If we take the incoming momenta of the modulus to be $q$, and the outgoing momenta of $\psi$ and $\overline{\chi}$ to be $k$ and $k'$, respectively, we have a vertex factor of $i g \gamma^\mu ( g_V - g_A \gamma_5 ) k_\mu$.
The matrix element is then 
\begin{align}
    i \mathcal{M}
    &=
    \overline{u}^s(k)
    \left(
        i 
        g 
        \slashed{k}
        (g_V - g_A \gamma_5)
    \right) 
    v^{s'}(k') 
    \label{matrixElementGaugino}
\end{align}

Summing over outgoing spins, the squared matrix element is then 
\begin{align*}
    \left| 
        \mathcal{M}
    \right|^2
    &=
    |g|^2
    \text{Tr}
    \left[
        \left(
            \slashed{k'}
            +
            m_\chi
        \right)
        (g_V + g_A \gamma_5)
        \slashed{k}
        \left(
            \slashed{k}
            + 
            m_\psi
        \right)
        \slashed{k}
        (g_V - g_A \gamma_5)
    \right]
    \\
    &=
    m_\psi^2
    |g|^2
    \text{Tr}
    \left[
        \left(
            \slashed{k'}
            +
            m_\chi
        \right)
        (g_V + g_A \gamma_5)
        \left(
            \slashed{k}
            + 
            m_\psi
        \right)
        (g_V - g_A \gamma_5)
    \right]
    \\
    &=
    m_\psi^2
    |g|^2
    \text{Tr}
    \left[
        (g_V - g_A \gamma_5)
        \slashed{k'}
        \left(
            \slashed{k}
            + 
            m_\psi
        \right)
        (g_V - g_A \gamma_5)
    \right]
    +
    4
    m_\psi^3
    m_\chi
    |g|^2
    \left(
        g_V^2
        - 
        g_A^2
    \right)
    \\
    &=
    4
    m_\psi^2
    |g|^2
    \left(
        g_V^2
        + 
        g_A^2
    \right)
    k'_\mu 
    k^\mu
    +
    4
    m_\psi^3
    m_\chi
    |g|^2
    \left(
        g_V^2
        - 
        g_A^2
    \right)
\end{align*}
With our momenta convention, we see that $k_\mu + k'_\mu = q_\mu$, so $2 k'_\mu k^\mu = m_\phi^2 - m_\psi^2 - m_\chi^2$.
The final squared matrix element is then 
\begin{align}
    \left| 
        \mathcal{M}
    \right|^2
    &=
    2
    m_\psi^2
    m_\phi^2
    |g|^2
    \left[
        g_V^2
        \left(
            1
            - 
            \left(
                \frac{
                    m_\psi
                    - 
                    m_\chi
                }{m_\phi}
            \right)^2
        \right)
        +
        g_A^2
        \left(
            1
            - 
            \left(
                \frac{
                    m_\psi
                    +
                    m_\chi
                }{m_\phi}
            \right)^2
        \right)
    \right]
    \label{matrixElementSqrGaugino}
\end{align}

As this is a 2-body decay with no angular dependence, we have the decay width immediately:
\begin{align}
    \Gamma_{\phi \overline{\chi} \psi}
    &=
    \frac{|g|^2 m_\psi^2 m_\phi }{8 \pi} 
    \left[
        g_V^2 \left( 1 - \left( \frac{ m_\psi - m_{\overline{\chi}} }{ m_\phi } \right)^2 \right)
        +
        g_A^2 \left( 1 - \left( \frac{m_\psi + m_{\overline{\chi}}}{ m_\phi } \right)^2 \right)
    \right]
    \lambda^{1/2}\left( 
        1, \frac{m_\psi^2}{m_\phi^2}, \frac{m_{\overline{\chi}}^2}{m_\phi^2}
    \right)
\end{align}
where we used the general formula 
\begin{align}
    \Gamma_{\phi \rightarrow 1 + 2}
    &=
    \frac{
        S 
        |\mathcal{M}|^2
    }{
        16
        \pi 
        m_\phi
    } 
    \lambda^{1/2}
    \left(
        1,
        \frac{
            m_1^2
        }{ 
            m_\phi^2
        },
        \frac{
            m_2^2
        }{ 
            m_\phi^2
        }
    \right)
\end{align}
with $S$ being the combinatorial factor for identical particles ($S=1$ in this case).
The decay of $\phi \rightarrow \chi \psi$ (summing over both $\overline{\psi} \chi$ and $\overline{\chi} \psi$ final states) is then 
\begin{align}
    \Gamma_{\phi \chi \psi}
    &=
    \frac{|g|^2 ( m_\psi^2 + m_{\chi}^2 ) m_\phi }{8 \pi} 
    \left[
        g_V^2 \left( 1 - \left( \frac{ m_\psi - m_{\overline{\chi}} }{ m_\phi } \right)^2 \right)
        +
        g_A^2 \left( 1 - \left( \frac{m_\psi + m_{\overline{\chi}}}{ m_\phi } \right)^2 \right)
    \right]
    \lambda^{1/2}\left( 
        1, \frac{m_\psi^2}{m_\phi^2}, \frac{m_{\chi}^2}{m_\phi^2}
    \right)    
\end{align}

\subsubsection{3-point interactions}

We first look at the 3-point interactions between the modulus and a gaugino pair.
The last pair of 3-point interaction terms of Eq. \ref{modulusGauginoComponents} give us our neutralino interactions:
\begin{alignat}{2}
    \mathcal{L}_{\phi \overline{\widetilde{Z}}_i \widetilde{Z}_j}
    &=
    \frac{4 \lambda_{SU(2)}}{\sqrt{2} M_*} 
    &&\frac{i}{2}
    \phi_R \left(
        \overline{\lambda}_3 \slashed{\partial} \lambda_3
    \right)
    +
    \frac{4 \lambda_{U(1)}}{\sqrt{2} M_*} 
    \frac{i}{2}
    \phi_R \left(
        \overline{\lambda}_0 \slashed{\partial} \lambda_0
    \right) \nonumber
    \\
    &\supset
    \frac{4 }{\sqrt{2} M_*} 
    &&\frac{i}{2}
    (-i)^{\theta_i}
    (i)^{\theta_j}
    \left(
        \lambda_{SU(2)}
        v_3^{(i)} v_3^{(j)}
        +
        \lambda_{U(1)}
        v_4^{(i)} v_4^{(j)}
    \right)
    \left(
        \phi_R 
        \overline{\widetilde{Z}}_i
        \gamma^\mu
        (\gamma_5)^{\theta_i + \theta_j} 
        \partial_\mu \widetilde{Z}_j
    \right)
\end{alignat}
The corresponding matrix element is then 
\begin{alignat*}{2}
    i \mathcal{M}_{\phi_R \rightarrow \overline{ \widetilde{Z} }_i \widetilde{Z}_j }
    &=&&
    \overline{u}^s (k)
    \left(
        i
        \frac{4}{\sqrt{2} M_*}
        \frac{i}{2}
        \left(
            -i
        \right)^{\theta_i}
        \left(
            i
        \right)^{\theta_j}
        \left(
            \lambda_{SU(2)}
            v_3^{(i)} v_3^{(j)}
            +
            \lambda_{U(1)}
            v_4^{(i)} v_4^{(j)}
        \right)
        \slashed{k}
        \left(
            g_V
            -
            g_A 
            \gamma_5
        \right)
    \right)
    v^{s'}(k')
\end{alignat*}
where $g_V = 0 \, (1)$ and $g_A = 1 \, (0)$ if $\theta_i + \theta_j$ is odd (even). 
The squared matrix elements for $\phi \rightarrow \widetilde{Z}_i \widetilde{Z}_j$ (summing over both $\widetilde{Z}_i + \overline{\widetilde{Z}}_j$ and $\overline{\widetilde{Z}}_i + \widetilde{Z}_j$ final states) is then, 
\begin{align}
    \left| 
        \mathcal{M}_{\phi_R \rightarrow \widetilde{Z}_i \widetilde{Z}_i }
    \right|^2
    &=
    4
    \frac{
        m_{ \widetilde{Z}_i }^2
        m_\phi^2
    }{
        M_*^2
    }
    \left(
        \lambda_{SU(2)}
        \left( 
            v_3^{(i)}
        \right)^2
        +
        \lambda_{U(1)}
        \left( 
            v_4^{(i)}
        \right)^2
    \right)^2
    \\
    \left| 
        \mathcal{M}_{\phi_R \rightarrow \widetilde{Z}_i \widetilde{Z}_j }
    \right|^2
    &=
    4
    \frac{
        \left(
            m_{ \widetilde{Z}_i }^2
            +
            m_{ \widetilde{Z}_j }^2
        \right)
        m_\phi^2
    }{
        M_*^2
    }
    \left(
        \lambda_{SU(2)}
        v_3^{(i)} v_3^{(j)}
        +
        \lambda_{U(1)}
        v_4^{(i)} v_4^{(j)}
    \right)^2
    \left(
        1
        -
        \left(
            \frac{
                m_{ \widetilde{Z}_j }
                - 
                m_{ \widetilde{Z}_i }
            }{
                m_\phi
            }
        \right)^2
    \right)
\end{align}
where we note that the $g_V$ and $g_A$ convention defined above simply flips the sign of the mass cross-term if one (not both) of the masses is negative, and hence we assume that all masses above are taken with their implicit absolute values.
The decay widths for these processes then are:
\begin{alignat}{2}
    \Gamma_{\phi_R \rightarrow \widetilde{Z}_i \widetilde{Z}_i}
    &=&&
    \frac{
        \left(
            \lambda_{SU(2)}
            \left( 
                v_3^{(i)}
            \right)^2
            +
            \lambda_{U(1)}
            \left( 
                v_4^{(i)}
            \right)^2
        \right)^2
    }{
        4
        \pi 
    } 
    \frac{
        m_{ \widetilde{Z}_i }^2
        m_\phi
    }{
        M_*^2
    }
    \lambda^{1/2}
    \left(
        1,
        \frac{
            m_{\widetilde{Z}_i}^2
        }{ 
            m_\phi^2
        },
        \frac{
            m_{\widetilde{Z}_i}^2
        }{ 
            m_\phi^2
        }
    \right)
    \\
    \Gamma_{\phi_R \rightarrow \widetilde{Z}_i \widetilde{Z}_j}
    &=&&
    \frac{
        \left(
            \lambda_{SU(2)}
            v_3^{(i)} v_3^{(j)}
            +
            \lambda_{U(1)}
            v_4^{(i)} v_4^{(j)}
        \right)^2
    }{
        4
        \pi 
    } 
    \nonumber
    \\ &
    && \times
    \frac{
        \left(
            m_{ \widetilde{Z}_i }^2
            +
            m_{ \widetilde{Z}_j }^2
        \right)
        m_\phi
    }{
        M_*^2
    }
    \left(
        1
        -
        \left(
            \frac{
                m_{ \widetilde{Z}_j }
                - 
                m_{ \widetilde{Z}_i }
            }{
                m_\phi
            }
        \right)^2
    \right)
    \lambda^{1/2}
    \left(
        1,
        \frac{
            m_{\widetilde{Z}_i}^2
        }{ 
            m_\phi^2
        },
        \frac{
            m_{\widetilde{Z}_j}^2
        }{ 
            m_\phi^2
        }
    \right)
\end{alignat}

The first 3-point interaction term of Eq. \ref{modulusGauginoComponents} then gives us our chargino interactions:
\begin{alignat*}{2}
    \mathcal{L}_{\phi \overline{\widetilde{W}}_i \widetilde{W}_j}
    &=
    \frac{4 \lambda_{SU(2)}}{\sqrt{2} M_*} 
    &&i
    \phi_R \left(
        \overline{\lambda} 
        \slashed{\partial} 
        \lambda
    \right)
\end{alignat*}
To evaluate this in the mass eigenstate basis, we see that:
\begin{alignat*}{2}
    \overline{\lambda} \gamma^\mu \lambda
    &=
    \overline{\lambda} \gamma^\mu &&\left[
        \theta_{x} \cos \gamma_L (\gamma_5)^{\theta_{\widetilde{W}_2}} P_L \widetilde{W}_2
        +
        \sin \gamma_L (\gamma_5)^{\theta_{\widetilde{W}_1}} P_L \widetilde{W}_1
    \right]
    \\ &
    &&+
    \overline{\lambda} \gamma^\mu \left[
        \theta_{y} \cos \gamma_R (\gamma_5)^{\theta_{\widetilde{W}_2}} P_R \widetilde{W}_2 
        +
        \sin \gamma_R (\gamma_5)^{\theta_{\widetilde{W}_1}} P_R \widetilde{W}_1    
    \right]
    \\
    &=&&
    \left[
        \theta_{x} \cos \gamma_L \overline{\widetilde{W}}_2 (-\gamma_5)^{\theta_{\widetilde{W}_2}} P_R
    \right]
    \gamma^\mu \left[
        \theta_{x} \cos \gamma_L (\gamma_5)^{\theta_{\widetilde{W}_2}} P_L \widetilde{W}_2
        +
        \sin \gamma_L (\gamma_5)^{\theta_{\widetilde{W}_1}} P_L \widetilde{W}_1
    \right]
    \\ &
    &&+
    \left[
        \sin \gamma_L \overline{\widetilde{W}}_1 (-\gamma_5)^{\theta_{\widetilde{W}_1}} P_R
    \right]
    \gamma^\mu \left[
        \theta_{x} \cos \gamma_L (\gamma_5)^{\theta_{\widetilde{W}_2}} P_L \widetilde{W}_2
        +
        \sin \gamma_L (\gamma_5)^{\theta_{\widetilde{W}_1}} P_L \widetilde{W}_1
    \right]
    \\ &
    &&+
    \left[ 
        \theta_{y} \cos \gamma_R \overline{\widetilde{W}}_2 (-\gamma_5)^{\theta_{\widetilde{W}_2}} P_L
    \right]
    \gamma^\mu \left[
        \theta_{y} \cos \gamma_R (\gamma_5)^{\theta_{\widetilde{W}_2}} P_R \widetilde{W}_2 
        +
        \sin \gamma_R (\gamma_5)^{\theta_{\widetilde{W}_1}} P_R \widetilde{W}_1    
    \right]
    \\ &
    &&+
    \left[
        \sin \gamma_R \overline{\widetilde{W}}_1 (-\gamma_5)^{\theta_{\widetilde{W}_1}} P_L    
    \right]
    \gamma^\mu \left[
        \theta_{y} \cos \gamma_R (\gamma_5)^{\theta_{\widetilde{W}_2}} P_R \widetilde{W}_2 
        +
        \sin \gamma_R (\gamma_5)^{\theta_{\widetilde{W}_1}} P_R \widetilde{W}_1    
    \right]
    \\
    &=&&
    \left[
        \overline{\widetilde{W}}_1
        \gamma^\mu 
        \left(
            x_c
            -
            y_c
            \gamma_5        
        \right)
        \widetilde{W}_1
    \right]
    +
    \left[
        \overline{\widetilde{W}}_2
        \gamma^\mu 
        \left(
            x_s
            -
            y_s
            \gamma_5
        \right)
        \widetilde{W}_2
    \right]
    +
    \Bigg[
        \overline{\widetilde{W}}_1
        \gamma^\mu 
        \left( 
            x
            -
            y
            \gamma_5    
        \right)
        \widetilde{W}_2
        +
        \overline{\widetilde{W}}_2
        \gamma^\mu 
        \left(
            x
            -
            y
            \gamma_5
        \right)
        \widetilde{W}_1
    \Bigg]
\end{alignat*}
where we made the definitions 
\begin{align}
    x_c
    &\equiv 
    \frac{1}{2}
    \left( 
        \sin^2 \gamma_L 
        +
        \sin^2 \gamma_R 
    \right)
    \\
    y_c
    &\equiv 
    \frac{1}{2}
    \left( 
        \sin^2 \gamma_L 
        -
        \sin^2 \gamma_R
    \right)
    \\
    x_s
    &\equiv 
    \frac{1}{2}
    \left( 
        \cos^2 \gamma_L 
        +
        \cos^2 \gamma_R 
    \right)
    \\
    y_s
    &\equiv 
    \frac{1}{2}
    \left( 
        \cos^2 \gamma_L
        -
        \cos^2 \gamma_R 
    \right)
    \\
    x
    &\equiv
    \frac{1}{2}
    \left(
        (-1)^{\theta_{\widetilde{W}_1} + \theta_{\widetilde{W}_2}} 
        \theta_{x} \sin \gamma_L \cos \gamma_L 
        +
        \theta_{y} \sin \gamma_R \cos \gamma_R 
    \right)
    \\
    y 
    &\equiv
    \frac{1}{2}
    \left(
        (-1)^{\theta_{\widetilde{W}_1} + \theta_{\widetilde{W}_2}} 
        \theta_{x} \sin \gamma_L \cos \gamma_L 
        -
        \theta_{y} \sin \gamma_R \cos \gamma_R 
    \right)
\end{align}
We therefore have the interaction terms for the charginos:
\begin{alignat}{2}
    \mathcal{L}_{\phi \overline{\widetilde{W}}_1 \widetilde{W}_1}
    &=
    \frac{4 \lambda_{SU(2)}}{\sqrt{2} M_*} 
    &&i
    \left(
        \phi_R 
        \overline{\widetilde{W}}_1
        \gamma^\mu 
        \left(
            x_c
            -
            y_c
            \gamma_5
        \right)
        \partial_\mu
        \widetilde{W}_1
    \right)
    \\
    \mathcal{L}_{\phi \overline{\widetilde{W}}_2 \widetilde{W}_2}
    &=
    \frac{4 \lambda_{SU(2)}}{\sqrt{2} M_*} 
    &&i
    \left(
        \phi_R
        \overline{\widetilde{W}}_2
        \gamma^\mu 
        \left(
            x_s
            -
            y_s
            \gamma_5
        \right)
        \partial_\mu
        \widetilde{W}_2
    \right)
    \\
    \mathcal{L}_{\phi \overline{\widetilde{W}}_1 \widetilde{W}_2}
    &=
    \frac{4 \lambda_{SU(2)}}{\sqrt{2} M_*} 
    &&i
    \Big(
        \phi_R 
        \overline{\widetilde{W}}_1
        \gamma^\mu 
        \left( 
            x
            -
            y
            \gamma_5    
        \right)
        \partial_\mu
        \widetilde{W}_2
        +
        \phi_R 
        \overline{\widetilde{W}}_2
        \gamma^\mu 
        \left(
            x
            -
            y
            \gamma_5
        \right)
        \partial_\mu
        \widetilde{W}_1
    \Big)
\end{alignat}
These have then the corresponding matrix elements
\begin{align}
    i \mathcal{M}_{\phi_R \rightarrow \overline{\widetilde{W}}_1 \widetilde{W}_1}
    &=
    \frac{4 \lambda_{SU(2)}}{\sqrt{2} M_*}
    \overline{u}^s(k)
    \left(
        i 
        \slashed{k}
        (
            x_c 
            - 
            y_c 
            \gamma_5
        )
    \right) 
    v^{s'}(k') 
    \\
    i \mathcal{M}_{\phi_R \rightarrow \overline{\widetilde{W}}_2 \widetilde{W}_2}
    &=
    \frac{4 \lambda_{SU(2)}}{\sqrt{2} M_*}
    \overline{u}^s(k)
    \left(
        i 
        \slashed{k}
        (
            x_s 
            - 
            y_s 
            \gamma_5
        )
    \right) 
    v^{s'}(k') 
    \\
    i \mathcal{M}_{\phi_R \rightarrow \overline{\widetilde{W}}_1 \widetilde{W}_2}
    &=
    \frac{4 \lambda_{SU(2)}}{\sqrt{2} M_*}
    \overline{u}^s(k)
    \left(
        i 
        \slashed{k}
        (
            x
            - 
            y
            \gamma_5
        )
    \right) 
    v^{s'}(k') 
\end{align}
where the matrix element for $\phi_R \rightarrow \overline{\widetilde{W}}_2 \widetilde{W}_1$ is identical to $\phi_R \rightarrow \overline{\widetilde{W}}_1 \widetilde{W}_2$.
Assuming no additional contributions to these decay channels, the squared matrix elements are then
\begin{alignat}{2}
    \left| 
        \mathcal{M}_{\phi_R \rightarrow \overline{\widetilde{W}}_1 \widetilde{W}_1}
    \right|^2
    &=&&
    16 \lambda_{SU(2)}^2
    \frac{
        m_{\widetilde{W}_1}^2
        m_\phi^2
    }{
        M_*^2
    }
    \left[
        x_c^2
        +
        y_c^2
        \left(
            1
            - 
            4
            \left(
                \frac{
                    m_{\widetilde{W}_1}
                }{m_\phi}
            \right)^2
        \right)
    \right]
    \\
    \left| 
        \mathcal{M}_{\phi_R \rightarrow \overline{\widetilde{W}}_2 \widetilde{W}_2}
    \right|^2
    &=&&
    16 \lambda_{SU(2)}^2
    \frac{
        m_{\widetilde{W}_2}^2
        m_\phi^2
    }{M_*^2}
    \left[
        x_s^2
        +
        y_s^2
        \left(
            1
            - 
            4
            \left(
                \frac{
                    m_{\widetilde{W}_2}
                }{m_\phi}
            \right)^2
        \right)
    \right]
    \\
    \left| 
        \mathcal{M}_{\phi_R \rightarrow \widetilde{W}_1 \widetilde{W}_2}
    \right|^2
    &=&&
    16 \lambda_{SU(2)}^2
    \frac{
        \left(
            m_{\widetilde{W}_1}^2
            +
            m_{\widetilde{W}_2}^2
        \right)
        m_\phi^2
    }{M_*^2}
    \nonumber
    \\ &
    && \times
    \left[
        x^2
        \left(
            1
            - 
            \left(
                \frac{
                    m_{\widetilde{W}_2}
                    - 
                    m_{\widetilde{W}_1}
                }{m_\phi}
            \right)^2
        \right)
        +
        y^2
        \left(
            1
            - 
            \left(
                \frac{
                    m_{\widetilde{W}_2}
                    +
                    m_{\widetilde{W}_1}
                }{m_\phi}
            \right)^2
        \right)
    \right]
\end{alignat}
where we summed over both $\phi \rightarrow \overline{\widetilde{W}}_1 \widetilde{W}_2$ and $\phi \rightarrow \overline{\widetilde{W}}_2 \widetilde{W}_1$ final states for the last entry.
These then have the corresponding decay widths 
\begin{alignat}{2}
    \Gamma_{\phi_R \rightarrow \overline{\widetilde{W}}_1 \widetilde{W}_1}
    &=&&
    \frac{
        \lambda_{SU(2)}^2
    }{
        \pi 
    } 
    \frac{
        m_{\widetilde{W}_1}^2
        m_\phi
    }{
        M_*^2
    }
    \left[
        x_c^2
        +
        y_c^2
        \left(
            1
            - 
            4
            \left(
                \frac{
                    m_{\widetilde{W}_1}
                }{m_\phi}
            \right)^2
        \right)
    \right]
    \lambda^{1/2}
    \left(
        1,
        \frac{
            m_{\widetilde{W}_1}^2
        }{ 
            m_\phi^2
        },
        \frac{
            m_{\widetilde{W}_1}^2
        }{ 
            m_\phi^2
        }
    \right)
    \\
    \Gamma_{\phi_R \rightarrow \overline{\widetilde{W}}_2 \widetilde{W}_2}
    &=&&
    \frac{
        \lambda_{SU(2)}^2
    }{
        \pi 
    } 
    \frac{
        m_{\widetilde{W}_2}^2
        m_\phi
    }{M_*^2}
    \left[
        x_s^2
        +
        y_s^2
        \left(
            1
            - 
            4
            \left(
                \frac{
                    m_{\widetilde{W}_2}
                }{m_\phi}
            \right)^2
        \right)
    \right]
    \lambda^{1/2}
    \left(
        1,
        \frac{
            m_{\widetilde{W}_2}^2
        }{ 
            m_\phi^2
        },
        \frac{
            m_{\widetilde{W}_2}^2
        }{ 
            m_\phi^2
        }
    \right)
    \\
    \Gamma_{\phi_R \rightarrow \widetilde{W}_1 \widetilde{W}_2}
    &=&&
    \frac{
        \lambda_{SU(2)}^2
    }{
        \pi 
    } 
    \frac{
        \left(
            m_{\widetilde{W}_1}^2
            +
            m_{\widetilde{W}_2}^2
        \right)
        m_\phi
    }{M_*^2}
    \nonumber
    \\ &
    && \times
    \left[
        x^2
        \left(
            1
            - 
            \left(
                \frac{
                    m_{\widetilde{W}_2}
                    - 
                    m_{\widetilde{W}_1}
                }{m_\phi}
            \right)^2
        \right)
        +
        y^2
        \left(
            1
            - 
            \left(
                \frac{
                    m_{\widetilde{W}_2}
                    +
                    m_{\widetilde{W}_1}
                }{m_\phi}
            \right)^2
        \right)
    \right]
    \lambda^{1/2}
    \left(
        1,
        \frac{
            m_{\widetilde{W}_1}^2
        }{ 
            m_\phi^2
        },
        \frac{
            m_{\widetilde{W}_2}^2
        }{ 
            m_\phi^2
        }
    \right)
\end{alignat}

\subsection{Modulus decay into electroweak vector bosons}

We now look at the decay of the modulus into the electroweak vector bosons.
Starting with the second term of Eq.~\ref{modulusGaugeGauginoLagrangian}, we can insert the $SU(2)_L$ and $U(1)_Y$ gauge fields:
\begin{alignat*}{2}
    \mathcal{L}_G
    &\supset
    -\frac{4\lambda_{SU(2)}}{\sqrt{2} M_*} &&
    \phi_R
    \left(
        \frac{1}{4} F_{\mu \nu A} F_A^{\mu \nu}
    \right)
    -
    \frac{4\lambda_{U(1)}}{\sqrt{2} M_*} 
    \phi_R
    \left(
        \frac{1}{4} B_{\mu \nu } B^{\mu \nu}
    \right)
    \\
    &=
    -\frac{4\lambda_{SU(2)}}{\sqrt{2} M_*} && \frac{1}{2} 
    \phi_R
    \left(
        \partial_\mu W_{\nu i} 
        \partial^\mu W_i^\nu 
        -
        \partial_\mu W_{\nu i} 
        \partial^\nu W_i^\mu 
        -
        2
        g \epsilon_{ijk} 
        (\partial_\mu W_{\nu i})
        W_j^\mu W_k^\nu
        + 
        \frac{1}{2}
        g^2 \epsilon_{ijk} \epsilon_{imn} 
        W_{\mu j} W_{\nu k}
        W_m^\mu W_n^\nu
    \right)
    \\ &
    &&-
    \frac{4\lambda_{U(1)}}{\sqrt{2} M_*} \frac{1}{2} 
    \phi_R
    \left(
        \partial_\mu B_\nu 
        \partial^\mu B^\nu 
        -
        \partial_\nu B_\mu
        \partial^\mu B^\nu 
    \right)
\end{alignat*}
We see that there are 3-point, 4-point, and 5-point interactions between the modulus and gauge fields.

\subsubsection{Relevant matrix element formulae for electroweak vector bosons}
In this section, we work out the general interactions that will be relevant for the electroweak vector bosons.
All of these formulae are worked out in the rest frame of the modulus.

We start with the 3-point interaction formulae.
The first general interaction is of the form:
\begin{align}
    \mathcal{L}_I
    &=
    -
    g
    \phi_R 
    \left(
        \partial_\mu 
        A_\nu^a
        \partial^\mu 
        B^{\nu \, b}
        -
        \partial_\mu 
        A_\nu^a
        \partial^\nu 
        B^{\mu \, b}
    \right)
    \\
    &=
    -
    g
    \phi_R 
    \left(
        \partial_\mu 
        A_\nu^a
        \partial_\rho
        B_\sigma^{b}
        g^{\mu \rho} 
        g^{\nu \sigma}
        -
        \partial_\mu 
        A_\nu^a
        \partial_\rho
        B_\sigma^{b}
        g^{\mu \sigma} 
        g^{\nu \rho}
    \right)
\end{align}
which corresponds to the diagram:
\newline
\begin{center}
    \begin{tikzpicture}
        \begin{feynman}
            \vertex (a) {\(\phi\)} ;
            \vertex [right=of a] (b) ;
            \vertex [below right=of b] (f1) {\( A_\nu^{a} \)};
            \vertex [above right=of b] (f2) {\( B_\sigma^{b} \)};
            \diagram* {
                (a) -- [scalar, momentum'={\(q\)}] (b),
                (b) -- [boson, momentum'={\(k'\)}] (f1), 
                (b) -- [boson] (f2),
            };
        \end{feynman}
    \end{tikzpicture}
\end{center}
Taking the incoming momenta of the modulus to be $q$, and the outgoing momenta of $A_\nu^{a}$ and $B_\sigma^{b}$ to be $k$ and $k'$, respectively, we have a vertex factor from the first term of $-ig ( i k'_\mu ) ( i k_\rho ) g^{\mu \rho} g^{\nu \sigma} = + i g k'_\mu k^\mu g^{\nu \sigma}$.
Since the second interaction term is of the same form but with the exchange $g^{\mu \rho} g^{\nu \sigma} \rightarrow -g^{\mu \sigma} g^{\nu \rho}$, we have its vertex factor of $+ ig (i k'_\mu) (i k_\rho) g^{\mu \sigma} g^{\nu \rho} = -i g k'^\sigma k^\nu $.
If both vector bosons are identical, each vertex factor should be multiplied by an extra factor of 2.
The associated matrix element is then 
\begin{align*}
    i \mathcal{M}^{\nu \sigma}
    \epsilon_\sigma^* (k) 
    \epsilon_\nu^* (k')
    &=
    i g 
    \left(
        k'_\mu k^\mu g^{\nu \sigma}
        -
        k'^\sigma k^\nu 
    \right)
    \epsilon_\sigma^* (k) 
    \epsilon_\nu^* (k')
\end{align*}
With our momenta conventions, assuming no additional contributions to this decay channel, the squared matrix element is then 
\begin{align*}
    \left| 
        \mathcal{M}^{\nu \sigma}
        \epsilon_\sigma^* (k) 
        \epsilon_\nu^* (k')
    \right|^2
    &=
    g^2
    \left(
        k'_\mu k^\mu g^{\nu \sigma}
        -
        k'^\sigma k^\nu 
    \right)
    \epsilon_\sigma^* (k) 
    \epsilon_\nu^* (k')
    \left(
        k'_\alpha k^\alpha g^{\beta \gamma}
        -
        k'^\gamma k^\beta
    \right)
    \epsilon_\gamma (k) 
    \epsilon_\beta (k')
\end{align*}
and summing over polarizations (assuming both $A$ and $B$ are massive, so we use $\sum_\lambda \epsilon_\mu^* \epsilon_\nu = - g_{\mu \nu} + \frac{p^\mu p^\nu}{m^2}$), this becomes
\begin{alignat*}{2}
    \left| 
        \mathcal{M}^{\nu \sigma}
        \epsilon_\sigma^* (k) 
        \epsilon_\nu^* (k')
    \right|^2
    &=&&
    g^2
    \left(
        k'_\mu 
        k^\mu 
        g^{\nu \sigma}
        -
        k'^\sigma 
        k^\nu 
    \right)
    \left(
        k'_\alpha 
        k^\alpha 
        g^{\beta \gamma}
        -
        k'^\gamma
        k^\beta
    \right)
    \left(
        g_{\sigma \gamma} 
        -
        \frac{
            k_\sigma 
            k_\gamma
        }{m_A^2}
    \right)
    \left(
        g_{\nu \beta} 
        -
        \frac{
            k'_\nu 
            k'_\beta
        }{m_B^2}
    \right)
    \\
    &=&&
    g^2
    \left(
        k'_\mu 
        k^\mu 
        g^{\nu \sigma}
        -
        k'^\sigma 
        k^\nu 
    \right)
    \left(
        k'_\alpha 
        k^\alpha 
        \delta_\sigma^\beta
        -
        k'_\sigma
        k^\beta
    \right)
    \left(
        g_{\nu \beta} 
        -
        \frac{
            k'_\nu 
            k'_\beta
        }{m_B^2}
    \right)
    \\
    &=&&
    g^2
    \left(
        2
        (
            k'_\mu 
            k^\mu
        )^2 
        +
        k'_\sigma
        k'^\sigma 
        k_\nu
        k^\nu 
    \right)
\end{alignat*}
In the rest frame of the modulus, we can evaluate $2 k'_\mu k^\mu = m_\phi^2 - m_A^2 - m_B^2$, giving us our squared matrix element: 
\begin{align} \label{matrixElementSqrDiMassiveBoson}
    \left| 
        \mathcal{M}^{\nu \sigma}
        \epsilon_\sigma^* (k) 
        \epsilon_\nu^* (k')
    \right|^2
    &=
    \frac{
        g^2
        m_\phi^4
    }{2}
    \left(
        \left(
            1
            -
            \frac{
                m_A^2
                +
                m_B^2    
            }{
                m_\phi^2
            }
        \right)^2 
        +
        2
        \frac{
            m_A^2
            m_B^2    
        }{
            m_\phi^4
        }
    \right)    
\end{align}
We also evaluate the squared matrix element if one of either $A$ or $B$ is massless.
We'll take $B$ to be massless (and from an Abelian group) and $A$ to be massive, where we now use $\sum_\lambda \epsilon_\mu^* \epsilon_\nu = -g_{\mu \nu}$ for the massless boson:
\begin{alignat*}{2}
    \left| 
        \mathcal{M}^{\nu \sigma}
        \epsilon_\sigma^* (k) 
        \epsilon_\nu^* (k')
    \right|^2
    &=&&
    g^2
    \left(
        k'_\mu 
        k^\mu 
        g^{\nu \sigma}
        -
        k'^\sigma 
        k^\nu 
    \right)
    \left(
        k'_\alpha 
        k^\alpha 
        g^{\beta \gamma}
        -
        k'^\gamma
        k^\beta
    \right)
    \left(
        g_{\sigma \gamma} 
        -
        \frac{
            k_\sigma 
            k_\gamma
        }{m_A^2}
    \right)
    \left(
        g_{\nu \beta} 
    \right)
    \\
    &=&&
    g^2
    \left(
        2
        \left(
            k'_\mu 
            k^\mu     
        \right)^2
        +
        k'_\sigma
        k'^\sigma 
        k_\nu
        k^\nu 
    \right)
\end{alignat*}
The squared matrix element then is given by 
\begin{align} \label{matrixElementSqrMassiveMasslessBoson}
    \left| 
        \mathcal{M}^{\nu \sigma}
        \epsilon_\sigma^* (k) 
        \epsilon_\nu^* (k')
    \right|^2
    &=
    \frac{
        g^2
        m_\phi^4
    }{2}
    \left(
        1
        -
        \frac{
            m_A^2
        }{
            m_\phi^2
        }
    \right)^2 
\end{align}
If we now take both $A$ and $B$ to be massless, the squared matrix element is 
\begin{alignat*}{2}
    \left| 
        \mathcal{M}^{\nu \sigma}
        \epsilon_\sigma^* (k) 
        \epsilon_\nu^* (k')
    \right|^2
    &=&&
    g^2
    \left(
        k'_\mu 
        k^\mu 
        g^{\nu \sigma}
        -
        k'^\sigma 
        k^\nu 
    \right)
    \left(
        k'_\alpha 
        k^\alpha 
        g^{\beta \gamma}
        -
        k'^\gamma
        k^\beta
    \right)
    \left(
        g_{\sigma \gamma} 
    \right)
    \left(
        g_{\nu \beta} 
    \right)    
    \\
    &=&&
    g^2
    \left(
        2
        k'_\mu 
        k^\mu 
        k'_\alpha 
        k^\alpha 
        +
        k'_\sigma
        k'^\sigma 
        k_\nu
        k^\nu 
    \right)
\end{alignat*}
and hence the squared matrix element is 
\begin{align} \label{matrixElementSqrDiMasslessBoson}
    \left| 
        \mathcal{M}^{\nu \sigma}
        \epsilon_\sigma^* (k) 
        \epsilon_\nu^* (k')
    \right|^2
    &=
    \frac{
        g^2
        m_\phi^4
    }{2}
\end{align}
For Eq's~\ref{matrixElementSqrDiMassiveBoson}, \ref{matrixElementSqrMassiveMasslessBoson}, and \ref{matrixElementSqrDiMasslessBoson}, again if $A$ and $B$ are identical, the entire squared matrix element must be multiplied by an additional factor of $4$.

\subsubsection{3-point interactions}
We first look at the 3-point interactions between the modulus and a gauge boson pair.
The relevant terms are then 
\begin{alignat*}{2}
    \mathcal{L}_{\phi V V}
    &=
    -\frac{4\lambda_{SU(2)}}{\sqrt{2} M_*} && \frac{1}{2} 
    \phi_R
    \left(
        \partial_\mu W_{\nu i} 
        \partial^\mu W_i^\nu 
        -
        \partial_\mu W_{\nu i} 
        \partial^\nu W_i^\mu
    \right)
    -
    \frac{4\lambda_{U(1)}}{\sqrt{2} M_*} \frac{1}{2} 
    \phi_R
    \left(
        \partial_\mu B_\nu 
        \partial^\mu B^\nu 
        -
        \partial_\nu B_\mu
        \partial^\mu B^\nu 
    \right)
\end{alignat*}
Isolating the charged vector bosons, we have the interaction between the modulus and the $W^\pm$ bosons:
\begin{alignat}{2}
    \mathcal{L}_{\phi W^+ W^-}
    &=
    -\frac{4\lambda_{SU(2)}}{\sqrt{2} M_*} && \frac{1}{2} 
    \phi_R
    \left(
        \partial_\mu W_{\nu 1} 
        \partial^\mu W_1^\nu 
        -
        \partial_\mu W_{\nu 1} 
        \partial^\nu W_1^\mu
        +
        \partial_\mu W_{\nu 2} 
        \partial^\mu W_2^\nu 
        -
        \partial_\mu W_{\nu 2} 
        \partial^\nu W_2^\mu
    \right) \nonumber
    \\
    &=
    -\frac{4\lambda_{SU(2)}}{\sqrt{2} M_*} &&
    \phi_R
    \left(
        \partial_\mu W_{\nu}^+ 
        \partial^\mu W^{\nu -} 
        -
        \partial_\mu W_{\nu}^+ 
        \partial^\nu W^{\mu -} 
    \right)
\end{alignat}
We can then evaluate the squared matrix element from Eq. \ref{matrixElementSqrDiMassiveBoson}, assuming no additional contributions to this decay channel:
\begin{align}
    \left| 
        \mathcal{M}_{ \phi_R \rightarrow W^+ W^- }^{\nu \sigma}
        \epsilon_\sigma^* (k) 
        \epsilon_\nu^* (k')
    \right|^2
    &=
    \frac{
        4
        \lambda_{SU(2)}^2
        m_\phi^4
    }{M_*^2}
    \left(
        \left(
            1
            -
            \frac{
                2
                m_W^2
            }{
                m_\phi^2
            }
        \right)^2 
        +
        2
        \frac{
            m_W^4
        }{
            m_\phi^4
        }
    \right)
\end{align}
We then have the decay width to $W^+ W^-$ pairs:
\begin{align}
    \Gamma_{\phi_R \rightarrow W^+ W^-}
    &=
    \frac{
        \lambda_{SU(2)}^2
    }{
        4
        \pi 
    } 
    \frac{
        m_\phi^3
    }{M_*^2}
    \left(
        1
        -
        4
        \frac{
            m_W^2
        }{
            m_\phi^2
        }
        +
        6
        \frac{
            m_W^4
        }{
            m_\phi^4
        }
    \right)
    \lambda^{1/2}
    \left(
        1,
        \frac{
            m_W^2
        }{ 
            m_\phi^2
        },
        \frac{
            m_W^2
        }{ 
            m_\phi^2
        }
    \right)
\end{align}

We now turn to the neutral vector bosons:
\begin{alignat*}{2}
    \mathcal{L}_{\phi VV}
    &\supset&&
    -
    \frac{4\lambda_{SU(2)}}{\sqrt{2} M_*}
    \frac{1}{2} 
    \phi_R
    \left(
        \partial_\mu W_{\nu 3} 
        \partial^\mu W_3^\nu 
        -
        \partial_\mu W_{\nu 3} 
        \partial^\nu W_3^\mu
    \right)
    -
    \frac{4\lambda_{U(1)}}{\sqrt{2} M_*} \frac{1}{2} 
    \phi_R
    \left(
        \partial_\mu B_\nu 
        \partial^\mu B^\nu 
        -
        \partial_\nu B_\mu
        \partial^\mu B^\nu 
    \right)
    \\
    &=&&
    -
    \frac{2\lambda_{SU(2)}}{\sqrt{2} M_*} 
    \phi_R
    \Big(
        \partial_\mu ( \sin \theta_W A_\nu - \cos \theta_W Z_\nu^0 ) 
        \partial^\mu ( \sin \theta_W A^\nu - \cos \theta_W Z^{\nu 0} )
        \\ &
        &&-
        \partial_\mu ( \sin \theta_W A_\nu - \cos \theta_W Z_\nu^0 )
        \partial^\nu ( \sin \theta_W A^\mu - \cos \theta_W Z^{\mu 0} )
    \Big)
    \\ &
    &&-
    \frac{2\lambda_{U(1)}}{\sqrt{2} M_*}
    \phi_R
    \Big(
        \partial_\mu ( \cos \theta_W A_\nu + \sin \theta_W Z_{\nu}^0 )
        \partial^\mu ( \cos \theta_W A^\nu + \sin \theta_W Z^{\nu 0} ) 
        \\ &
        &&-
        \partial_\nu ( \cos \theta_W A_\mu + \sin \theta_W Z_{\mu}^0 )
        \partial^\mu ( \cos \theta_W A^\nu + \sin \theta_W Z^{\nu 0} )
    \Big)
\end{alignat*}
From this, we can extract the $\phi Z^0 Z^0$ interaction:
\begin{alignat}{2}
    \mathcal{L}_{\phi Z^0 Z^0}
    &=&&
    -
    \frac{2\lambda_{SU(2)}}{\sqrt{2} M_*}
    \phi_R
    \cos^2 \theta_W 
    \Big(
        \partial_\mu Z_\nu^0 
        \partial^\mu Z^{\nu 0}
        -
        \partial_\mu Z_\nu^0 
        \partial^\nu Z^{\mu 0}
    \Big) \nonumber
    \\ &
    &&-
    \frac{2\lambda_{U(1)}}{\sqrt{2} M_*} 
    \phi_R
    \sin^2 \theta_W 
    \Big(
        \partial_\mu Z_{\nu}^0
        \partial^\mu Z^{\nu 0} 
        -
        \partial_\nu Z_{\mu}^0
        \partial^\mu Z^{\nu 0}
    \Big) \nonumber
    \\
    &=&&
    -
    \frac{2}{\sqrt{2} M_*} 
    \left( 
        \lambda_{SU(2)}
        \cos^2 \theta_W 
        +
        \lambda_{U(1)}
        \sin^2 \theta_W 
    \right)
    \phi_R
    \Big(
        \partial_\mu Z_\nu^0 
        \partial^\mu Z^{\nu 0}
        -
        \partial_\mu Z_\nu^0 
        \partial^\nu Z^{\mu 0}
    \Big)
\end{alignat}
The squared matrix element, assuming no other contributions to the decay, is taken from Eq. \ref{matrixElementSqrDiMassiveBoson}:
\begin{align}
    \left| 
        \mathcal{M}_{ \phi_R \rightarrow Z^0 Z^0 }^{\nu \sigma}
        \epsilon_\sigma^* (k) 
        \epsilon_\nu^* (k')
    \right|^2
    &=
    \frac{
        4
        \left(
            \lambda_{SU(2)}
            \cos^2 \theta_W 
            +
            \lambda_{U(1)}
            \sin^2 \theta_W 
        \right)^2
        m_\phi^4
    }{M_*^2} 
    \left(
        \left(
            1
            -
            \frac{
                2
                m_Z^2
            }{
                m_\phi^2
            }
        \right)^2 
        +
        2
        \frac{
            m_Z^4
        }{
            m_\phi^4
        }
    \right)
\end{align}
The decay width to $Z^0$ pairs is then given by:
\begin{align}
    \Gamma_{\phi_R \rightarrow Z^0 Z^0}
    &=
    \frac{
        \left(
            \lambda_{SU(2)}
            \cos^2 \theta_W 
            +
            \lambda_{U(1)}
            \sin^2 \theta_W 
        \right)^2
    }{
        8
        \pi 
    } 
    \frac{
        m_\phi^3
    }{M_*^2} 
    \left(
        1
        -
        4
        \frac{
            m_Z^2
        }{
            m_\phi^2
        }
        +
        6
        \frac{
            m_Z^4
        }{
            m_\phi^4
        }
    \right)
    \lambda^{1/2}
    \left(
        1,
        \frac{
            m_Z^2
        }{ 
            m_\phi^2
        },
        \frac{
            m_Z^2
        }{ 
            m_\phi^2
        }
    \right)
\end{align}
where we divide by an additional factor of 2 since the final state particles are indistinguishable.

The $\phi \gamma \gamma$ interaction follows similarly, with the interaction term:
\begin{alignat}{2}
    \mathcal{L}_{\phi \gamma \gamma}
    &=&&
    -
    \frac{2\lambda_{SU(2)}}{\sqrt{2} M_*}
    \phi_R
    \sin^2 \theta_W
    \Big(
        \partial_\mu A_\nu 
        \partial^\mu A^\nu
        -
        \partial_\mu A_\nu 
        \partial^\nu A^\mu
    \Big) \nonumber
    \\ &
    &&-
    \frac{2\lambda_{U(1)}}{\sqrt{2} M_*} 
    \phi_R
    \cos^2 \theta_W
    \Big(
        \partial_\mu A_\nu
        \partial^\mu A^\nu
        -
        \partial_\nu A_\mu
        \partial^\mu A^\nu
    \Big) \nonumber
    \\
    &=&&
    -
    \frac{2}{\sqrt{2} M_*} 
    \left(
        \lambda_{SU(2)}
        \sin^2 \theta_W
        +
        \lambda_{U(1)}
        \cos^2 \theta_W
    \right)
    \phi_R
    \Big(
        \partial_\mu A_\nu 
        \partial^\mu A^\nu
        -
        \partial_\mu A_\nu 
        \partial^\nu A^\mu
    \Big)
\end{alignat}
which has an associated squared matrix element from Eq. \ref{matrixElementSqrDiMasslessBoson}:
\begin{align}
    \left| 
        \mathcal{M}_{ \phi_R \rightarrow \gamma \gamma }^{\nu \sigma}
        \epsilon_\sigma^* (k) 
        \epsilon_\nu^* (k')
    \right|^2
    &=
    \frac{
        4
        m_\phi^4
    }{M_*^2} 
    \left(
        \lambda_{SU(2)}
        \sin^2 \theta_W
        +
        \lambda_{U(1)}
        \cos^2 \theta_W
    \right)^2
\end{align}
The decay width to photon pairs is then given by:
\begin{align}
    \Gamma_{\phi_R \rightarrow \gamma \gamma}
    &=
    \frac{
        \left(
            \lambda_{SU(2)}
            \sin^2 \theta_W
            +
            \lambda_{U(1)}
            \cos^2 \theta_W
        \right)^2
    }{
        8
        \pi 
    } 
    \frac{
        m_\phi^3
    }{M_*^2} 
\end{align}
where we again divide by an additional factor of 2 as the final state particles are indistinguishable.

Finally, we also need to include the $\phi Z^0 \gamma$ interaction:
\begin{alignat}{2}
    \mathcal{L}_{\phi Z^0 \gamma}
    &=&&
    \frac{4\lambda_{SU(2)}}{\sqrt{2} M_*}
    \sin \theta_W \cos \theta_W
    \phi_R
    \Big(
        \partial_\mu Z_{\nu}^0
        \partial^\mu A^\nu
        -
        \partial_\mu A_\nu
        \partial^\nu Z^{\mu 0}
    \Big) \nonumber
    \\ &
    &&-
    \frac{4\lambda_{U(1)}}{\sqrt{2} M_*}
    \sin \theta_W 
    \cos \theta_W 
    \phi_R
    \Big(
        \partial_\mu Z_{\nu}^0
        \partial^\mu A^\nu
        -
        \partial_\nu Z_{\mu}^0
        \partial^\mu A^\nu
    \Big) \nonumber
    \\
    &=&&
    \frac{2}{\sqrt{2} M_*}
    \sin 2 \theta_W 
    \left( 
        \lambda_{SU(2)}
        -
        \lambda_{U(1)}
    \right)
    \phi_R
    \Big(
        \partial_\mu Z_{\nu}^0
        \partial^\mu A^\nu
        -
        \partial_\mu A_\nu
        \partial^\nu Z^{\mu 0}
    \Big)
\end{alignat}
The corresponding matrix element from Eq. \ref{matrixElementSqrMassiveMasslessBoson} is 
\begin{align}
    \left| 
        \mathcal{M}_{ \phi_R \rightarrow \gamma Z^0 }^{\nu \sigma}
        \epsilon_\sigma^* (k) 
        \epsilon_\nu^* (k')
    \right|^2
    &=
    \frac{
        m_\phi^4
    }{
        M_*^2
    }
    \sin^2 2 \theta_W 
    \left(
        \lambda_{SU(2)}
        -
        \lambda_{U(1)}
    \right)^2
    \left(
        1
        -
        \frac{
            m_Z^2
        }{
            m_\phi^2
        }
    \right)^2 
\end{align}
Note that, as expected, if the couplings between $SU(2)_L$ and $U(1)_Y$ are equal, the $\phi Z^0 \gamma$ interaction disappears, and the $\phi \gamma \gamma$ and $\phi Z^0 Z^0$ interactions are independent of $\theta_W$.
The decay width to the $Z^0 \gamma$ state is then:
\begin{align}
    \Gamma_{\phi_R \rightarrow Z^0 \gamma}
    &=
    \frac{
        \sin^2 2 \theta_W 
        \left(
            \lambda_{SU(2)}
            -
            \lambda_{U(1)}
        \right)^2
    }{
        16
        \pi 
    } 
    \frac{
        m_\phi^3
    }{
        M_*^2
    }
    \left(
        1
        -
        \frac{
            m_Z^2
        }{
            m_\phi^2
        }
    \right)^3
\end{align}
where we used $\lambda^{1/2}
\left(
    1,
    0,
    \frac{
        m_Z^2
    }{ 
        m_\phi^2
    }
\right)
=
\left( 
    1
    -
    \frac{
        m_Z^2
    }{ 
        m_\phi^2
    }
\right)
$.

\subsection{Modulus decay into gluinos}

The modulus decay into gluinos is given by the first term in Eq. \ref{modulusGaugeGauginoLagrangian}:
\begin{align}
    \mathcal{L}_{\phi \widetilde{g} \widetilde{g}}
    &=
    \frac{4 \lambda_{SU(3)}}{\sqrt{2} M_*}
    \phi_R
    \frac{i}{2}
    \overline{\lambda}_A
    \slashed{D}_{AC}
    \lambda_C
    \nonumber
    \\
    &=
    \frac{2 \lambda_{SU(3)}}{\sqrt{2} M_*}
    \phi_R
    \left(
        i
        \overline{\widetilde{g}}
        \gamma^\mu 
        \partial_\mu 
        \widetilde{g}
        - 
        \frac{g_s}{2}
        \overline{\widetilde{g}}
        \gamma^\mu 
        \lambda_A
        G_{A \mu}
        \widetilde{g}
    \right)
    \label{modulusGluinoLagrangian}
\end{align}

\subsubsection{3-point interactions}

We look first at the 3-point interactions between the modulus and a gluino pair.
The interaction term relevant for this process is 
\begin{align*}
    \mathcal{L}_{\phi \widetilde{g} \widetilde{g}}
    &=
    i
    \frac{2 \lambda_{SU(3)}}{\sqrt{2} M_*}
    \phi_R
    \overline{\widetilde{g}}_A
    \gamma^\mu 
    \partial_\mu 
    \widetilde{g}_A
\end{align*}
Since there is no symmetry breaking in the $SU(3)$ sector, we can immediately write down the matrix element to gluinos using Eq. \ref{matrixElementGaugino}:
\begin{align}
    i \mathcal{M}_{\phi_R \rightarrow \overline{\widetilde{g}}_A \widetilde{g}_A }
    &=
    -
    \overline{u}^s (k)
    \left(
        \frac{2 \lambda_{SU(3)}}{\sqrt{2} M_*}
        \slashed{k}
        v^{s'} (k')
    \right)
\end{align}
and so we get the squared matrix element from Eq. \ref{matrixElementSqrGaugino}:
\begin{align}
    \left| 
        \mathcal{M}_{\phi_R \rightarrow \overline{\widetilde{g}} \widetilde{g} }
    \right|^2
    &=
    32
    \lambda_{SU(3)}^2
    \frac{ 
        m_{\widetilde{g}}^2 \,
        m_\phi^2
    }{
        M_*^2
    }
\end{align}
The additional factor of $8$ comes from summing over the color indices, as while the gluinos are a color octet, specific color charges are of no interest for this work.
The computation of the decay width, including a factor $1/2$ for identical final 
state particles, gives
\begin{align}
    \Gamma_{\phi_R \rightarrow \widetilde{g} \widetilde{g}}
    &=
    \frac{
        \lambda_{SU(3)}^2
    }{
        \pi 
    } 
    \frac{ 
        m_{\widetilde{g}}^2 \,
        m_\phi
    }{
        M_*^2
    }
    \lambda^{1/2}
    \left(
        1,
        \frac{
            m_{\widetilde{g}}^2
        }{ 
            m_\phi^2
        },
        \frac{
            m_{\widetilde{g}}^2
        }{ 
            m_\phi^2
        }
    \right)
\end{align}

\subsection{Modulus decay into gluons}

We can get the modulus decay into gluons from the second term in Eq. \ref{modulusGaugeGauginoLagrangian}:
\begin{alignat}{2}
    \mathcal{L}_{\phi g g}
    &=&&
    -
    \frac{\lambda_{SU(3)}}{\sqrt{2} M_*}
    \phi_R
    G_{\mu \nu A}
    G_A^{\mu \nu}
    \nonumber
    \\
    &=&&
    -
    \frac{2\lambda_{SU(3)}}{\sqrt{2} M_*}
    \phi_R
    \Big(
        \partial_\mu 
        G_{A \nu}
        \partial^\mu 
        G_{A}^\nu
        -
        \partial_\mu 
        G_{A \nu}
        \partial^\nu 
        G_{A}^\mu
        +
        g_s
        f_{ABC}
        \left(
            \partial_\nu 
            G_{A \mu}
            -
            \partial_\mu 
            G_{A \nu}
        \right)
        G_{B}^\mu
        G_{C}^\nu
        \nonumber
        \\ &
        &&+
        \frac{g_s^2}{2}
        f_{ABC}
        f_{ADE}
        G_{B \mu}
        G_{C \nu}
        G_{D}^\mu
        G_{E}^\nu
    \Big)
    \label{modulusGluonLagrangian}
\end{alignat}

\subsubsection{Relevant matrix element formulae for gluons}
In this section, we work out the general interactions that will be relevant for the gluons.
All of these formulae are worked out in the rest frame of the modulus.

We start with the 3-point interaction formulae.
The first general interaction is of the form:
\begin{align}
    \mathcal{L}_I
    &=
    -
    g
    \phi_R 
    \left(
        \partial_\mu 
        G_\nu^A
        \partial^\mu 
        G^{\nu \, A}
        -
        \partial_\mu 
        G_\nu^A
        \partial^\nu 
        G^{\mu \, A}
    \right)
    \\
    &=
    -
    g
    \phi_R 
    \left(
        \partial_\mu 
        G_\nu^A
        \partial_\rho
        G_\sigma^{A}
        g^{\mu \rho} 
        g^{\nu \sigma}
        -
        \partial_\mu 
        G_\nu^A
        \partial_\rho
        G_\sigma^{A}
        g^{\mu \sigma} 
        g^{\nu \rho}
    \right)
\end{align}
which corresponds to the diagram:
\newline
\begin{center}
    \begin{tikzpicture}
        \begin{feynman}
            \vertex (a) {\(\phi\)} ;
            \vertex [right=of a] (b) ;
            \vertex [below right=of b] (f1) {\( G_\nu^{A} \)};
            \vertex [above right=of b] (f2) {\( G_\sigma^{A} \)};
            \diagram* {
                (a) -- [scalar, momentum'={\(q\)}] (b),
                (b) -- [boson, momentum'={\(k'\)}] (f1), 
                (b) -- [boson] (f2),
            };
        \end{feynman}
    \end{tikzpicture}
\end{center}
Taking the incoming momenta of the modulus to be $q$, and the outgoing momenta of $G_\nu^{A}$ and $G_\sigma^{A}$ to be $k$ and $k'$, respectively, we have a vertex factor from the first term of $-ig ( i k'_\mu ) ( i k_\rho ) g^{\mu \rho} g^{\nu \sigma} = + i g k'_\mu k^\mu g^{\nu \sigma}$.
Since the second interaction term is of the same form but with the exchange $g^{\mu \rho} g^{\nu \sigma} \rightarrow -g^{\mu \sigma} g^{\nu \rho}$, we have its vertex factor of $+ ig (i k'_\mu) (i k_\rho) g^{\mu \sigma} g^{\nu \rho} = -i g k'^\sigma k^\nu $.
Because the gluons have identical color charge, each vertex factor should be multiplied by an extra factor of 2.
The associated matrix element is then 
\begin{align*}
    i \mathcal{M}^{\nu \sigma}
    \epsilon_\sigma^* (k) 
    \epsilon_\nu^* (k')
    &=
    2
    i g 
    \left(
        k'_\mu k^\mu g^{\nu \sigma}
        -
        k'^\sigma k^\nu 
    \right)
    \epsilon_\sigma^* (k) 
    \epsilon_\nu^* (k')
\end{align*}
Since we're dealing with bosons associated with a non-Abelian symmetry, we have to be careful about the polarization sums here.
We first note that only the transverse polarizations are physical, and since the gluons are back-to-back in the rest frame of the modulus, we must have $k'^\sigma \epsilon_\sigma^*(k) = k^\nu \epsilon_\nu^*(k') = 0$.
On summing over the polarizations, we have also $\sum_i \epsilon_i^\mu \epsilon_i^{\nu *} = - g_{\mu \nu} + \frac{k_\mu \overline{k}_\nu + k_\nu \overline{k}_\mu}{k \cdot \overline{k}}$ where we defined $\overline{k} \equiv ( k, 0, 0, -k )$ assuming $k$ is propagating in the $+z$ direction.

This simplifies the physical matrix element to 
\begin{align*}
    i \mathcal{M}^{\nu \sigma}
    \epsilon_\sigma^* (k) 
    \epsilon_\nu^* (k')
    &=
    2
    i g 
    k'_\mu k^\mu g^{\nu \sigma}
    \epsilon_\sigma^* (k) 
    \epsilon_\nu^* (k')
\end{align*}
and the squared matrix element becomes 
\begin{align*}
    \left|
        \mathcal{M}^{\nu \sigma}
        \epsilon_\sigma^* (k) 
        \epsilon_\nu^* (k')
    \right|^2
    &=
    4
    g^2 
    \left(
        \frac{1}{2}
        m_\phi^2
    \right)^2
    g^{\nu \sigma}
    g^{\alpha \beta}
    \sum \limits_{\text{pol.}}
    \epsilon_\sigma^* (k) 
    \epsilon_\alpha (k) 
    \epsilon_\nu^* (k')
    \epsilon_\beta (k')
    \\
    &=
    g^2 
    m_\phi^4
    \left(
        g_{\beta}^{\sigma}
        g_{\sigma}^{\beta}
        -
        \frac{
            k^\sigma \overline{k}_\sigma 
            + 
            k_\sigma \overline{k}^\sigma
        }{
            k \cdot \overline{k}
        }
        - 
        \frac{
            k'_\sigma \overline{k'}^\sigma 
            + 
            k'^\sigma \overline{k'}_\sigma
        }{
            k' \cdot \overline{k'}
        }
        + 
        \frac{
            \left(
                k^\beta \overline{k}_\sigma 
                + 
                k_\sigma \overline{k}^\beta
            \right)
            \left(
                k'_\beta \overline{k'}^\sigma 
                + 
                k'^\sigma \overline{k'}_\beta
            \right)
        }{
            \left(
                k \cdot \overline{k}
            \right)
            \left(
                k' \cdot \overline{k'}
            \right)
        }
    \right)
    \\
    &=
    g^2 
    m_\phi^4
    \left(
        g_{\beta}^{\sigma}
        g_{\sigma}^{\beta}
        -
        2
        - 
        2
        + 
        2
        \frac{
            \left(
                2 k k'
            \right)^2
        }{
            \left(
                2 k^2
            \right)
            \left(
                2 k'^2
            \right)
        }
    \right)
    \\
    &=
    2
    g^2 
    m_\phi^4
\end{align*}
which matches our expectation based on the $\phi \rightarrow \gamma \gamma$ matrix element. 

\subsubsection{3-point interactions}
We look first at the 3-point interactions between the modulus and gluons.  
The interaction term relevant for this process is 
\begin{align*}
    \mathcal{L}_{\phi g g}
    &=
    -
    \frac{2\lambda_{SU(3)}}{\sqrt{2} M_*}
    \phi_R
    \left(
        \partial_\mu 
        G_{A \nu}
        \partial^\mu 
        G_{A}^\nu
        -
        \partial_\mu 
        G_{A \nu}
        \partial^\nu 
        G_{A}^\mu
    \right)
\end{align*}
This corresponds to a squared matrix element 
\begin{align}
    \left| 
        \mathcal{M}_{\phi_R \rightarrow g g}^{\nu \sigma}
        \epsilon_\sigma^* (k)
        \epsilon_\nu^* (k')
    \right|^2
    &=
    32
    \lambda_{SU(3)}^2
    \frac{
        m_\phi^4
    }{
        M_*^2
    }
\end{align}
where, again, the extra factor of 8 comes from summing over all possible final gluons.
This gives us a total decay width of 
\begin{align}
    \Gamma_{\phi_R \rightarrow g g}
    &=
    \frac{
        \lambda_{SU(3)}^2
    }{
        \pi 
    } 
    \frac{
        m_\phi^3
    }{
        M_*^2
    }
\end{align}
where we divide by an extra factor of 2 as the final state particles are indistinguishable.

\subsection{Decay into Higgs fields}
\label{app:HF}

We have the interaction term for the modulus-Higgs sector \cite{Moroi:1999zb} 
\begin{align*}
    \mathcal{L}_H
    &=
    \frac{\lambda_H}{M_*} 
    \int d^4 \theta \,
    \hat{\phi}
    \hat{H}_d^{*} 
    \hat{H}_u^{*}
    +
    \text{h.c.}    
\end{align*}
We focus first on the subset of terms $\hat{\phi} \hat{h}_d^{0*} \hat{h}_u^{0*} + \text{h.c.}$ and obtain the remaining results by the substitutions $\hat{h}_{u}^{0*} \rightarrow \hat{h}_{u}^{+*}$ and $\hat{h}_d^{0*} \rightarrow \hat{h}_d^{-*}$.
Expanding as per Ref.~\cite{Baer:2006rs} Eqs. 5.34 and 5.37, we need to first evaluate the superspace integral (noting that we disregard the modulino and the $\mathcal{F}_\phi$ terms):
\begin{alignat*}{2}
    \mathcal{L}_{h^0}
    &\supset&&
    \frac{\lambda_H}{M_*} 
    \phi
    \int d^4 \theta \,
    \left[
        \frac{1}{8} 
        \left(
            \bar{\theta} \gamma_5 \theta
        \right)^2
        \left(
            h_d^{0*} 
            \partial_\mu \partial^\mu
            h_u^{0*}
            +
            h_u^{0*} 
            \partial_\mu \partial^\mu
            h_d^{0*}
        \right)
        -
        \frac{1}{4} 
        \left(
            \bar{\theta} \gamma_5 \gamma_\mu \theta
            \bar{\theta} \gamma_5 \gamma_\nu \theta
        \right)
        \partial^\mu h_d^{0*}
        \partial^\nu h_u^{0*}
    \right]
    \\ &
    &&+ 
    \frac{\lambda_H}{M_*} 
    \phi
    \int d^4 \theta \,
    \left[
        i \bar{\psi}_{h_d^0} \theta_R
        \bar{\theta} \gamma_5 \theta \bar{\theta} \slashed{\partial} \psi_{h_u^0, R}
        +
        i \bar{\theta} \gamma_5 \theta \bar{\theta} \slashed{\partial} \psi_{h_d^0, R} \bar{\psi}_{h_u^0} \theta_R
        -
        \bar{\theta} \theta_R \mathcal{F}_{h_d}^\dagger 
        \bar{\theta} \theta_R \mathcal{F}_{h_u}^\dagger 
    \right]
    \\ &
    &&+
    \frac{\lambda_H}{M_*} 
    \frac{1}{4} 
    \int d^4 \theta \,
    \left(
        \bar{\theta} \gamma_5 \gamma_\mu \theta
        \bar{\theta} \gamma_5 \gamma_\nu \theta
    \right)
    \partial^\mu \phi 
    \left[
        \partial^\nu( h_d^{0*} )
        h_u^{0*}
        +
        h_d^{0*}
        \partial^\nu( h_u^{0*} )
    \right]
    \\ &
    &&-
    i
    \frac{\lambda_H}{M_*} 
    \int d^4 \theta \,
    \left(
        \bar{\theta} \gamma_5 \gamma_\mu \theta
    \right)
    \partial^\mu \phi 
    \left[
        \bar{\psi}_{h_d^0} \theta_R \bar{\psi}_{h_u^0} \theta_R
    \right]
    \\ &
    &&+
    \frac{\lambda_H}{M_*} 
    \frac{1}{8} 
    \int d^4 \theta \,
    \left(
        \bar{\theta} \gamma_5 \theta
    \right)^2 
    \partial_\mu \partial^\mu \phi 
    \left[
        h_d^{0*} h_u^{0*}
    \right]
    +
    \text{h.c.}
\end{alignat*}
Using the Grassmann variable identities, integrating some terms by parts, leveraging the Majorana nature of $\theta$ (taking e.g. $\bar{\psi} \theta = \bar{\theta} \psi$), and rearranging, we can rewrite this as 
\begin{alignat*}{2}
    \mathcal{L}_{h^0}
    &=&&
    \frac{1}{8} 
    \frac{\lambda_H}{M_*} 
    \int d^4 \theta \,
    \left(
        \bar{\theta} \gamma_5 \theta
    \right)^2
    \left[
        \phi \,
        \partial_\mu \partial^\mu 
        -
        2
        \partial_\mu \phi  \,
        \partial^\mu
        +
        \partial_\mu \partial^\mu ( \phi )
    \right]
    h_d^{0*} 
    h_u^{0*}
    \\ &
    &&+ 
    \frac{\lambda_H}{M_*} 
    \int d^4 \theta \,
    \phi
    \left[
        i \bar{\psi}_{h_d^0} \theta_R
        \bar{\theta} \gamma_5 \theta \bar{\theta} \slashed{\partial} \psi_{h_u^0, R}
        +
        i \bar{\theta} \gamma_5 \theta \bar{\theta} \slashed{\partial} \psi_{h_d^0, R} \bar{\psi}_{h_u^0} \theta_R
        -
        \bar{\theta} \theta_R \mathcal{F}_{h_d}^\dagger 
        \bar{\theta} \theta_R \mathcal{F}_{h_u}^\dagger 
    \right]
    \\ &
    &&-
    i
    \frac{\lambda_H}{M_*} 
    \int d^4 \theta \,
    \left(
        \bar{\theta} \gamma_5 \gamma_\mu \theta
    \right)
    \partial^\mu \phi 
    \left[
        \bar{\psi}_{h_d^0} 
        P_R
        \theta
        \bar{\theta}
        P_R
        \psi_{h_u^0} 
    \right]
    +
    \text{h.c.}
    \\
    &=&&
    -
    \frac{1}{4} 
    \frac{\lambda_H}{M_*} 
    \left[
        \phi \,
        \partial_\mu \partial^\mu 
        -
        2
        \partial_\mu \phi  \,
        \partial^\mu
        +
        \partial_\mu \partial^\mu ( \phi )
    \right]
    h_d^{0*} 
    h_u^{0*}
    +
    \text{h.c.}
    \\ &
    &&+ 
    \frac{i}{2}
    \frac{\lambda_H}{M_*} 
    \phi
    \left[
        \bar{\psi}_{h_d^0, L}
        \gamma_5
        \slashed{\partial}
        \psi_{h_u^0, R}
        +
        \bar{\psi}_{h_u^0, L} 
        \gamma_5
        \slashed{\partial}
        \psi_{h_d^0, R}
    \right]
    +
    \frac{i}{2}
    \frac{\lambda_H}{M_*} 
    \phi^*
    \left[
        \partial^\mu
        \bar{\psi}_{h_u^0, R}
        \gamma_\mu
        \gamma_5
        \psi_{h_d^0, L}
        +
        \partial^\mu
        \bar{\psi}_{h_d^0, R}
        \gamma_\mu
        \gamma_5
        \psi_{h_u^0, L} 
    \right]
\end{alignat*}
This Lagrangian contains some redundancies in the scalar interactions, so we need to exploit surface term identities to simplify this expression.
We can take the total derivatives:
\begin{align*}
    \partial_\mu \partial^\mu 
    \left(
        \phi 
        h_d^{0*} 
        h_u^{0*}
    \right)
    &=
    \partial_\mu
    \left(
        \partial^\mu ( \phi ) 
        h_d^{0*} 
        h_u^{0*}
        +
        \phi 
        \partial^\mu 
        \left( 
            h_d^{0*}
            h_u^{0*}
        \right)
    \right)
    \\
    &=
    \partial_\mu \partial^\mu ( \phi ) 
    h_d^{0*} 
    h_u^{0*}
    +
    \partial^\mu ( \phi ) 
    \partial_\mu ( h_d^{0*} )
    h_u^{0*}
    +
    \partial^\mu ( \phi ) 
    h_d^{0*} 
    \partial_\mu h_u^{0*}
    +
    \partial_\mu
    \left(
        \phi 
        \partial^\mu 
        \left( 
            h_d^{0*}
            h_u^{0*}
        \right)
    \right)
    \\
    &=
    \partial_\mu \phi 
    \partial^\mu 
    \left( 
        h_d^{0*}
        h_u^{0*}
    \right)
    +
    \phi 
    \partial_\mu \partial^\mu 
    \left( 
        h_d^{0*}
        h_u^{0*}
    \right)
    +
    \partial_\mu
    \left(
        \partial^\mu ( \phi ) 
        h_d^{0*} 
        h_u^{0*}
    \right)
\end{align*}
and, on ignoring the surface terms, we arrive at the following identities:
\begin{align*}
    \partial_\mu \partial^\mu ( \phi ) 
    h_d^{0*} 
    h_u^{0*}
    &=
    -
    \partial^\mu ( \phi ) 
    \partial_\mu 
    \left( 
        h_d^{0*} 
        h_u^{0*}
    \right)
    =
    \phi 
    \partial_\mu \partial^\mu 
    \left( 
        h_d^{0*}
        h_u^{0*}
    \right)
\end{align*}
We also realize that the entire second line of the Lagrangian vanishes, due to the separate chirality of the spinors in each interaction.
Hence, the modulus should only interact with the scalar components of each of the Higgs superfields.
This then allows us to rewrite the Lagrangian as 
\begin{alignat*}{2}
    \mathcal{L}_{h^0}
    &=&&
    -
    \frac{\lambda_H}{M_*} 
    \left[
        h_d^{0*} 
        h_u^{0*}
        \partial_\mu \partial^\mu \phi
        +
        h_d^{0}
        h_u^{0}
        \partial_\mu \partial^\mu \phi^*
    \right]
\end{alignat*}
Finally, we separate the fields into their real and imaginary components:
\begin{alignat}{2} \label{neutralHiggsLagrangianGaugeBasis}
    \mathcal{L}_{h^0}
    &=&&
    -
    \frac{\lambda_H}{M_* \sqrt{2}} 
    \left[
        \left(
            h_{d, R}^0
            h_{u, R}^0
            -
            h_{d, I}^0
            h_{u, I}^0
        \right)
        \partial_\mu \partial^\mu 
        \phi_R
        +
        \left(
            h_{d, I}^0
            h_{u, R}^0
            +
            h_{d, R}^0
            h_{u, I}^0
        \right)
        \partial_\mu \partial^\mu 
        \phi_I
    \right]
\end{alignat}
We can write down the result for the charged Higgs sectors by the aforementioned replacements $h_{u}^{0} \rightarrow h_{u}^+$ and $h_{d}^{0} \rightarrow h_{d}^-$:
\begin{alignat}{2} \label{chargedHiggsLagrangianGaugeBasis}
    \mathcal{L}_{h^\pm}
    &=&&
    -
    \frac{\lambda_H}{M_* \sqrt{2}} 
    \left[
        \left(
            h_{d, R}^-
            h_{u, R}^+
            -
            h_{d, I}^-
            h_{u, I}^+
        \right)
        \partial_\mu \partial^\mu 
        \phi_R
        +
        \left(
            h_{d, I}^-
            h_{u, R}^+
            +
            h_{d, R}^-
            h_{u, I}^+
        \right)
        \partial_\mu \partial^\mu 
        \phi_I
    \right]
\end{alignat}

\subsubsection{Relevant matrix element formulae for Higgs bosons}
In this section, we work out the general interactions that will be relevant for the Higgs bosons.
All of these formulae are worked out in the rest frame of the modulus.

We start with the 3-point interaction formulae.
The first general interaction is of the form:
\begin{align}
    \mathcal{L}_I
    &=
    -
    g
    \partial_\mu 
    \partial^\mu 
    \phi_R \,
    h_i h_j
\end{align}
which corresponds to the diagram:
\newline
\begin{center}
    \begin{tikzpicture}
        \begin{feynman}
            \vertex (a) {\(\phi\)} ;
            \vertex [right=of a] (b) ;
            \vertex [below right=of b] (f1) {\( h_j \)};
            \vertex [above right=of b] (f2) {\( h_i \)};
            \diagram* {
                (a) -- [scalar, momentum'={\(q\)}] (b),
                (b) -- [scalar, momentum'={\(k'\)}] (f1), 
                (b) -- [scalar] (f2),
            };
        \end{feynman}
    \end{tikzpicture}
\end{center}
Taking the incoming momenta of the modulus to be $q$, and the outgoing momenta of $h_i$ and $h_j$ to be $k$ and $k'$, respectively, we have a vertex factor of $-i g m_\phi^2$.
The corresponding matrix element is then 
\begin{align*}
    i \mathcal{M}
    &=
    -i g m_\phi^2
\end{align*}
which must be also multiplied by an additional factor of 2 if $h_i = h_j$.
Assuming no additional contributions to this decay channel, the squared matrix element is 
\begin{align} \label{matrixElementHiggs}
    \left|
        \mathcal{M}
    \right|^2
    &=
    g^2
    m_\phi^4 
\end{align}
which, again, must be multiplied by an additional factor of 4 if $h_i = h_j$.

\subsubsection{Neutral Higgs interactions}
We first start with the neutral Higgs sector interactions.
We write down the term for the real component of the modulus interacting with the neutral Higgs from \ref{neutralHiggsLagrangianGaugeBasis}:
\begin{alignat*}{2}
    \mathcal{L}_{\phi_R h h}
    &\supset&&
    -
    \frac{\lambda_H}{M_* \sqrt{2}} 
    \left(
        h_{d, R}^0
        h_{u, R}^0
        -
        h_{d, I}^0
        h_{u, I}^0
    \right)
    \partial_\mu \partial^\mu 
    \phi_R
\end{alignat*}
We then move to the mass eigenstate basis with the following transformations:
\begin{align*}
    \begin{pmatrix}
        h_{u, R}^0 \\
        h_{d, R}^0
    \end{pmatrix}
    &=
    \begin{pmatrix}
        \cos \alpha && -\sin \alpha \\
        \sin \alpha && \cos \alpha
    \end{pmatrix}
    \begin{pmatrix}
        h \\
        H
    \end{pmatrix}
    \\
    \begin{pmatrix}
        h_{u, I}^0 \\
        h_{d, I}^0
    \end{pmatrix}
    &=
    \begin{pmatrix}
        \sin \beta && \cos \beta \\
        -\cos \beta && \sin \beta
    \end{pmatrix}
    \begin{pmatrix}
        G^0 \\
        A
    \end{pmatrix}
\end{align*}
This gives us a Lagrangian of the form 
\begin{alignat*}{2}
    \mathcal{L}_{\phi_R h h}
    &\supset&&
    -
    \frac{\lambda_H}{M_* \sqrt{2}} 
    \left(
        \sin \alpha
        \cos \alpha 
        \, h^2 
        +
        \left(
            \cos^2 \alpha 
            -
            \sin^2 \alpha
        \right)
        h H 
        -
        \sin \alpha 
        \cos \alpha
        \, H^2
    \right)
    \partial_\mu \partial^\mu 
    \phi_R 
    \\ &
    &&+
    \frac{\lambda_H}{M_* \sqrt{2}}     
    \left(
        -
        \cos \beta
        \sin \beta 
        \, (G^0)^2 
        +
        \left(
            \sin^2 \beta 
            -
            \cos^2 \beta
        \right)
        \, A G^0 
        +
        \sin \beta
        \cos \beta 
        \, A^2
    \right)
    \partial_\mu \partial^\mu 
    \phi_R    
    \\
    &=&&
    -
    \frac{\lambda_H}{M_* \sqrt{2}} 
    \left(
        \frac{1}{2}
        \sin (2 \alpha)
        \, h^2 
        +
        \cos ( 2 \alpha )
        h H 
        -
        \frac{1}{2}
        \sin (2 \alpha)
        \, H^2
    \right)
    \partial_\mu \partial^\mu 
    \phi_R 
    \\ &
    &&+
    \frac{\lambda_H}{M_* \sqrt{2}}     
    \left(
        -
        \frac{1}{2}
        \sin (2 \beta)
        \, (G^0)^2 
        -
        \cos ( 2 \beta )
        \, A G^0 
        +
        \frac{1}{2}
        \sin (2 \beta)
        \, A^2
    \right)
    \partial_\mu \partial^\mu 
    \phi_R    
\end{alignat*}
We can extract from this the interaction between the modulus and the light Higgs pair:
\begin{align}
    \mathcal{L}_{\phi_R h h}
    &=
    -
    \frac{\lambda_H}{2 \sqrt{2} M_*} 
    \sin (2 \alpha)
    \, h^2 
    \partial_\mu \partial^\mu 
    \phi_R 
\end{align}
The corresponding squared matrix element from Eq. \ref{matrixElementHiggs} is then 
\begin{align}
    \left| 
        \mathcal{M}_{\phi_R \rightarrow h h}
    \right|^2
    &=
    \frac{
        \lambda_H^2 m_\phi^4
    }{
        2 M_*^2
    } 
    \sin^2 (2 \alpha)
\end{align}
This gives us the decay width to light Higgs pairs:
\begin{align}
    \Gamma_{\phi_R \rightarrow h h}
    &=
    \frac{
        \lambda_H^2 
    }{
        64
        \pi 
    } 
    \frac{
        m_\phi^3
    }{
        M_*^2
    } 
    \sin^2 (2 \alpha)
    \lambda^{1/2}
    \left(
        1,
        \frac{
            m_h^2
        }{ 
            m_\phi^2
        },
        \frac{
            m_h^2
        }{ 
            m_\phi^2
        }
    \right)
\end{align}
where we divide by an additional factor of 2 due to the final state particles being indistinguishable.

Likewise, the interaction between the modulus, the light Higgs, and the heavy Higgs is given by
\begin{align}
    \mathcal{L}_{\phi_R h H}
    &=
    -
    \frac{\lambda_H}{\sqrt{2} M_*} 
    \cos ( 2 \alpha )
    h H 
    \partial_\mu \partial^\mu 
    \phi_R
\end{align}
which has a squared matrix element of 
\begin{align}
    \left| 
        \mathcal{M}_{\phi_R \rightarrow h H}
    \right|^2
    &=
    \frac{
        \lambda_H^2 m_\phi^4
    }{
        2 M_*^2
    } 
    \cos^2 (2 \alpha)
\end{align}
The decay width for this process is then given by
\begin{align}
    \Gamma_{\phi_R \rightarrow h H}
    &=
    \frac{
        \lambda_H^2 
    }{
        32
        \pi 
    } 
    \frac{
        m_\phi^3
    }{
        M_*^2
    } 
    \cos^2 (2 \alpha)
    \lambda^{1/2}
    \left(
        1,
        \frac{
            m_h^2
        }{ 
            m_\phi^2
        },
        \frac{
            m_H^2
        }{ 
            m_\phi^2
        }
    \right)
\end{align}

Finally, the interaction between the modulus and the heavy Higgs pair is:
\begin{align}
    \mathcal{L}_{\phi_R H H}
    &=
    \frac{\lambda_H}{ 2\sqrt{2} M_*} 
    \sin (2 \alpha)
    \, H^2
    \partial_\mu \partial^\mu 
    \phi_R 
\end{align}
which has corresponding squared matrix element
\begin{align}
    \left| 
        \mathcal{M}_{\phi_R \rightarrow H H}
    \right|^2
    &=
    \frac{
        \lambda_H^2 m_\phi^4
    }{
        2 M_*^2
    } 
    \sin^2 (2 \alpha)
\end{align}
The decay width to heavy Higgs pairs is then given by:
\begin{align}
    \Gamma_{\phi_R \rightarrow H H}
    &=
    \frac{
        \lambda_H^2 
    }{
        64
        \pi 
    } 
    \frac{
        m_\phi^3
    }{
        M_*^2
    } 
    \sin^2 (2 \alpha)
    \lambda^{1/2}
    \left(
        1,
        \frac{
            m_H^2
        }{ 
            m_\phi^2
        },
        \frac{
            m_H^2
        }{ 
            m_\phi^2
        }
    \right)
\end{align}
where we again divide by an additional factor of 2 since the final state particles are indistinguishable.

In the CP-odd sector, we have the decay to the pseudoscalar Higgs given by 
\begin{align}
    \mathcal{L}_{\phi_R A A}
    &=
    \frac{\lambda_H}{2 \sqrt{2} M_*}     
    \sin (2 \beta)
    \, A^2
    \partial_\mu \partial^\mu 
    \phi_R
\end{align}
The corresponding squared matrix element is 
\begin{align}
    \left| 
        \mathcal{M}_{\phi_R \rightarrow A A}
    \right|^2
    &=
    \frac{
        \lambda_H^2 m_\phi^4
    }{
        2 M_*^2
    } 
    \sin^2 (2 \beta)    
\end{align}
The decay width to pseudoscalar Higgs pairs is then:
\begin{align}
    \Gamma_{\phi_R \rightarrow A A}
    &=
    \frac{
        \lambda_H^2 
    }{
        64
        \pi 
    } 
    \frac{
        m_\phi^3
    }{
        M_*^2
    } 
    \sin^2 (2 \beta)
    \lambda^{1/2}
    \left(
        1,
        \frac{
            m_A^2
        }{ 
            m_\phi^2
        },
        \frac{
            m_A^2
        }{ 
            m_\phi^2
        }
    \right)
\end{align}
where we once again divide by an additional factor of 2 for indistinguishable final state particles.

We can similarly write down the term for the imaginary component of the modulus interacting with the neutral Higgs from Eq. \ref{neutralHiggsLagrangianGaugeBasis}:
\begin{alignat*}{2}
    \mathcal{L}_{\phi_I hh}
    &=&&
    -
    \frac{\lambda_H}{M_* \sqrt{2}} 
    \left(
        h_{d, I}^0
        h_{u, R}^0
        +
        h_{d, R}^0
        h_{u, I}^0
    \right)
    \partial_\mu \partial^\mu 
    \phi_I
\end{alignat*}
and move to the mass eigenstate basis:
\begin{alignat*}{2}
    \mathcal{L}_{\phi_I hh}
    &=&&
    -
    \frac{\lambda_H}{M_* \sqrt{2}} 
        \left(
        \sin \alpha 
        \sin \beta 
        - 
        \cos \beta 
        \cos \alpha 
    \right)
    G^0 
    h 
    \partial_\mu \partial^\mu 
    \phi_I
    -
    \frac{\lambda_H}{M_* \sqrt{2}} 
    \left(
        \sin \beta 
        \cos \alpha 
        +
        \sin \alpha 
        \cos \beta 
    \right)
    h 
    A
    \partial_\mu \partial^\mu 
    \phi_I
    \\ &
    &&-
    \frac{\lambda_H}{M_* \sqrt{2}} 
    \left(
        \sin \alpha 
        \cos \beta 
        +
        \cos \alpha 
        \sin \beta 
    \right)
    H
    G^0 
    \partial_\mu \partial^\mu 
    \phi_I
    -
    \frac{\lambda_H}{M_* \sqrt{2}} 
    \left(
        \cos \alpha 
        \cos \beta
        - 
        \sin \alpha 
        \sin \beta     
    \right)
    H
    A
    \partial_\mu \partial^\mu 
    \phi_I
    \\
    &=&&
    \frac{\lambda_H}{M_* \sqrt{2}} 
    \cos (\alpha + \beta) \,
    G^0 
    h \,
    \partial_\mu \partial^\mu 
    \phi_I
    -
    \frac{\lambda_H}{M_* \sqrt{2}} 
    \sin ( \alpha + \beta ) \,
    h 
    A \,
    \partial_\mu \partial^\mu 
    \phi_I
    \\ &
    &&-
    \frac{\lambda_H}{M_* \sqrt{2}} 
    \sin ( \alpha + \beta ) \,
    H
    G^0 \,
    \partial_\mu \partial^\mu 
    \phi_I
    -
    \frac{\lambda_H}{M_* \sqrt{2}} 
    \cos ( \alpha + \beta ) \, 
    H
    A \,
    \partial_\mu \partial^\mu 
    \phi_I
\end{alignat*}
We extract the interaction between the modulus, the light Higgs, and the pseudoscalar Higgs:
\begin{align}
    \mathcal{L}_{\phi_I h A}
    &=
    -
    \frac{\lambda_H}{M_* \sqrt{2}} 
    \sin ( \alpha + \beta ) \,
    h 
    A \,
    \partial_\mu \partial^\mu 
    \phi_I
\end{align}
The corresponding squared matrix element for this decay channel is 
\begin{align}
    \left| 
        \mathcal{M}_{\phi_I \rightarrow h A}
    \right|^2
    &=
    \frac{
        \lambda_H^2 m_\phi^4
    }{
        2 M_*^2
    } 
    \sin^2 ( \alpha + \beta)    
\end{align}
which then has a decay width of 
\begin{align}
    \Gamma_{\phi_I \rightarrow h A}
    &=
    \frac{
        \lambda_H^2 
    }{
        32
        \pi 
    } 
    \frac{
        m_\phi^3
    }{
        M_*^2
    } 
    \sin^2 ( \alpha + \beta)    
    \lambda^{1/2}
    \left(
        1,
        \frac{
            m_h^2
        }{ 
            m_\phi^2
        },
        \frac{
            m_A^2
        }{ 
            m_\phi^2
        }
    \right)
\end{align}

Likewise, the interaction between the modulus, the heavy Higgs, and the pseudoscalar Higgs is given by:
\begin{align}
    \mathcal{L}_{\phi_I HA}
    &=
    -
    \frac{\lambda_H}{M_* \sqrt{2}} 
    \cos ( \alpha + \beta ) \, 
    H
    A \,
    \partial_\mu \partial^\mu 
    \phi_I    
\end{align}
which has a corresponding matrix element
\begin{align}
    \left| 
        \mathcal{M}_{\phi_I \rightarrow H A}
    \right|^2
    &=
    \frac{
        \lambda_H^2 m_\phi^4
    }{
        2 M_*^2
    } 
    \cos^2 ( \alpha + \beta)    
\end{align}
The decay width is given by:
\begin{align}
    \Gamma_{\phi_I \rightarrow H A}
    &=
    \frac{
        \lambda_H^2 
    }{
        32
        \pi 
    } 
    \frac{
        m_\phi^3
    }{
        M_*^2
    } 
    \cos^2 ( \alpha + \beta)
    \lambda^{1/2}
    \left(
        1,
        \frac{
            m_H^2
        }{ 
            m_\phi^2
        },
        \frac{
            m_A^2
        }{ 
            m_\phi^2
        }
    \right)
\end{align}

\subsubsection{Charged Higgs interactions}

We find the charged Higgs interactions similarly to the neutral Higgs interactions.
Taking the interactions with the real component of the modulus, 
we have from Eq. \ref{chargedHiggsLagrangianGaugeBasis} the term 
\begin{alignat*}{2}
    \mathcal{L}_{\phi_R h^+ h^-}
    &=&&
    -
    \frac{\lambda_H}{M_* \sqrt{2}} 
    \left(
        h_{d, R}^-
        h_{u, R}^+
        -
        h_{d, I}^-
        h_{u, I}^+
    \right)
    \partial_\mu \partial^\mu 
    \phi_R
    \\
    &=&&
    -
    \frac{\lambda_H}{M_* \sqrt{2}} 
    \left(
        h_{d}^{- *}
        h_{u}^{+ *}
        +
        h_{d}^{-}
        h_{u}^{+}
    \right)
    \partial_\mu \partial^\mu 
    \phi_R
\end{alignat*}
We move to the mass eigenstate basis with the following transformations:
\begin{align*}
    \begin{pmatrix}
        h_{d}^{-*} \\
        h_{u}^{+}
    \end{pmatrix}
    &=
    \begin{pmatrix}
        \cos \beta && -\sin \beta \\
        \sin \beta && \cos \beta
    \end{pmatrix}
    \begin{pmatrix}
        G^+ \\
        H^+
    \end{pmatrix}
\end{align*}
which gives us the following Lagrangian:
\begin{align}
    \mathcal{L}_{\phi_R h^+ h^-}
    &=&&
    -
    \frac{\lambda_H}{M_* \sqrt{2}} 
    \left(
        \sin 2 \beta 
        G^+ 
        G^-
        +
        \cos 2 \beta \,
        G^-
        H^+
        +
        \cos 2 \beta \,
        G^+ 
        H^-
        -
        \sin 2 \beta 
        H^+
        H^-
    \right)
    \partial_\mu \partial^\mu 
    \phi_R
\end{align}

We can now extract the interaction between the modulus and the charged Higgs pair:
\begin{align}
    \mathcal{L}_{\phi_R H^+ H^-}
    &=
    \frac{\lambda_H}{M_* \sqrt{2}} 
    \left(
        \sin 2 \beta 
        H^+
        H^-
    \right)
    \partial_\mu \partial^\mu 
    \phi_R    
\end{align}
which has, from Eq. \ref{matrixElementHiggs}, the squared matrix element
\begin{align}
    \left| 
        \mathcal{M}_{\phi_R \rightarrow H^+ H^-}
    \right|^2
    &=
    \frac{
        \lambda_H^2 m_\phi^4
    }{
        2 M_*^2
    } 
    \sin^2 ( 2 \beta)
\end{align}
The decay width to charged Higgs pairs is given by 
\begin{align}
    \Gamma_{\phi_R \rightarrow H^+ H^-}
    &=
    \frac{
        \lambda_H^2 
    }{
        32
        \pi 
    } 
    \frac{
        m_\phi^3
    }{
        M_*^2
    } 
    \sin^2 ( 2 \beta)
    \lambda^{1/2}
    \left(
        1,
        \frac{
            m_{H^\pm}^2
        }{ 
            m_\phi^2
        },
        \frac{
            m_{H^\pm}^2
        }{ 
            m_\phi^2
        }
    \right)
\end{align}

We follow similarly for the interactions with the imaginary component of the modulus.
From Eq. \ref{chargedHiggsLagrangianGaugeBasis}, we have the term 
\begin{alignat*}{2}
    \mathcal{L}_{\phi_I h^+ h^-}
    &=&&
    -
    \frac{\lambda_H}{M_* \sqrt{2}} 
    \left(
        h_{d, I}^-
        h_{u, R}^+
        +
        h_{d, R}^-
        h_{u, I}^+
    \right)
    \partial_\mu \partial^\mu 
    \phi_I
    \\
    &=&&
    -
    \frac{\lambda_H}{M_* \sqrt{2}} 
    i
    \left(
        h_{d}^{- *}
        h_{u}^{+ *}
        -
        h_{d}^{-}
        h_{u}^{+}
    \right)
    \partial_\mu \partial^\mu 
    \phi_I
\end{alignat*}
We again move to the mass basis with the same transformations as above:
\begin{align}
    \mathcal{L}_{\phi_I h^+ h^-}
    &=
    -
    \frac{\lambda_H}{M_* \sqrt{2}} 
    i
    \left(
        G^+ 
        H^-
        -
        G^- 
        H^+
    \right)
    \partial_\mu \partial^\mu 
    \phi_I
\end{align}

\subsection{Decay into matter fields}
\label{app:MF}

We have the operator relevant for fermion interactions (Ref. \cite{Moroi:1999zb})
\begin{align*}
    \mathcal{L}_Q
    &= 
    \int d^4 \theta \, \frac{\lambda_{Q}}{M_*} \hat{\phi} \hat{Q}^\dagger \hat{Q}
    +
    \text{h.c.}
\end{align*}
Ignoring the modulino, we have then the following interaction Lagrangian:
\begin{alignat*}{2}
    \mathcal{L}_Q
    &=&&
    \frac{\lambda_{Q^* Q}}{M_*} 
    \int d^4 \theta \, 
    \left(
        \overline{\theta} \gamma_5 \theta
    \right)^2
    \left[
        \phi \,
        \{
            \frac{1}{8}
            Q^\dagger \partial_\mu \partial^\mu Q 
            +
            \frac{1}{8}
            \partial_\mu \partial^\mu ( Q^\dagger ) Q 
            -
            \frac{1}{4}
            \partial_\mu Q^\dagger \partial^\mu Q
            -
            \frac{1}{2}
            \frac{i}{2}
            \overline{\psi}_Q \slashed{\partial} \psi_Q
            -
            \frac{1}{2}
            \mathcal{F}_Q^\dagger \mathcal{F}_Q
        \}
    \right]
    \\ &
    &&-
    \frac{\lambda_{Q^* Q}}{M_*} 
    \int d^4 \theta \, 
    \frac{1}{4} 
    \left(
        \overline{\theta} \gamma_5 \gamma_\mu \theta
        \overline{\theta} \gamma_5 \gamma_\nu \theta 
    \right)
    \left[
        \partial^\mu \phi \,
        Q^\dagger
        \partial^\nu Q
        -
        \partial^\mu \phi \,
        \partial^\nu ( Q^\dagger )
        Q
    \right]
    \\ &
    &&+
    \frac{\lambda_{Q^* Q}}{M_*} 
    \int d^4 \theta \, 
    \left[
        \frac{1}{8}
        \left(
            \overline{\theta} \gamma_5 \theta
        \right)^2
        Q^\dagger Q \,
        \partial_\mu \partial^\mu \phi
    \right]
    +
    \text{h.c.}
    \\
    &=&&
    \frac{\lambda_{Q^* Q}}{M_*}
    \frac{1}{4}
    \left[
        -
        \phi \,
        Q^\dagger \partial_\mu \partial^\mu Q 
        -
        \phi \,
        \partial_\mu \partial^\mu ( Q^\dagger ) Q 
        +
        2
        \phi \,
        \partial_\mu Q^\dagger \partial^\mu Q
        -
        2
        \partial_\mu \phi \,
        Q^\dagger
        \partial^\mu Q
        +
        2
        \partial_\mu \phi \,
        \partial^\mu ( Q^\dagger )
        Q
        -
        Q^\dagger Q \,
        \partial_\mu \partial^\mu \phi
    \right]
    \\ &
    &&+
    \frac{\lambda_{Q^* Q}}{M_*} 
    \left[
        \frac{i}{2}
        \phi \,
        \overline{\psi}_Q \slashed{\partial} \psi_Q
        +
        \phi \,
        \mathcal{F}_Q^\dagger \mathcal{F}_Q
    \right]
    +
    \text{h.c.}
\end{alignat*}
This expression has some redundancies in the scalar components, and we can exploit surface terms to simplify this Lagrangian.
Taking the total derivative and distributing differentials in various permutations (so as to end up with surface terms that can be neglected), we have:
\begin{align*}
    \partial_\mu \partial^\mu 
    \left(
        \phi Q^\dagger Q
    \right)
    &=
    \partial_\mu
    \left(
        \partial^\mu \phi 
        Q^\dagger 
        Q
        +
        \phi 
        \partial^\mu Q^\dagger 
        Q
        +
        \phi 
        Q^\dagger 
        \partial^\mu Q
    \right)
    \\
    &=
    \partial_\mu \partial^\mu (\phi) 
    Q^\dagger 
    Q
    +
    \partial^\mu (\phi) 
    \partial_\mu ( Q^\dagger )
    Q
    +
    \partial^\mu (\phi) 
    Q^\dagger 
    \partial_\mu Q
    +
    \partial_\mu
    \left(
        \phi 
        \partial^\mu Q^\dagger 
        Q
        +
        \phi 
        Q^\dagger 
        \partial^\mu Q
    \right)
    \\
    &=
    \partial_\mu ( \phi )
    \partial^\mu ( Q^\dagger )
    Q
    +
    \phi 
    \partial_\mu \partial^\mu ( Q^\dagger )
    Q
    +
    \phi 
    \partial^\mu ( Q^\dagger )
    \partial_\mu Q
    +
    \partial_\mu
    \left(
        \partial^\mu \phi 
        Q^\dagger 
        Q
        +
        \phi 
        Q^\dagger 
        \partial^\mu Q
    \right)
    \\
    &=
    \partial_\mu ( \phi )
    Q^\dagger 
    \partial^\mu Q
    +
    \phi 
    \partial_\mu ( Q^\dagger )
    \partial^\mu Q
    +
    \phi 
    Q^\dagger 
    \partial_\mu \partial^\mu Q
    +
    \partial_\mu
    \left(
        \partial^\mu \phi 
        Q^\dagger 
        Q
        +
        \phi 
        \partial^\mu Q^\dagger 
        Q
    \right)
\end{align*}
Putting these together, we arrive at the following identity which allows us to replace the terms with mixed differentials:
\begin{align*}
    Q^\dagger Q \,
    \partial_\mu \partial^\mu \phi
    -
    3
    \partial_\mu \partial^\mu Q^\dagger \,
    \phi 
    Q
    +
    \phi 
    Q^\dagger 
    \partial_\mu \partial^\mu Q
    &=
    2
    \partial_\mu ( \phi )
    \partial^\mu ( Q^\dagger )
    Q
    +
    2
    \phi 
    \partial^\mu ( Q^\dagger )
    \partial_\mu Q    
    -
    2
    \partial^\mu (\phi) 
    Q^\dagger 
    \partial_\mu Q    
\end{align*}
This then allows us to rewrite the Lagrangian as:
\begin{alignat}{2}\label{modulusMatterLagrangian}
    \mathcal{L}_Q
    &=&&
    -
    \frac{\lambda_{Q^* Q}}{M_*}
    \phi \,
    \partial_\mu \partial^\mu ( Q^\dagger ) \,
    Q
    +
    \frac{\lambda_{Q^* Q}}{M_*} 
    \left[
        \frac{i}{2}
        \phi \,
        \overline{\psi}_Q \slashed{\partial} \psi_Q
        +
        \phi \,
        \mathcal{F}_Q^\dagger \mathcal{F}_Q
    \right]
    +
    \text{h.c.}
\end{alignat}
We neglect the couplings to the $\mathcal{F}_Q$ fields for now, as they will be 4-body decay terms.

\subsubsection{Relevant matrix element formulae for sfermions}

In this section, we work out the general interactions that will be relevant for the sfermions.
All of these formulae are worked out in the rest frame of the modulus.

We start with the 3-point interaction formulae.
The first general interaction is of the form:
\begin{align}
    \mathcal{L}_I
    &=
    -
    g
    \phi_R 
    \left(
        \Phi_1
        \partial_\mu
        \partial^\mu
        \Phi_2^\dagger
    \right)
\end{align}
which corresponds to the diagram:
\newline
\begin{center}
    \begin{tikzpicture}
        \begin{feynman}
            \vertex (a) {\(\phi\)} ;
            \vertex [right=of a] (b) ;
            \vertex [below right=of b] (f1) {\( \Phi_2^\dagger \)};
            \vertex [above right=of b] (f2) {\( \Phi_1 \)};
            \diagram* {
                (a) -- [scalar, momentum'={\(q\)}] (b),
                (b) -- [anti charged scalar, momentum'={\(k'\)}] (f1), 
                (b) -- [charged scalar] (f2),
            };
        \end{feynman}
    \end{tikzpicture}
\end{center}
Since we have a real scalar decaying to complex scalars, the tree-level matrix element is simply the vertex factor.
The vertex factor here is then $-i g (i k'^\mu)^2 = + i g m_{\Phi_2}^2$ and hence the matrix element is 
\begin{align} \label{matrixElementSfermion}
    i \mathcal{M}_{\phi \rightarrow \Phi_1 \Phi_2^\dagger}
    &=
    i g
    m_{\Phi_2}^2
\end{align}
Note that if the interaction Lagrangian has a second derivative on the $\Phi$ term instead of $\Phi^\dagger$, the mass in the matrix element will simply be relabeled to correspond to the field with the second derivative coupling.

\subsubsection{Relevant matrix element formulae for fermions}

In this section, we work out the general interactions that will be relevant for the fermions.
All of these formulae are worked out in the rest frame of the modulus.

We start with the 3-point interaction formulae.
The first general interaction is of the form:
\begin{align}
    \mathcal{L}_I
    &=
    g
    i
    \phi_R 
    \left(
        \overline{\psi} 
        P_R 
        \slashed{\partial}
        P_L 
        \psi
    \right)
    \\
    &=
    \frac{g}{2}
    i
    \phi_R 
    \left(
        \overline{\psi} 
        \gamma^\mu
        \left(
            1
            -
            \gamma_5
        \right)
        \partial_\mu
        \psi
    \right)
\end{align}
which corresponds to the diagram:
\newline
\begin{center}
    \begin{tikzpicture}
        \begin{feynman}
            \vertex (a) {\(\phi\)} ;
            \vertex [right=of a] (b) ;
            \vertex [below right=of b] (f1) {\( \overline{ \psi } \)};
            \vertex [above right=of b] (f2) {\( \psi \)};
            \diagram* {
                (a) -- [scalar, momentum'={\(q\)}] (b),
                (b) -- [anti fermion, momentum'={\(k'\)}] (f1), 
                (b) -- [fermion] (f2),
            };
        \end{feynman}
    \end{tikzpicture}
\end{center}
Taking the incoming momenta of the modulus to be $q$, and the outgoing momenta of $\psi$ and $\overline{\psi}$ to be $k$ and $k'$, respectively, we have a vertex factor of $i \frac{g}{2} \gamma^\mu ( 1 - \gamma_5 ) k_\mu$.
The matrix element is then 
\begin{align*} \label{matrixElementFermion1}
    i \mathcal{M}
    &=
    \overline{u}^s(k)
    \left(
        i 
        \frac{g}{2} 
        \gamma^\mu 
        (1 - \gamma_5) 
        k_\mu 
    \right) 
    v^{s'}(k') 
\end{align*}

The second general interaction is rather similar:
\begin{align}
    \mathcal{L}_I 
    &=
    g 
    i 
    \phi_R 
    \left(
        \partial_\mu 
        \overline{\psi} 
        P_R 
        \gamma^\mu 
        P_L 
        \psi
    \right)
    \\
    &=
    \frac{g}{2}
    i 
    \phi_R 
    \left(
        \partial_\mu 
        \overline{\psi} 
        \gamma^\mu 
        \left(
            1
            -
            \gamma_5
        \right)
        \psi
    \right)
\end{align}
which corresponds to the same diagram as above, albeit with a different vertex factor.
Taking again the incoming momenta of the modulus to be $q$, and the outgoing momenta of $\psi$ and $\overline{\psi}$ to be $k$ and $k'$, respectively, we have a vertex factor of $-i \frac{g}{2} k'_\mu \gamma^\mu ( 1 - \gamma_5 ) $.
The matrix element is then 
\begin{align*} \label{matrixElementFermion2}
    i \mathcal{M}
    &=
    \overline{u}^s(k)
    \left(
        i 
        \frac{g}{2} 
        k'_\mu 
        \gamma^\mu 
        (1 - \gamma_5) 
    \right) 
    v^{s'}(k') 
\end{align*}

There are similar interactions as well, which interchange $P_L \leftrightarrow P_R$.
These interactions have identical matrix elements as above, but with the replacement $(1 - \gamma_5) \rightarrow (1 + \gamma_5)$.

It is worthwhile to also work out the combination:
\begin{align}
    \mathcal{L}_I
    &=
    g
    i
    \phi_R 
    \left(
        \overline{\psi} 
        P_R 
        \slashed{\partial}
        P_L 
        \psi
        -
        \partial_\mu 
        \overline{\psi} 
        P_R 
        \gamma^\mu 
        P_L 
        \psi
    \right)
\end{align}
The matrix element for this combination reduces to 
\begin{align} \label{matrixElementFermionCombo1}
    i \mathcal{M}
    &=
    \overline{u}^s(k)
    \left(
        i 
        \frac{g}{2} 
        \gamma^\mu 
        (1 - \gamma_5) 
        \left(
            k_\mu 
            -
            k'_\mu
        \right)
    \right) 
    v^{s'}(k') 
\end{align}

If there are no other contributions to this $\phi_R \rightarrow \overline{\psi} \psi $ decay process, we can evaluate the squared matrix element for the combination above:
\begin{align*}
    | \mathcal{M} |^2
    &=
    \frac{|g|^2}{4} 
    \overline{v}^{s'}(k')
    (1 + \gamma_5)
    \gamma^\mu
    \left(
        k_\mu 
        -
        k'_\mu
    \right)
    u^s(k)
    \overline{u}^s(k)
    \gamma^\mu 
    (1 - \gamma_5) 
    \left(
        k_\mu 
        -
        k'_\mu
    \right)
    v^{s'}(k')
\end{align*}
Summing over outgoing spins, we have then:
\begin{alignat*}{2}
    | \mathcal{M} |^2
    &=&&
    \frac{|g|^2}{4} 
    \text{Tr} 
    \left[
        \left(
            \slashed{k'}
            -
            m_\psi
        \right)
        (1 + \gamma_5)
        \left(
            \slashed{k}
            -
            \slashed{k'}
        \right)
        \left(
            \slashed{k} 
            +
            m_\psi
        \right)
        \left(
            \slashed{k}
            -
            \slashed{k'}
        \right)
        (1 - \gamma_5) 
    \right]
    \\
    &=&&
    \frac{|g|^2}{4} 
    \text{Tr} 
    \left[
        \left(
            \slashed{k'}
            -
            m_\psi
        \right)
        (1 + \gamma_5)
        \left(
            m_\psi^2
            +
            m_\psi
            \left(
                \slashed{k}
                -
                \slashed{k'}
            \right)
            -
            \slashed{k'}
            \slashed{k} 
        \right)
        \left(
            \slashed{k}
            -
            \slashed{k'}
        \right)
        (1 - \gamma_5) 
    \right]
    \\
    &=&&
    \frac{|g|^2}{4} 
    \text{Tr} 
    \left[
        \left(
            \slashed{k'}
            -
            m_\psi
        \right)
        \left(
            m_\psi^2
            -
            \slashed{k'}
            \slashed{k} 
        \right)
        \left(
            \slashed{k}
            -
            \slashed{k'}
        \right)
    \right]
    -
    \frac{|g|^2}{4} 
    \text{Tr} 
    \left[
        \left(
            \slashed{k'}
            -
            m_\psi
        \right)
        \left(
            m_\psi^2
            -
            \slashed{k'}
            \slashed{k} 
        \right)
        \left(
            \slashed{k}
            -
            \slashed{k'}
        \right)
        \gamma_5
    \right]
    \\ &
    &&+
    \frac{|g|^2}{4} 
    \text{Tr} 
    \left[
        \left(
            \slashed{k'}
            -
            m_\psi
        \right)
        \gamma_5
        \left(
            m_\psi^2
            -
            \slashed{k'}
            \slashed{k} 
        \right)
        \left(
            \slashed{k}
            -
            \slashed{k'}
        \right)
    \right]
    -
    \frac{|g|^2}{4} 
    \text{Tr} 
    \left[
        \left(
            \slashed{k'}
            -
            m_\psi
        \right)
        \gamma_5
        \left(
            m_\psi^2
            -
            \slashed{k'}
            \slashed{k} 
        \right)
        \left(
            \slashed{k}
            -
            \slashed{k'}
        \right)
        \gamma_5
    \right]
    \\
    &=&&
    \frac{|g|^2}{4} 
    m_\psi^2
    \text{Tr} 
    \left[
        \left(
            \slashed{k}
            \slashed{k'}
            -
            m_\psi^2
        \right)
    \right]
    -
    \frac{|g|^2}{4} 
    \text{Tr} 
    \left[
        \slashed{k'}
        \slashed{k} 
        \left(
            \slashed{k}
            \slashed{k'}
            -
            m_\psi^2
        \right)
    \right]
    +
    \frac{|g|^2}{4} 
    m_\psi^2
    \text{Tr} 
    \left[
        \left(
            \slashed{k'}
            -
            \slashed{k}
        \right)
        \left(
            \slashed{k}
            -
            \slashed{k'}
        \right)
    \right]
    \\
    &=&&
    4
    |g|^2
    m_\psi^2 \,
    k_\mu 
    k'^\mu
    -
    4
    |g|^2
    m_\psi^4
\end{alignat*}
Noting that (with our assigned momenta), we have $k'^\mu = q^\mu - k^\mu$, in the rest frame we have then $2 k_\mu k'^\mu = m_\phi^2 - 2 m_\psi^2$.
Therefore, the final squared matrix element in the rest frame of the modulus is 
\begin{align} \label{matrixElementSqrFermionCombo1}
    |\mathcal{M}|^2
    &=
    2
    |g|^2 
    m_\psi^2
    m_\phi^2
    \left(
        1 
        -
        4 
        \frac{m_\psi^2}{m_\phi^2}
    \right)
\end{align}

Finally, we should work out the combination:
\begin{align}
    \mathcal{L}_I
    &=
    g_1
    i
    \phi_R 
    \left(
        \overline{\psi} 
        P_R 
        \slashed{\partial}
        P_L 
        \psi
        -
        \partial_\mu 
        \overline{\psi} 
        P_R 
        \gamma^\mu 
        P_L 
        \psi
    \right)
    +
    g_2
    i
    \phi_R 
    \left(
        \overline{\psi} 
        P_L 
        \slashed{\partial}
        P_R 
        \psi
        -
        \partial_\mu 
        \overline{\psi} 
        P_L 
        \gamma^\mu 
        P_R
        \psi
    \right)
\end{align}
The matrix element for this decay channel is then (keeping the same momenta conventions)
\begin{align} \label{matrixElementFermionCombo2}
    i \mathcal{M}
    &=
    \overline{u}^s(k)
    \left(
        i 
        \frac{g_1}{2} 
        \gamma^\mu 
        (1 - \gamma_5) 
        \left(
            k_\mu 
            -
            k'_\mu
        \right)
        +
        i 
        \frac{g_2}{2} 
        \gamma^\mu 
        (1 + \gamma_5) 
        \left(
            k_\mu 
            -
            k'_\mu
        \right)
    \right) 
    v^{s'}(k') 
    \nonumber
    \\
    &=
    \overline{u}^s(k)
    \left(
        i 
        \frac{g_1 + g_2}{2} 
        \gamma^\mu 
        \left(
            k_\mu 
            -
            k'_\mu
        \right)
        -
        i 
        \frac{g_1 - g_2}{2} 
        \gamma^\mu 
        \gamma_5
        \left(
            k_\mu 
            -
            k'_\mu
        \right)
    \right) 
    v^{s'}(k') 
\end{align}
Again assuming no other contributions to this channel, we can evaluate the squared matrix element for this combination (making the definitions $a \equiv \frac{1}{2} (g_1 + g_2)$ and $b \equiv \frac{1}{2} (g_1 - g_2)$):
\begin{align*}
    | \mathcal{M} |^2
    &=
    \overline{v}^{s'}(k')
    \left(
        a
        \gamma^\mu 
        -
        b
        \gamma^\mu 
        \gamma_5
    \right) 
    \left(
        k_\mu 
        -
        k'_\mu
    \right)
    u^s(k)
    \overline{u}^s(k)
    \left(
        a
        \gamma^\mu 
        -
        b
        \gamma^\mu 
        \gamma_5
    \right) 
    \left(
        k_\mu 
        -
        k'_\mu
    \right)
    v^{s'}(k') 
\end{align*}
Summing over spins, we get 
\begin{alignat*}{2}
    | \mathcal{M} |^2
    &=&&
    \text{Tr}
    \left[
        \left(
            a
            +
            b
            \gamma_5
        \right) 
        \left(
            \slashed{k} 
            -
            \slashed{k'}
        \right)
        \left(
            \slashed{k}
            +
            m_\psi
        \right)
        \left(
            \slashed{k} 
            -
            \slashed{k'}
        \right)
        \left(
            a
            -
            b
            \gamma_5
        \right) 
        \left(
            \slashed{k'}
            -
            m_\psi
        \right)
    \right]
    \\
    &=&&
    \text{Tr}
    \left[
        \left(
            \slashed{k'}
            \slashed{k} 
            +
            m_\psi
            \left(
                \slashed{k'}
                -
                \slashed{k}
            \right)
            -
            m_\psi^2
        \right)
        \left(
            a
            -
            b
            \gamma_5
        \right) 
        \left(
            m_\psi^2
            +
            m_\psi
            \left(
                \slashed{k} 
                -
                \slashed{k'}
            \right)
            -
            \slashed{k}
            \slashed{k'}
        \right)
        \left(
            a
            -
            b
            \gamma_5
        \right) 
    \right]
    \\
    &=&&
    \text{Tr}
    \left[
        \left(
            \slashed{k'}
            \slashed{k} 
            -
            m_\psi^2
        \right)
        \left(
            a
            -
            b
            \gamma_5
        \right) 
        \left(
            m_\psi^2
            -
            \slashed{k}
            \slashed{k'}
        \right)
        \left(
            a
            -
            b
            \gamma_5
        \right) 
    \right]
    +
    m_\psi^2
    \text{Tr}
    \left[
        \left(
            \slashed{k'}
            -
            \slashed{k}
        \right)
        \left(
            a
            -
            b
            \gamma_5
        \right) 
        \left(
            \slashed{k} 
            -
            \slashed{k'}
        \right)
        \left(
            a
            -
            b
            \gamma_5
        \right) 
    \right]
    \\
    &=&&
    a
    m_\psi^2
    \text{Tr}
    \left[
        \slashed{k'}
        \slashed{k} 
        \left(
            a
            -
            b
            \gamma_5
        \right) 
    \right]
    -
    4
    a^2
    m_\psi^4
    -
    b
    m_\psi^2
    \text{Tr}
    \left[
        \slashed{k'}
        \slashed{k} 
        \gamma_5
        \left(
            a
            -
            b
            \gamma_5
        \right) 
    \right]
    +
    b
    \text{Tr}
    \left[
        \slashed{k'}
        \slashed{k} 
        \gamma_5
        \slashed{k}
        \slashed{k'}
        \left(
            a
            -
            b
            \gamma_5
        \right) 
    \right]
    \\ &
    &&-
    m_\psi^4
    \text{Tr}
    \left[
        \left(
            a
            -
            b
            \gamma_5
        \right) 
        \left(
            a
            -
            b
            \gamma_5
        \right) 
    \right]
    +
    m_\psi^2
    \text{Tr}
    \left[
        \left(
            a
            -
            b
            \gamma_5
        \right) 
        \slashed{k}
        \slashed{k'}
        \left(
            a
            -
            b
            \gamma_5
        \right) 
    \right]
    \\ &
    &&+
    m_\psi^2
    \text{Tr}
    \left[
        \left(
            \slashed{k'}
            \slashed{k} 
            +
            \slashed{k}
            \slashed{k'}
            -
            2
            m_\psi^2
        \right)
        \left(
            a
            +
            b
            \gamma_5
        \right) 
        \left(
            a
            -
            b
            \gamma_5
        \right) 
    \right]
    \\
    &=&&
    2
    \left(
        a^2
        +
        b^2
    \right)
    m_\psi^2
    \text{Tr}
    \left[
        \slashed{k'}
        \slashed{k} 
    \right]
    -
    8
    \left(
        a^2
        +
        b^2
    \right)
    m_\psi^4
    +
    2
    \left(
        a^2
        -
        b^2
    \right) 
    m_\psi^2
    \text{Tr}
    \left[
        \slashed{k'}
        \slashed{k} 
    \right]
    +
    8
    \left(
        a^2
        -
        b^2
    \right) 
    m_\psi^4
    \\
    &=&&
    16
    a^2
    m_\psi^2
    k'_\mu 
    k^\mu
    -
    16
    b^2
    m_\psi^4
\end{alignat*}
Again, we can substitute $2 k'_\mu k^\mu = m_\phi^2 - 2 m_\psi^2$ and plug back in $a$ and $b$ to get the final squared matrix element in the rest frame of the modulus:
\begin{align} \label{matrixElementSqrFermionCombo2}
    | \mathcal{M} |^2
    &=
    2
    \left(
        g_1^2 
        +
        g_2^2
    \right)
    m_\psi^2
    m_\phi^2 
    \left(
        1
        -
        4 
        \frac{
            m_\psi^2
        }{
            m_\phi^2
        }
    \right)
    +
    4
    g_1
    g_2
    m_\psi^2
    m_\phi^2 
\end{align}

\subsection{Modulus decay into squarks}

We take the term from Eq. \ref{modulusMatterLagrangian}
\begin{align*}
    \mathcal{L}_{\widetilde{Q}}
    &=
    -
    \frac{\lambda_{Q^* Q}}{M_*}
    \left[
        \phi \,
        \partial_\mu \partial^\mu ( \widetilde{Q}^\dagger )
        \widetilde{Q}
        +
        \phi^* \,
        \widetilde{Q}^\dagger
        \partial_\mu \partial^\mu \widetilde{Q}
    \right]
    \\
    &=
    -
    \frac{\lambda_{Q^* Q}}{M_* \sqrt{2}}
    \left[
        \phi_R \,
        \left(
            \partial_\mu \partial^\mu ( \widetilde{Q}^\dagger )
            \widetilde{Q}
            +
            \widetilde{Q}^\dagger
            \partial_\mu \partial^\mu \widetilde{Q}
        \right)
        +
        i
        \phi_I \,
        \left(
            \partial_\mu \partial^\mu ( \widetilde{Q}^\dagger )
            \widetilde{Q}
            -
            \widetilde{Q}^\dagger
            \partial_\mu \partial^\mu \widetilde{Q}
        \right)
    \right]
\end{align*}

Starting with the right-handed quark singlets (note the abuse of notation, where the subscript $R$ refers to either right-handed or real part based on the context), we have 
\begin{align*}
    \mathcal{L}_{\widetilde{U}_R}
    &=
    -
    \frac{\lambda_{U_i^* U_i}}{M_* \sqrt{2}}
    \left[
        \phi_R \,
        \left(
            \partial_\mu \partial^\mu ( \widetilde{u}_{R,i}^\dagger )
            \widetilde{u}_{R,i}
            +    
            \widetilde{u}_{R,i}^\dagger
            \partial_\mu \partial^\mu \widetilde{u}_{R,i}
        \right)
        +
        i
        \phi_I \,
        \left(
            \widetilde{u}_{R,i} \partial_\mu \partial^\mu \widetilde{u}_{R,i}^\dagger
            -
            \widetilde{u}_{R,i}^\dagger \partial_\mu \partial^\mu \widetilde{u}_{R,i}
        \right)
    \right]
\end{align*}
and similarly for the down-type quark singlets.
This is then approximately the mass eigenstate Lagrangian for the right-handed $u,d,s,c$ squarks as we neglect mixing effects.

Using the mixing matrix for $f \in \{ b,t \}$
\begin{align*}
    \begin{pmatrix}
        \widetilde{f}_L \\
        \widetilde{f}_R
    \end{pmatrix}
    &=
    \begin{pmatrix}
        \cos \theta_f && \sin \theta_f \\
        - \sin \theta_f && \cos \theta_f
    \end{pmatrix}
    \begin{pmatrix}
        \widetilde{f}_1 \\
        \widetilde{f}_2
    \end{pmatrix}
\end{align*}
we then have the following Lagrangian in the mass eigenstate basis for the 3rd squark generations where mixing is relevant:
\begin{alignat*}{2}
    \mathcal{L}_{\widetilde{t}_R}
    &=&&
    -
    \frac{\lambda_{U_t^* U_t}}{M_* \sqrt{2}}
    \left[
        \sin^2 \theta_t 
        \phi_R \,
        \left(
            \widetilde{t}_1 
            \partial_\mu 
            \partial^\mu 
            \widetilde{t}_1^\dagger
            +
            \widetilde{t}_1^\dagger
            \partial_\mu 
            \partial^\mu 
            \widetilde{t}_1
        \right)
        + 
        \cos^2 \theta_t 
        \phi_R \,
        \left(
            \widetilde{t}_2
            \partial_\mu 
            \partial^\mu 
            \widetilde{t}_2^\dagger
            +
            \widetilde{t}_2^\dagger
            \partial_\mu 
            \partial^\mu 
            \widetilde{t}_2
        \right)
    \right] 
    \\ &
    &&+
    \frac{\lambda_{U_t^* U_t}}{M_* \sqrt{2}}
    \sin \theta_t \cos \theta_t
    \left[
        \phi_R \,
        \left(
            \widetilde{t}_1
            \partial_\mu 
            \partial^\mu 
            \widetilde{t}_2^\dagger
            +
            \widetilde{t}_2^\dagger
            \partial_\mu 
            \partial^\mu 
            \widetilde{t}_1
        \right)
        +
        \phi_R \,
        \left(
            \widetilde{t}_2
            \partial_\mu 
            \partial^\mu 
            \widetilde{t}_1^\dagger
            +
            \widetilde{t}_1^\dagger
            \partial_\mu 
            \partial^\mu 
            \widetilde{t}_2
        \right)
    \right] 
    \\ &
    &&-
    \frac{\lambda_{U_t^* U_t}}{M_* \sqrt{2}}
    \left[
        i
        \sin^2 \theta_t 
        \phi_I \,
        \left(
            \widetilde{t}_1
            \partial_\mu 
            \partial^\mu 
            \widetilde{t}_1^\dagger
            -
            \widetilde{t}_1^\dagger
            \partial_\mu 
            \partial^\mu 
            \widetilde{t}_1
        \right)
        +
        i
        \cos^2 \theta_t 
        \phi_I \,
        \left(
            \widetilde{t}_2
            \partial_\mu 
            \partial^\mu 
            \widetilde{t}_2^\dagger
            -
            \widetilde{t}_2^\dagger
            \partial_\mu 
            \partial^\mu 
            \widetilde{t}_2
        \right)
    \right] 
    \\ &
    &&-
    \frac{\lambda_{U_t^* U_t}}{M_* \sqrt{2}}
    \left[
        i
        \sin \theta_t 
        \cos \theta_t 
        \phi_I \,
        \left(
            \widetilde{t}_1^\dagger
            \partial_\mu 
            \partial^\mu 
            \widetilde{t}_2
            -
            \widetilde{t}_1
            \partial_\mu 
            \partial^\mu 
            \widetilde{t}_2^\dagger
            +
            \widetilde{t}_2^\dagger
            \partial_\mu 
            \partial^\mu 
            \widetilde{t}_1
            -
            \widetilde{t}_2
            \partial_\mu 
            \partial^\mu 
            \widetilde{t}_1^\dagger
        \right)
    \right]
\end{alignat*}

We now evaluate the left-handed quark doublets.
This gives us 
\begin{alignat*}{2}
    \mathcal{L}_{\widetilde{Q}_L}
    &=&&
    -
    \frac{\lambda_{Q_i^* Q_i}}{M_* \sqrt{2}}
    \left[
        \phi_R \,
        \left(
            \widetilde{u}_{L,i}
            \partial_\mu 
            \partial^\mu 
            \widetilde{u}_{L,i}^\dagger 
            +
            \widetilde{u}_{L,i}^\dagger 
            \partial_\mu 
            \partial^\mu 
            \widetilde{u}_{L,i}
        \right)
        +
        i
        \phi_I \,
        \left(
            \widetilde{u}_{L,i} 
            \partial_\mu 
            \partial^\mu 
            \widetilde{u}_{L,i}^\dagger
            -
            \widetilde{u}_{L,i}^\dagger 
            \partial_\mu 
            \partial^\mu 
            \widetilde{u}_{L,i}
        \right)
    \right]
    \\ &
    &&-
    \frac{\lambda_{Q_i^* Q_i}}{M_* \sqrt{2}}
    \left[
        \phi_R \,
        \left(
            \widetilde{d}_{L,i}
            \partial_\mu 
            \partial^\mu 
            \widetilde{d}_{L,i}^\dagger 
            +
            \widetilde{d}_{L,i}^\dagger 
            \partial_\mu 
            \partial^\mu 
            \widetilde{d}_{L,i}
        \right)
        +
        i
        \phi_I \,
        \left(
            \widetilde{d}_{L,i} 
            \partial_\mu 
            \partial^\mu 
            \widetilde{d}_{L,i}^\dagger
            -
            \widetilde{d}_{L,i}^\dagger 
            \partial_\mu 
            \partial^\mu 
            \widetilde{d}_{L,i}
        \right)
    \right]
\end{alignat*}
where the coupling is the same between the up and down type left-handed squarks (within the same generation).
This is approximately the mass eigenstate Lagrangian for the left-handed $u,d,s,c$ squarks.

Using the left-handed mixing similar to before for $f\in \{ b,t \}$, we then have the following Lagrangian in the mass eigenstate basis for 3rd generation squarks:
\begin{alignat*}{2}
    \mathcal{L}_{\widetilde{t}_L}
    &=&&
    -
    \frac{\lambda_{Q_3^* Q_3}}{M_* \sqrt{2}}
    \left[
        \cos^2 \theta_t 
        \phi_R \,
        \left(
            \widetilde{t}_1 
            \partial_\mu 
            \partial^\mu 
            \widetilde{t}_1^\dagger
            +
            \widetilde{t}_1^\dagger
            \partial_\mu 
            \partial^\mu 
            \widetilde{t}_1 
        \right)
        +
        \sin^2 \theta_t 
        \phi_R \,
        \left(
            \widetilde{t}_2
            \partial_\mu 
            \partial^\mu 
            \widetilde{t}_2^\dagger
            +
            \widetilde{t}_2^\dagger
            \partial_\mu 
            \partial^\mu 
            \widetilde{t}_2
        \right)
    \right]
    \\ &
    &&-
    \frac{\lambda_{Q_3^* Q_3}}{M_* \sqrt{2}}
    \sin \theta_t 
    \cos \theta_t 
    \left[
        \phi_R \,
        \left(
            \widetilde{t}_2
            \partial_\mu 
            \partial^\mu 
            \widetilde{t}_1^\dagger
            +
            \widetilde{t}_1^\dagger
            \partial_\mu 
            \partial^\mu 
            \widetilde{t}_2
        \right)
        +
        \phi_R \,
        \left(
            \widetilde{t}_1 
            \partial_\mu 
            \partial^\mu 
            \widetilde{t}_2^\dagger
            +
            \widetilde{t}_2^\dagger
            \partial_\mu 
            \partial^\mu 
            \widetilde{t}_1 
        \right)
    \right]
    \\ &
    &&-
    \frac{\lambda_{Q_3^* Q_3}}{M_* \sqrt{2}}
    \left[
        i
        \cos^2 \theta_t 
        \phi_I \,
        \left(
            \widetilde{t}_1 
            \partial_\mu 
            \partial^\mu 
            \widetilde{t}_1^\dagger
            -
            \widetilde{t}_1^\dagger
            \partial_\mu 
            \partial^\mu 
            \widetilde{t}_1 
        \right) 
        +
        i
        \sin^2 \theta_t 
        \phi_I \,
        \left(    
            \widetilde{t}_2
            \partial_\mu 
            \partial^\mu 
            \widetilde{t}_2^\dagger
            -
            \widetilde{t}_2^\dagger
            \partial_\mu 
            \partial^\mu 
            \widetilde{t}_2
        \right) 
    \right] 
    \\ &
    &&+
    \frac{\lambda_{Q_3^* Q_3}}{M_* \sqrt{2}}
    \left[
        i
        \sin \theta_t 
        \cos \theta_t 
        \phi_I \,
        \left(
            \widetilde{t}_1^\dagger
            \partial_\mu \partial^\mu 
            \widetilde{t}_2
            -
            \widetilde{t}_1 
            \partial_\mu \partial^\mu 
            \widetilde{t}_2^\dagger
            +
            \widetilde{t}_2^\dagger
            \partial_\mu \partial^\mu 
            \widetilde{t}_1 
            -
            \widetilde{t}_2
            \partial_\mu \partial^\mu 
            \widetilde{t}_1^\dagger
        \right)
    \right]
\end{alignat*}
The left and right handed 3rd generation squarks then combine to give the following interactions with $\phi_R$:
\begin{align}
    \mathcal{L}_{\phi_R \widetilde{t}_1 \widetilde{t}_1}
    &=
    -
    \frac{
        \lambda_{U_t^* U_t}
        \sin^2 \theta_t 
        +
        \lambda_{Q_3^* Q_3}
        \cos^2 \theta_t 
    }{M_* \sqrt{2}}
    \phi_R \,
    \left(
        \widetilde{t}_1 
        \partial_\mu 
        \partial^\mu 
        \widetilde{t}_1^\dagger
        +
        \widetilde{t}_1^\dagger
        \partial_\mu 
        \partial^\mu 
        \widetilde{t}_1 
    \right)
    \\
    \mathcal{L}_{\phi_R \widetilde{t}_2 \widetilde{t}_2}
    &=
    -
    \frac{
        \lambda_{U_t^* U_t}
        \cos^2 \theta_t 
        +
        \lambda_{Q_3^* Q_3}
        \sin^2 \theta_t 
    }{M_* \sqrt{2}}
    \phi_R \,
    \left(
        \widetilde{t}_2
        \partial_\mu 
        \partial^\mu 
        \widetilde{t}_2^\dagger
        +
        \widetilde{t}_2^\dagger
        \partial_\mu 
        \partial^\mu 
        \widetilde{t}_2
    \right)
    \\
    \mathcal{L}_{\phi_R \widetilde{t}_1 \widetilde{t}_2}
    &=
    -
    \frac{
        \lambda_{Q_3^* Q_3}
        -
        \lambda_{U_t^* U_t}
    }{M_* 2 \sqrt{2}}
    \sin 2 \theta_t \,
    \phi_R \,
    \left(
        \widetilde{t}_2
        \partial_\mu 
        \partial^\mu 
        \widetilde{t}_1^\dagger
        +
        \widetilde{t}_1^\dagger
        \partial_\mu 
        \partial^\mu 
        \widetilde{t}_2
        +
        \widetilde{t}_1 
        \partial_\mu 
        \partial^\mu 
        \widetilde{t}_2^\dagger
        +
        \widetilde{t}_2^\dagger
        \partial_\mu 
        \partial^\mu 
        \widetilde{t}_1 
    \right)
\end{align}
with corresponding matrix elements 
\begin{align}
    i \mathcal{M}_{\phi_R \rightarrow \widetilde{t}_1 \widetilde{t}_1}
    &=
    i
    \frac{
        \lambda_{U_t^* U_t}
        \sin^2 \theta_t 
        +
        \lambda_{Q_3^* Q_3}
        \cos^2 \theta_t 
    }{M_* \sqrt{2}}
    \left(
        2
        m_{\widetilde{t}_1}^2
    \right)
    \\
    i \mathcal{M}_{\phi_R \rightarrow \widetilde{t}_2 \widetilde{t}_2}
    &=
    i
    \frac{
        \lambda_{U_t^* U_t}
        \cos^2 \theta_t 
        +
        \lambda_{Q_3^* Q_3}
        \sin^2 \theta_t 
    }{M_* \sqrt{2}}
    \left(
        2
        m_{\widetilde{t}_2}^2
    \right)
    \\
    i \mathcal{M}_{\phi_R \rightarrow \widetilde{t}_1 \widetilde{t}_2}
    &=
    i
    \frac{
        \lambda_{Q_3^* Q_3}
        -
        \lambda_{U_t^* U_t}
    }{M_* 2 \sqrt{2}}
    \sin 2 \theta_t \,
    \left(
        2
        m_{\widetilde{t}_1}^2
        +
        2
        m_{\widetilde{t}_2}^2
    \right)
\end{align}
Assuming there are no additional contributions to the $\phi \rightarrow \overline{\widetilde{t}}_i \widetilde{t}_j$ decay channels, we can write down the squared matrix elements:
\begin{align}
    \left| 
        \mathcal{M}_{\phi_R \rightarrow \widetilde{t}_1 \widetilde{t}_1} 
    \right|^2
    &=
    \left(
        2
        \lambda_{U_t^* U_t}^2
        \sin^4 \theta_t
        +
        2
        \lambda_{Q_3^* Q_3}^2
        \cos^4 \theta_t 
        +
        \lambda_{U_t^* U_t}
        \lambda_{Q_3^* Q_3}
        \sin^2 2 \theta_t 
    \right)
    \frac{
        m_{\widetilde{t}_1}^4
    }{M_*^2}
    \\
    \left| 
        \mathcal{M}_{\phi_R \rightarrow \widetilde{t}_2 \widetilde{t}_2}
    \right|^2
    &=
    \left(
        2
        \lambda_{U_t^* U_t}^2
        \cos^4 \theta_t 
        +
        2
        \lambda_{Q_3^* Q_3}^2
        \sin^4 \theta_t 
        +
        \lambda_{U_t^* U_t}
        \lambda_{Q_3^* Q_3}
        \sin^2 2 \theta_t         
    \right)
    \frac{
        m_{\widetilde{t}_2}^4
    }{M_*^2}
    \\
    \left| 
        \mathcal{M}_{\phi_R \rightarrow \widetilde{t}_1 \widetilde{t}_2} 
    \right|^2
    &=
    \left(
        \lambda_{Q_3^* Q_3}^2
        +
        \lambda_{U_t^* U_t}^2
        -
        2
        \lambda_{Q_3^* Q_3}
        \lambda_{U_t^* U_t}
    \right)
    \sin^2 2 \theta_t \,
    \frac{
        m_{\widetilde{t}_1}^4
        +
        m_{\widetilde{t}_2}^4
        +
        2
        m_{\widetilde{t}_1}^2
        m_{\widetilde{t}_2}^2
    }{2 M_*^2}
\end{align}
These processes then have the associated decay widths:
\begin{align}
    \Gamma_{\phi_R \rightarrow \widetilde{t}_1 \widetilde{t}_1}
    &=
    \frac{
        \left(
            2
            \lambda_{U_t^* U_t}^2
            \sin^4 \theta_t
            +
            2
            \lambda_{Q_3^* Q_3}^2
            \cos^4 \theta_t 
            +
            \lambda_{U_t^* U_t}
            \lambda_{Q_3^* Q_3}
            \sin^2 2 \theta_t 
        \right)
    }{
        16
        \pi 
    } 
    \frac{
        m_{\widetilde{t}_1}^4
    }{
        m_\phi
        M_*^2
    }
    \lambda^{1/2}
    \left(
        1,
        \frac{
            m_{\widetilde{t}_1}^2
        }{ 
            m_\phi^2
        },
        \frac{
            m_{\widetilde{t}_1}^2
        }{ 
            m_\phi^2
        }
    \right)
    \\
    \Gamma_{\phi_R \rightarrow \widetilde{t}_2 \widetilde{t}_2}
    &=
    \frac{
        \left(
            2
            \lambda_{U_t^* U_t}^2
            \cos^4 \theta_t 
            +
            2
            \lambda_{Q_3^* Q_3}^2
            \sin^4 \theta_t 
            +
            \lambda_{U_t^* U_t}
            \lambda_{Q_3^* Q_3}
            \sin^2 2 \theta_t
        \right)
    }{
        16
        \pi 
    } 
    \frac{
        m_{\widetilde{t}_2}^4
    }{
        m_\phi
        M_*^2
    }
    \lambda^{1/2}
    \left(
        1,
        \frac{
            m_{\widetilde{t}_2}^2
        }{ 
            m_\phi^2
        },
        \frac{
            m_{\widetilde{t}_2}^2
        }{ 
            m_\phi^2
        }
    \right)
    \\
    \Gamma_{\phi_R \rightarrow \widetilde{t}_1 \widetilde{t}_2}
    &=
    \frac{
        \left(
            \lambda_{Q_3^* Q_3}^2
            +
            \lambda_{U_t^* U_t}^2
            -
            2
            \lambda_{Q_3^* Q_3}
            \lambda_{U_t^* U_t}
        \right)
    }{
        32
        \pi 
    } 
    \sin^2 2 \theta_t \,
    \frac{
        m_{\widetilde{t}_1}^4
        +
        m_{\widetilde{t}_2}^4
        +
        2
        m_{\widetilde{t}_1}^2
        m_{\widetilde{t}_2}^2
    }{
        m_\phi
        M_*^2
    }
    \lambda^{1/2}
    \left(
        1,
        \frac{
            m_{\widetilde{t}_1}^2
        }{ 
            m_\phi^2
        },
        \frac{
            m_{\widetilde{t}_2}^2
        }{ 
            m_\phi^2
        }
    \right)
\end{align}
Note that we do not divide by an additional factor of 2 here, as these final states are complex scalars and hence are distinguishable due to their associated charge.

The interactions with $\phi_I$ are given by:
\begin{alignat}{2}
    \mathcal{L}_{\phi_I \widetilde{t}_1 \widetilde{t}_1}
    &=&&
    -
    \frac{
        \lambda_{Q_3^* Q_3}
        \cos^2 \theta_t 
        +
        \lambda_{U_t^* U_t}
        \sin^2 \theta_t 
    }{M_* \sqrt{2}}
    i
    \phi_I \,
    \left(
        \widetilde{t}_1 
        \partial_\mu \partial^\mu 
        \widetilde{t}_1^\dagger
        -
        \widetilde{t}_1^\dagger
        \partial_\mu \partial^\mu 
        \widetilde{t}_1 
    \right) 
    \\
    \mathcal{L}_{\phi_I \widetilde{t}_2 \widetilde{t}_2}
    &=&&
    -
    \frac{
        \lambda_{Q_3^* Q_3}
        \sin^2 \theta_t 
        +
        \lambda_{U_t^* U_t}
        \cos^2 \theta_t 
    }{M_* \sqrt{2}}
    i
    \phi_I \,
    \left(
        \widetilde{t}_2 
        \partial_\mu \partial^\mu 
        \widetilde{t}_2^\dagger
        -
        \widetilde{t}_2^\dagger
        \partial_\mu \partial^\mu 
        \widetilde{t}_2 
    \right) 
    \\
    \mathcal{L}_{\phi_I \widetilde{t}_1 \widetilde{t}_2}
    &=&&
    -
    \frac{
        \lambda_{U_t^* U_t}
        -
        \lambda_{Q_3^* Q_3}
    }{M_* 2 \sqrt{2}}
    \sin 2 \theta_t \,
    i
    \phi_I \,
    \left(
        \widetilde{t}_1^\dagger
        \partial_\mu \partial^\mu 
        \widetilde{t}_2
        -
        \widetilde{t}_1 
        \partial_\mu \partial^\mu 
        \widetilde{t}_2^\dagger
        +
        \widetilde{t}_2^\dagger
        \partial_\mu \partial^\mu 
        \widetilde{t}_1 
        -
        \widetilde{t}_2
        \partial_\mu \partial^\mu 
        \widetilde{t}_1^\dagger
    \right)
\end{alignat}
Note that the extra factor of $i$ in the interactions with $\phi_I$ takes care of the field interactions which are of the form $a - a^\dagger$ (and hence are purely imaginary).
Furthermore, note that all tree-level matrix elements are 0 for the decay of $\phi_I$ to the squarks.  

The interactions with the $\widetilde{b}_i$ squarks are identical, with the replacements $\widetilde{t}_i \rightarrow \widetilde{b}_i$, $\theta_t \rightarrow \theta_b$, $\lambda_{U_t^* U_t} \rightarrow \lambda_{D_b^* D_b}$, however the $\lambda_{Q_3^* Q_3}$ coupling is unchanged (as left-handed components belong to the same multiplet).
For completeness, we also recap the interactions for the first two generations.
The left-handed squarks are given by
\begin{alignat}{2}
    \mathcal{L}_{\phi_R \widetilde{q}_L \widetilde{q}_L}
    &=&&
    -
    \frac{
        \lambda_{Q_i^* Q_i}
    }{M_* \sqrt{2}}
    \phi_R \,
    \left(
        \widetilde{q}_L 
        \partial_\mu 
        \partial^\mu 
        \widetilde{q}_L^\dagger
        +
        \widetilde{q}_L^\dagger
        \partial_\mu 
        \partial^\mu 
        \widetilde{q}_L
    \right)
    \\
    \mathcal{L}_{\phi_I \widetilde{q}_L \widetilde{q}_L}
    &=&&
    -
    \frac{
        \lambda_{Q_i^* Q_i}
    }{M_* \sqrt{2}}
    i
    \phi_I \,
    \left(
        \widetilde{q}_{L} 
        \partial_\mu \partial^\mu 
        \widetilde{q}_{L}^\dagger
        -
        \widetilde{q}_{L}^\dagger 
        \partial_\mu \partial^\mu 
        \widetilde{q}_{L}
    \right)
\end{alignat}
with corresponding matrix elements 
\begin{align}
    i \mathcal{M}_{\phi_R \rightarrow \widetilde{q}_L \widetilde{q}_L}
    &=
    i
    \frac{
        \lambda_{Q_i^* Q_i}
    }{M_* \sqrt{2}}
    \left(
        2
        m_{\widetilde{q}_L}^2
    \right)
    \\
    i \mathcal{M}_{\phi_I \rightarrow \widetilde{q}_L \widetilde{q}_L}
    &=
    0
\end{align}
and squared matrix element 
\begin{align}
    \left|
        \mathcal{M}_{\phi_R \rightarrow \widetilde{q}_L \widetilde{q}_L}
    \right|^2
    &=
    2
    \lambda_{Q_i^* Q_i}^2
    \frac{
        m_{\widetilde{q}_L}^4
    }{M_*^2}
\end{align}
where $\lambda_{Q_i^* Q_i} = \lambda_{Q_1^* Q_1}$ if $\widetilde{q}_L \in \{ \widetilde{u}_L, \widetilde{d}_L \}$ and $\lambda_{Q_i^* Q_i} = \lambda_{Q_2^* Q_2}$ if $\widetilde{q}_L \in \{ \widetilde{c}_L, \widetilde{s}_L \}$.
The decay width to the left-handed squarks is then given by
\begin{align}
    \Gamma_{\phi_R \rightarrow \widetilde{q}_L \widetilde{q}_L}
    &=
    \frac{
        \lambda_{Q_i^* Q_i}^2
    }{
        8
        \pi 
    } 
    \frac{
        m_{\widetilde{q}_L}^4
    }{
        m_\phi
        M_*^2
    }
    \lambda^{1/2}
    \left(
        1,
        \frac{
            m_{\widetilde{q}_L}^2
        }{ 
            m_\phi^2
        },
        \frac{
            m_{\widetilde{q}_L}^2
        }{ 
            m_\phi^2
        }
    \right)
\end{align}

The right-handed squarks are given by (again, calling attention to the abuse of notation by using the subscript $R$ to refer to either real part or right-handed based on context):
\begin{alignat}{2}
    \mathcal{L}_{\phi_R \widetilde{q}_R \widetilde{q}_R}
    &=&&
    -
    \frac{
        \lambda_{U_{\widetilde{q}}^* U_{\widetilde{q}}}
    }{M_* \sqrt{2}}
    \phi_R \,
    \left(
        \widetilde{q}_R
        \partial_\mu 
        \partial^\mu 
        \widetilde{q}_R^\dagger
        +
        \widetilde{q}_R^\dagger
        \partial_\mu 
        \partial^\mu 
        \widetilde{q}_R
    \right)
    \\
    \mathcal{L}_{\phi_I \widetilde{q}_R \widetilde{q}_R}
    &=&&
    -
    \frac{
        \lambda_{U_{\widetilde{q}}^* U_{\widetilde{q}}}
    }{M_* \sqrt{2}}
    i
    \phi_I \,
    \left(
        \widetilde{q}_{R} 
        \partial_\mu \partial^\mu 
        \widetilde{q}_{R}^\dagger
        -
        \widetilde{q}_{R}^\dagger 
        \partial_\mu \partial^\mu 
        \widetilde{q}_{R}
    \right)
\end{alignat}
with corresponding matrix elements 
\begin{align}
    i \mathcal{M}_{\phi_R \rightarrow \widetilde{q}_R \widetilde{q}_R}
    &=
    i
    \frac{
        \lambda_{U_{\widetilde{q}}^* U_{\widetilde{q}}}
    }{M_* \sqrt{2}}
    \left(
        2
        m_{\widetilde{q}_R}^2
    \right)
    \\
    i \mathcal{M}_{\phi_I \rightarrow \widetilde{q}_R \widetilde{q}_R}
    &=
    0    
\end{align}
and squared matrix element 
\begin{align}
    \left|
        \mathcal{M}_{\phi_R \rightarrow \widetilde{q}_R \widetilde{q}_R}
    \right|^2
    &=
    2
    \lambda_{U_{\widetilde{q}}^* U_{\widetilde{q}}}^2
    \frac{
        m_{\widetilde{q}_R}^4
    }{M_*^2}
\end{align}
for up-type squarks, with the replacement $\lambda_{U_{\widetilde{q}}^* U_{\widetilde{q}}} \rightarrow \lambda_{D_{\widetilde{q}}^* D_{\widetilde{q}}}$ for down-type squarks.
The decay width to the right-handed (up-type) squarks is then given by
\begin{align}
    \Gamma_{\phi_R \rightarrow \widetilde{q}_R \widetilde{q}_R}
    &=
    \frac{
        \lambda_{U_{\widetilde{q}}^* U_{\widetilde{q}}}^2
    }{
        8
        \pi 
    } 
    \frac{
        m_{\widetilde{q}_R}^4
    }{
        m_\phi
        M_*^2
    }
    \lambda^{1/2}
    \left(
        1,
        \frac{
            m_{\widetilde{q}_R}^2
        }{ 
            m_\phi^2
        },
        \frac{
            m_{\widetilde{q}_R}^2
        }{ 
            m_\phi^2
        }
    \right)
\end{align}
noting the replacement for also the down-type right-handed squarks.

\subsection{Modulus decay into sleptons}

The modulus decays into sleptons follow from the same term from 
Eq. \ref{modulusMatterLagrangian} as the squarks.
We simply write down the interaction terms by direct analogy.

Since the neutrino masses have been neglected, the sneutrinos have no mixing and hence the interaction terms are given by 
\begin{alignat}{2}
    \mathcal{L}_{\phi_R \widetilde{\nu_i} \widetilde{\nu_i}}
    &=&&
    -
    \frac{
        \lambda_{L_i^* L_i}
    }{M_* \sqrt{2}}
    \phi_R \,
    \left(
        \widetilde{\nu_i}
        \partial_\mu 
        \partial^\mu 
        \widetilde{\nu_i}^\dagger
        +
        \widetilde{\nu_i}^\dagger
        \partial_\mu 
        \partial^\mu 
        \widetilde{\nu_i}
    \right)
    \\
    \mathcal{L}_{\phi_I \widetilde{\nu_i} \widetilde{\nu_i}}
    &=&&
    -
    \frac{
        \lambda_{L_i^* L_i}
    }{M_* \sqrt{2}}
    i
    \phi_I \,
    \left(
        \widetilde{\nu_i}
        \partial_\mu 
        \partial^\mu 
        \widetilde{\nu_i}^\dagger
        -
        \widetilde{\nu_i}^\dagger
        \partial_\mu 
        \partial^\mu 
        \widetilde{\nu_i}
    \right)
\end{alignat}
which have corresponding matrix elements 
\begin{align}
    i \mathcal{M}_{\phi_R \rightarrow \widetilde{\nu}_i \widetilde{\nu}_i}
    &=
    i
    \frac{
        \lambda_{L_i^* L_i}
    }{M_* \sqrt{2}}
    \left(
        2
        m_{\widetilde{\nu}_i}^2
    \right)
    \\
    i \mathcal{M}_{\phi_I \rightarrow \widetilde{\nu}_i \widetilde{\nu}_i}
    &=
    0 
\end{align}
and squared matrix element 
\begin{align}
    \left| 
        \mathcal{M}_{\phi_R \rightarrow \widetilde{\nu}_i \widetilde{\nu}_i}
    \right|^2
    &=
    2
    \lambda_{L_i^* L_i}^2
    \frac{
        m_{\widetilde{\nu}_i}^4
    }{M_*^2}
\end{align}
where $i \in \{ e, \mu, \tau \}$.
The decay width to sneutrino pairs is then given by 
\begin{align}
    \Gamma_{\phi_R \rightarrow \widetilde{\nu}_i \widetilde{\nu}_i}
    &=
    \frac{
        \lambda_{L_i^* L_i}^2
    }{
        8
        \pi 
    } 
    \frac{
        m_{\widetilde{\nu}_i}^4
    }{
        m_\phi
        M_*^2
    }
    \lambda^{1/2}
    \left(
        1,
        \frac{
            m_{\widetilde{\nu}_i}^2
        }{ 
            m_\phi^2
        },
        \frac{
            m_{\widetilde{\nu}_i}^2
        }{ 
            m_\phi^2
        }
    \right)
\end{align}

The first two generations of selectron-type sleptons follow:
\begin{alignat}{2}
    \mathcal{L}_{\phi_R \widetilde{e_i}_L \widetilde{e_i}_L}
    &=&&
    -
    \frac{
        \lambda_{L_i^* L_i}
    }{M_* \sqrt{2}}
    \phi_R \,
    \left(
        \widetilde{e_i}_{L}
        \partial_\mu 
        \partial^\mu 
        \widetilde{e_i}_{L}^\dagger 
        +
        \widetilde{e_i}_{L}^\dagger 
        \partial_\mu 
        \partial^\mu 
        \widetilde{e_i}_{L}
    \right)
    \\
    \mathcal{L}_{\phi_I \widetilde{e_i}_L \widetilde{e_i}_L}
    &=&&
    -
    \frac{
        \lambda_{L_i^* L_i}
    }{M_* \sqrt{2}}
    i
    \phi_I \,
    \left(
        \widetilde{e_i}_{L} 
        \partial_\mu \partial^\mu 
        \widetilde{e_i}_{L}^\dagger
        -
        \widetilde{e_i}_{L}^\dagger 
        \partial_\mu \partial^\mu 
        \widetilde{e_i}_{L}
    \right)
    \\
    \mathcal{L}_{\phi_R \widetilde{e_i}_R \widetilde{e_i}_R}
    &=&&
    -
    \frac{
        \lambda_{E_{i}^* E_{i}}
    }{M_* \sqrt{2}}
    \phi_R \,
    \left(
        \widetilde{e_i}_{R}
        \partial_\mu 
        \partial^\mu 
        \widetilde{e_i}_{R}^\dagger 
        +
        \widetilde{e_i}_{R}^\dagger 
        \partial_\mu 
        \partial^\mu 
        \widetilde{e_i}_{R}
    \right)
    \\
    \mathcal{L}_{\phi_I \widetilde{e_i}_R \widetilde{e_i}_R}
    &=&&
    -
    \frac{
        \lambda_{E_{i}^* E_{i}}
    }{M_* \sqrt{2}}
    i
    \phi_I \,
    \left(
        \widetilde{e_i}_{R} 
        \partial_\mu \partial^\mu 
        \widetilde{e_i}_{R}^\dagger
        -
        \widetilde{e_i}_{R}^\dagger 
        \partial_\mu \partial^\mu 
        \widetilde{e_i}_{R}
    \right)
\end{alignat}
which have corresponding matrix elements
\begin{align}
    i \mathcal{M}_{\phi_R \rightarrow \widetilde{e}_{iL} \widetilde{e}_{iL}}
    &=
    i
    \frac{
        \lambda_{L_i^* L_i}
    }{M_* \sqrt{2}}
    \left(
        2
        m_{\widetilde{e}_{iL}}^2
    \right)
    \\
    i \mathcal{M}_{\phi_I \rightarrow \widetilde{e}_{iL} \widetilde{e}_{iL}}
    &=
    0 
    \\
    i \mathcal{M}_{\phi_R \rightarrow \widetilde{e}_{iR} \widetilde{e}_{iR}}
    &=
    i
    \frac{
        \lambda_{E_i^* E_i}
    }{M_* \sqrt{2}}
    \left(
        2
        m_{\widetilde{e}_{iR}}^2
    \right)
    \\
    i \mathcal{M}_{\phi_R \rightarrow \widetilde{e}_{iR} \widetilde{e}_{iR}}
    &=
    0 
\end{align}
and squared matrix elements
\begin{align}
    \left|
        \mathcal{M}_{\phi_R \rightarrow \widetilde{e}_{iL} \widetilde{e}_{iL}}
    \right|^2
    &=
    2
    \lambda_{L_i^* L_i}^2
    \frac{
        m_{\widetilde{e}_{iL}}^4
    }{M_*^2}
    \\
    \left|
        \mathcal{M}_{\phi_R \rightarrow \widetilde{e}_{iR} \widetilde{e}_{iR}}
    \right|^2
    &=
    2
    \lambda_{E_i^* E_i}^2
    \frac{
        m_{\widetilde{e}_{iR}}^4
    }{M_*^2}
\end{align}
where here $i \in \{ e, \mu \}$.
The decay widths for these processes are then given by 
\begin{align}
    \Gamma_{\phi_R \rightarrow \widetilde{e}_{iL} \widetilde{e}_{iL}}
    &=
    \frac{
        \lambda_{L_i^* L_i}^2
    }{
        8
        \pi 
    } 
    \frac{
        m_{\widetilde{e}_{iL}}^4
    }{
        m_\phi
        M_*^2
    }
    \lambda^{1/2}
    \left(
        1,
        \frac{
            m_{\widetilde{e}_{iL}}^2
        }{ 
            m_\phi^2
        },
        \frac{
            m_{\widetilde{e}_{iL}}^2
        }{ 
            m_\phi^2
        }
    \right)
    \\
    \Gamma_{\phi_R \rightarrow \widetilde{e}_{iR} \widetilde{e}_{iR}}
    &=
    \frac{
        \lambda_{E_i^* E_i}^2
    }{
        8
        \pi 
    } 
    \frac{
        m_{\widetilde{e}_{iR}}^4
    }{
        m_\phi
        M_*^2
    }
    \lambda^{1/2}
    \left(
        1,
        \frac{
            m_{\widetilde{e}_{iR}}^2
        }{ 
            m_\phi^2
        },
        \frac{
            m_{\widetilde{e}_{iR}}^2
        }{ 
            m_\phi^2
        }
    \right)
\end{align}

For the staus, we assume mixing of the same form as for the stops, so the resulting interaction terms with $\phi_R$ are 
\begin{alignat}{2}
    \mathcal{L}_{\phi_R \widetilde{\tau}_1 \widetilde{\tau}_1}
    &=&&
    -
    \frac{
        \lambda_{E_\tau^* E_\tau}
        \sin^2 \theta_\tau 
        +
        \lambda_{L_\tau^* L_\tau}
        \cos^2 \theta_\tau
    }{M_* \sqrt{2}}
    \phi_R \,
    \left(
        \widetilde{\tau}_1 
        \partial_\mu 
        \partial^\mu 
        \widetilde{\tau}_1^\dagger
        +
        \widetilde{\tau}_1^\dagger
        \partial_\mu 
        \partial^\mu 
        \widetilde{\tau}_1 
    \right)
    \\
    \mathcal{L}_{\phi_R \widetilde{\tau}_2 \widetilde{\tau}_2}
    &=&&
    -
    \frac{
        \lambda_{E_\tau^* E_\tau}
        \cos^2 \theta_\tau
        +
        \lambda_{L_\tau^* L_\tau}
        \sin^2 \theta_\tau
    }{M_* \sqrt{2}}
    \phi_R \,
    \left(
        \widetilde{\tau}_2
        \partial_\mu 
        \partial^\mu 
        \widetilde{\tau}_2^\dagger
        +
        \widetilde{\tau}_2^\dagger
        \partial_\mu 
        \partial^\mu 
        \widetilde{\tau}_2
    \right)
    \\
    \mathcal{L}_{\phi_R \widetilde{\tau}_1 \widetilde{\tau}_2}
    &=&&
    -
    \frac{
        \lambda_{L_\tau^* L_\tau}
        -
        \lambda_{E_\tau^* E_\tau}
    }{M_* 2 \sqrt{2}}
    \sin 2 \theta_\tau \,
    \phi_R \,
    \left(
        \widetilde{\tau}_2
        \partial_\mu 
        \partial^\mu 
        \widetilde{\tau}_1^\dagger
        +
        \widetilde{\tau}_1^\dagger
        \partial_\mu 
        \partial^\mu 
        \widetilde{\tau}_2
        +
        \widetilde{\tau}_1
        \partial_\mu 
        \partial^\mu 
        \widetilde{\tau}_2^\dagger
        +
        \widetilde{\tau}_2^\dagger
        \partial_\mu 
        \partial^\mu 
        \widetilde{\tau}_1
    \right)
\end{alignat}
with corresponding matrix elements 
\begin{align}
    i \mathcal{M}_{\phi_R \rightarrow \widetilde{\tau}_{1} \widetilde{\tau}_{1}}
    &=
    i
    \frac{
        \lambda_{E_\tau^* E_\tau}
        \sin^2 \theta_\tau 
        +
        \lambda_{L_\tau^* L_\tau}
        \cos^2 \theta_\tau
    }{M_* \sqrt{2}}
    \left(
        2
        m_{\widetilde{\tau}_{1}}^2
    \right)
    \\
    i \mathcal{M}_{\phi_R \rightarrow \widetilde{\tau}_{2} \widetilde{\tau}_{2}}
    &=
    i
    \frac{
        \lambda_{E_\tau^* E_\tau}
        \cos^2 \theta_\tau
        +
        \lambda_{L_\tau^* L_\tau}
        \sin^2 \theta_\tau
    }{M_* \sqrt{2}}
    \left(
        2
        m_{\widetilde{\tau}_{2}}^2
    \right)
    \\
    i \mathcal{M}_{\phi_R \rightarrow \widetilde{\tau}_{1} \widetilde{\tau}_{2}}
    &=
    i
    \frac{
        \lambda_{L_\tau^* L_\tau}
        -
        \lambda_{E_\tau^* E_\tau}
    }{M_* 2 \sqrt{2}}
    \sin 2 \theta_\tau \,
    \left(
        2
        m_{\widetilde{\tau}_{1}}^2
        +
        2
        m_{\widetilde{\tau}_{2}}^2
    \right)    
\end{align}
The squared matrix elements are then 
\begin{align}
    \left| 
        \mathcal{M}_{\phi_R \rightarrow \widetilde{\tau}_{1} \widetilde{\tau}_{1}}
    \right|^2
    &=
    \left(
        2
        \lambda_{E_\tau^* E_\tau}^2
        \sin^4 \theta_\tau 
        +
        2
        \lambda_{L_\tau^* L_\tau}^2
        \cos^4 \theta_\tau
        +
        \lambda_{E_\tau^* E_\tau}
        \lambda_{L_\tau^* L_\tau}
        \sin^2 2 \theta_\tau 
    \right)
    \frac{
        m_{\widetilde{\tau}_{1}}^4
    }{
        M_*^2
    }
    \\
    \left|
        \mathcal{M}_{\phi_R \rightarrow \widetilde{\tau}_{2} \widetilde{\tau}_{2}}
    \right|^2
    &=
    \left(
        2
        \lambda_{E_\tau^* E_\tau}^2
        \cos^4 \theta_\tau
        +
        2
        \lambda_{L_\tau^* L_\tau}^2
        \sin^4 \theta_\tau
        +
        \lambda_{E_\tau^* E_\tau}
        \lambda_{L_\tau^* L_\tau}
        \sin^2 2\theta_\tau
    \right)
    \frac{
        m_{\widetilde{\tau}_{2}}^4
    }{
        M_*^2
    }
    \\
    \left|
        \mathcal{M}_{\phi_R \rightarrow \widetilde{\tau}_{1} \widetilde{\tau}_{2}}
    \right|^2
    &=
    \left(
        \lambda_{L_\tau^* L_\tau}^2
        +
        \lambda_{E_\tau^* E_\tau}^2
        -
        2
        \lambda_{L_\tau^* L_\tau}
        \lambda_{E_\tau^* E_\tau}
    \right)
    \frac{
        m_{\widetilde{\tau}_{1}}^4
        +
        m_{\widetilde{\tau}_{2}}^4
        +
        2
        m_{\widetilde{\tau}_{1}}^2
        m_{\widetilde{\tau}_{2}}^2
    }{
        2 M_*^2
    }
    \sin^2 2 \theta_\tau \,
\end{align}
The associated decay widths for these processes are then given by
\begin{align}
    \Gamma_{\phi_R \rightarrow \overline{\tau}_1 \overline{\tau}_1}
    &=
    \frac{
        \left(
            \lambda_{E_\tau^* E_\tau}
            \sin^2 \theta_\tau 
            +
            \lambda_{L_\tau^* L_\tau}
            \cos^2 \theta_\tau
        \right)^2
    }{
        8
        \pi 
    } 
    \frac{
        m_{\widetilde{\tau}_{1}}^4
    }{
        m_\phi
        M_*^2
    }
    \lambda^{1/2}
    \left(
        1,
        \frac{
            m_{\overline{\tau}_1}^2
        }{ 
            m_\phi^2
        },
        \frac{
            m_{\overline{\tau}_1}^2
        }{ 
            m_\phi^2
        }
    \right)
    \\
    \Gamma_{\phi_R \rightarrow \overline{\tau}_2 \overline{\tau}_2}
    &=
    \frac{
        \left(
            \lambda_{E_\tau^* E_\tau}
            \cos^2 \theta_\tau
            +
            \lambda_{L_\tau^* L_\tau}
            \sin^2 \theta_\tau
        \right)^2
    }{
        8
        \pi 
    } 
    \frac{
        m_{\widetilde{\tau}_{2}}^4
    }{
        m_\phi
        M_*^2
    }
    \lambda^{1/2}
    \left(
        1,
        \frac{
            m_{\overline{\tau}_2}^2
        }{ 
            m_\phi^2
        },
        \frac{
            m_{\overline{\tau}_2}^2
        }{ 
            m_\phi^2
        }
    \right)
    \\
    \Gamma_{\phi_R \rightarrow \overline{\tau}_1 \overline{\tau}_2}
    &=
    \frac{
        \left(
            \lambda_{L_\tau^* L_\tau}
            -
            \lambda_{E_\tau^* E_\tau}
        \right)^2
    }{
        32
        \pi 
    } 
    \frac{
        m_{\widetilde{\tau}_{1}}^4
        +
        m_{\widetilde{\tau}_{2}}^4
        +
        2
        m_{\widetilde{\tau}_{1}}^2
        m_{\widetilde{\tau}_{2}}^2
    }{
        m_\phi
        M_*^2
    }
    \sin^2 2 \theta_\tau \,
    \lambda^{1/2}
    \left(
        1,
        \frac{
            m_{\overline{\tau}_1}^2
        }{ 
            m_\phi^2
        },
        \frac{
            m_{\overline{\tau}_2}^2
        }{ 
            m_\phi^2
        }
    \right)
\end{align}

The interactions with $\phi_I$ are given by 
\begin{alignat}{2}
    \mathcal{L}_{\phi_I \widetilde{\tau}_1 \widetilde{\tau}_1}
    &=&&
    -
    \frac{
        \lambda_{L_\tau^* L_\tau}
        \cos^2 \theta_\tau
        +
        \lambda_{E_\tau^* E_\tau}
        \sin^2 \theta_\tau 
    }{M_* \sqrt{2}}
    i
    \phi_I \,
    \left(
        \widetilde{\tau}_1 
        \partial_\mu \partial^\mu 
        \widetilde{\tau}_1^\dagger
        -
        \widetilde{\tau}_1^\dagger
        \partial_\mu \partial^\mu 
        \widetilde{\tau}_1 
    \right) 
    \\
    \mathcal{L}_{\phi_I \widetilde{\tau}_2 \widetilde{\tau}_2}
    &=&&
    -
    \frac{
        \lambda_{L_\tau^* L_\tau}
        \sin^2 \theta_\tau
        +
        \lambda_{E_\tau^* E_\tau}
        \cos^2 \theta_\tau 
    }{M_* \sqrt{2}}
    i
    \phi_I \,
    \left(
        \widetilde{\tau}_2 
        \partial_\mu \partial^\mu 
        \widetilde{\tau}_2^\dagger
        -
        \widetilde{\tau}_2^\dagger
        \partial_\mu \partial^\mu 
        \widetilde{\tau}_2 
    \right) 
    \\
    \mathcal{L}_{\phi_I \widetilde{\tau}_1 \widetilde{\tau}_2}
    &=&&
    -
    \frac{
        \lambda_{E_\tau^* E_\tau}
        -
        \lambda_{L_\tau^* L_\tau}
    }{M_* 2 \sqrt{2}}
    \sin 2 \theta_\tau \,
    i
    \phi_I \,
    \left(
        \widetilde{\tau}_1^\dagger
        \partial_\mu \partial^\mu 
        \widetilde{\tau}_2
        -
        \widetilde{\tau}_1 
        \partial_\mu \partial^\mu 
        \widetilde{\tau}_2^\dagger
        +
        \widetilde{\tau}_2^\dagger
        \partial_\mu \partial^\mu 
        \widetilde{\tau}_1 
        -
        \widetilde{\tau}_2
        \partial_\mu \partial^\mu 
        \widetilde{\tau}_1^\dagger
    \right)
\end{alignat}
Again, all tree-level matrix elements vanish for the decay of $\phi_I$ to the sleptons.

\subsection{Modulus decay into quarks}

We take the term from Eq. \ref{modulusMatterLagrangian}
\begin{alignat*}{2}
    \mathcal{L}_q
    &=&&
    \frac{\lambda_{Q^* Q}}{M_*} 
    \left[
        \frac{i}{2} 
        \phi 
        \overline{\psi}_Q 
        \slashed{\partial}
        \psi_Q
        -
        \frac{i}{2} 
        \phi^\dagger
        \partial_\mu
        ( \overline{\psi}_Q )
        \gamma^\mu 
        \psi_Q
    \right]
    \\
    &=&&
    \frac{\lambda_{Q^* Q}}{M_*} 
    \frac{i}{2 \sqrt{2}}
    \phi_R
    \left[
        \overline{\psi}_Q 
        \gamma^\mu 
        \partial_\mu
        \psi_Q
        -
        \partial_\mu
        ( \overline{\psi}_Q )
        \gamma^\mu 
        \psi_Q
    \right] 
    -
    \frac{\lambda_{Q^* Q}}{M_*} 
    \frac{1}{2 \sqrt{2}}
    \phi_I
    \left[
        \overline{\psi}_Q 
        \gamma^\mu 
        \partial_\mu
        \psi_Q
        +
        \partial_\mu
        ( \overline{\psi}_Q )
        \gamma^\mu 
        \psi_Q
    \right]
    \\
    &=&&
    \frac{\lambda_{Q^* Q}}{M_*} 
    \frac{i}{\sqrt{2}}
    \phi_R
    \overline{\psi}_Q 
    \slashed{\partial}
    \psi_Q
\end{alignat*}
where in the last line, we use the Majorana spinor identity $\overline{\psi} \gamma_\mu \chi = - \overline{\chi} \gamma_\mu \psi$, which makes the interaction with $\phi_I$ vanish.

We can then write the full interaction piece for a single generation of both up and down type quarks:
\begin{alignat*}{2}
    \mathcal{L}_q
    &=&&
    \frac{\lambda_{Q_i^* Q_i}}{M_*} 
    \frac{i}{\sqrt{2}}
    \phi_R
    \overline{\psi}_{u_i}
    \slashed{\partial}
    \psi_{u_i}
    +
    \frac{\lambda_{Q_i^* Q_i}}{M_*} 
    \frac{i}{\sqrt{2}}
    \phi_R
    \overline{\psi}_{d_i}
    \slashed{\partial}
    \psi_{d_i}
    +
    \frac{\lambda_{U_q^* U_q}}{M_*} 
    \frac{i}{\sqrt{2}}
    \phi_R
    \overline{\psi}_{U_q}
    \slashed{\partial}
    \psi_{U_q}
    +
    \frac{\lambda_{D_q^* D_q}}{M_*} 
    \frac{i}{\sqrt{2}}
    \phi_R
    \overline{\psi}_{D_q}
    \slashed{\partial}
    \psi_{D_q}
    \\
    &\supset &&
    \frac{\lambda_{Q_i^* Q_i}}{M_*} 
    \frac{i}{\sqrt{2}}
    \phi_R
    \overline{\psi}_{u_i}
    \slashed{\partial}
    \psi_{u_i}
    +
    \frac{\lambda_{U_q^* U_q}}{M_*} 
    \frac{i}{\sqrt{2}}
    \phi_R
    \overline{\psi}_{U_q}
    \slashed{\partial}
    \psi_{U_q}
\end{alignat*}
To progress further in terms of the Dirac quark fields, we will need to use a few tricks, namely:
\begin{align*}
    P_L 
    q
    &=
    P_L
    \psi_q
    \\
    P_R
    q
    &=
    P_R
    \psi_{Q^c}
\end{align*}
and insert factors of $\mathbb{I} = P_L + P_R$.
Upon evaluating this out and leveraging some Majorana bilinear identities, we can rewrite an interaction piece as 
\begin{alignat*}{2}
    \mathcal{L}_q
    &\supset &&
    \frac{\lambda_{Q_i^* Q_i}}{M_*} 
    \frac{i}{\sqrt{2}}
    \phi_R
    \left[
        \overline{\psi}_{u_i}
        P_R
        \slashed{\partial}
        P_L 
        \psi_{u_i}
        +
        \overline{\psi}_{u_i}
        P_L
        \slashed{\partial}
        P_L 
        \psi_{u_i}
        +
        \overline{\psi}_{u_i}
        P_R
        \slashed{\partial}
        P_R
        \psi_{u_i}
        +
        \overline{\psi}_{u_i}
        P_L
        \slashed{\partial}
        P_R
        \psi_{u_i}
    \right]
    \\
    &=&&
    \frac{\lambda_{Q_i^* Q_i}}{M_*} 
    \frac{i}{\sqrt{2}}
    \phi_R
    \left[
        \overline{u}_{i}
        P_R
        \slashed{\partial}
        P_L 
        u_i
        -
        \partial_\mu
        \overline{u}_{i}
        P_R
        \gamma^\mu
        P_L
        u_i
    \right]
\end{alignat*}
Hence, for the up-type quarks, we have the full interaction piece involving the Dirac fields 
\begin{alignat}{2}
    \mathcal{L}_{\phi_R u_i u_i}
    &=&&
    \frac{\lambda_{Q_i^* Q_i}}{M_*} 
    \frac{i}{\sqrt{2}}
    \phi_R
    \left[
        \overline{u}_{i}
        P_R
        \slashed{\partial}
        P_L 
        u_i
        -
        \partial_\mu
        \overline{u}_{i}
        P_R
        \gamma^\mu
        P_L
        u_i
    \right]
    +
    \frac{\lambda_{U_q^* U_q}}{M_*} 
    \frac{i}{\sqrt{2}}
    \phi_R
    \left[
        \overline{u}_{i}
        P_L
        \slashed{\partial}
        P_R 
        u_i
        -
        \partial_\mu
        \overline{u}_{i}
        P_L
        \gamma^\mu
        P_R
        u_i
    \right]
\end{alignat}
The down-type quarks can be gotten directly with the replacements $\lambda_{U_q^* U_q} \rightarrow \lambda_{D_q^* D_q}$ and $u_i \rightarrow d_i$.
Note that, unless $\lambda_{U_q^* U_q} = \lambda_{D_q^* D_q} = \lambda_{Q_i^* Q_i}$, the different chiral components of the quark fields interact slightly differently.
The associated matrix element is then (from Eq. \ref{matrixElementFermionCombo2}, and using the subscript $q_u$ to distinguish the spinor components of the up-quark field)
\begin{align}
    i \mathcal{M}_{ \phi_R \rightarrow \overline{u}_i u_i }
    &=
    \overline{u}_{q_u}^s(k)
    \left(
        i 
        \frac{\lambda_{Q_i^* Q_i} + \lambda_{U_q^* U_q}}{2\sqrt{2} M_*} 
        \gamma^\mu 
        \left(
            k_\mu 
            -
            k'_\mu
        \right)
        -
        i 
        \frac{\lambda_{Q_i^* Q_i} - \lambda_{U_q^* U_q}}{2\sqrt{2} M_*} 
        \gamma^\mu 
        \gamma_5
        \left(
            k_\mu 
            -
            k'_\mu
        \right)
    \right) 
    v^{s'}_{q_u}(k') 
\end{align}
(where the associated matrix element for the down-type quarks can be retrieved from the replacements above).
Assuming no additional contributions to the $\phi \rightarrow \overline{u}_i u_i$ decay channel, we can write down the squared matrix element from Eq. \ref{matrixElementSqrFermionCombo2}:
\begin{align}
    | \mathcal{M}_{ \phi_R \rightarrow \overline{u}_i u_i } |^2
    &=
    \left(
        \frac{
            \lambda_{Q_i^* Q_i}^2 + \lambda_{U_q^* U_q}^2
        }{
            M_*^2
        }
    \right)
    m_{u_i}^2
    m_\phi^2
    \left(
        1
        -
        4
        \frac{m_{u_i}^2}{m_\phi^2}
    \right)
    +
    \left(
        \frac{
            2
            \lambda_{Q_i^* Q_i} \lambda_{U_q^* U_q}
        }{M_*^2}
    \right)
    m_{u_i}^2
    m_\phi^2
\end{align}
The decay width to up-type quark pairs is then given by 
\begin{align}
    \Gamma_{\phi_R \rightarrow \overline{u}_i u_i}
    &=
    \frac{
        \left(
            \left(
                \lambda_{Q_i^* Q_i}^2 + \lambda_{U_q^* U_q}^2
            \right)
            \left(
                1
                -
                4
                \frac{m_{u_i}^2}{m_\phi^2}
            \right)
            +
            2
            \lambda_{Q_i^* Q_i} \lambda_{U_q^* U_q}
        \right)
    }{
        16
        \pi
    } 
    \frac{
        m_{u_i}^2
        m_\phi
    }{
        M_*^2
    }
    \lambda^{1/2}
    \left(
        1,
        \frac{
            m_{u_i}^2
        }{ 
            m_\phi^2
        },
        \frac{
            m_{u_i}^2
        }{ 
            m_\phi^2
        }
    \right)
\end{align}

For the sake of completeness, we write the associated squared matrix element for the down-type quarks:
\begin{align}
    | \mathcal{M}_{ \phi_R \rightarrow \overline{d}_i d_i  } |^2
    &=
    \left(
        \frac{
            \lambda_{Q_i^* Q_i}^2 + \lambda_{D_q^* D_q}^2
        }{M_*^2}
    \right)
    m_{d_i}^2
    m_\phi^2
    \left(
        1
        -
        4
        \frac{m_{d_i}^2}{m_\phi^2}
    \right)
    +
    \left(
        \frac{
            2
            \lambda_{Q_i^* Q_i} \lambda_{D_q^* D_q}
        }{M_*^2}
    \right)
    m_{d_i}^2
    m_\phi^2
\end{align}
The decay width to down-type quark pairs is then given by 
\begin{align}
    \Gamma_{\phi_R \rightarrow \overline{d}_i d_i}
    &=
    \frac{
        \left(
            \left(
                \lambda_{Q_i^* Q_i}^2 + \lambda_{D_q^* D_q}^2
            \right)
            \left(
                1
                -
                4
                \frac{m_{d_i}^2}{m_\phi^2}
            \right)
            +
            2
            \lambda_{Q_i^* Q_i} \lambda_{D_q^* D_q}
        \right)
    }{
        16
        \pi 
    } 
    \frac{
        m_{d_i}^2
        m_\phi
    }{
        M_*^2
    }
    \lambda^{1/2}
    \left(
        1,
        \frac{
            m_{d_i}^2
        }{ 
            m_\phi^2
        },
        \frac{
            m_{d_i}^2
        }{ 
            m_\phi^2
        }
    \right)
\end{align}

\subsection{Modulus decay into leptons}

We proceed to calculate the widths to leptons similarly to the quarks.
We can immediately write down the interactions:
\begin{align*}
    \mathcal{L}_l
    &=
    \frac{\lambda_{L_i^* L_i}}{M_*}
    \frac{i}{\sqrt{2}}
    \phi_R
    \left[
        \overline{\psi}_{e_i}
        \slashed{\partial}
        \psi_{e_i}
    \right]
    +
    \frac{\lambda_{L_i^* L_i}}{M_*}
    \frac{i}{\sqrt{2}}
    \phi_R
    \left[
        \overline{\psi}_{\nu_{e_i}}
        \slashed{\partial}
        \psi_{\nu_{e_i}}
    \right]
    +
    \frac{\lambda_{E_i^* E_i}}{M_*}
    \frac{i}{\sqrt{2}}
    \phi_R
    \left[
        \overline{\psi}_{E_i}
        \slashed{\partial}
        \psi_{E_i}
    \right]
\end{align*}
We can again write down the Dirac fields as 
\begin{align*}
    P_L 
    e_i 
    &=
    P_L 
    \psi_{e_i}
    \\
    P_R 
    e_i 
    &=
    P_R
    \psi_{E_i^c}
\end{align*}
and, inserting factors of $\mathbb{I} = P_L + P_R$, we have 
\begin{align*}
    \mathcal{L}_l
    &\supset
    \frac{\lambda_{L_i^* L_i}}{M_*}
    \frac{i}{\sqrt{2}}
    \phi_R
    \left[
        \overline{\psi}_{e_i}
        P_L
        \slashed{\partial}
        P_L
        \psi_{e_i}
        +
        \overline{\psi}_{e_i}
        P_R
        \slashed{\partial}
        P_L
        \psi_{e_i}
        +
        \overline{\psi}_{e_i}
        P_L
        \slashed{\partial}
        P_R
        \psi_{e_i}
        +
        \overline{\psi}_{e_i}
        P_R
        \slashed{\partial}
        P_R
        \psi_{e_i}
    \right]
    \\
    &=
    \frac{\lambda_{L_i^* L_i}}{M_*}
    \frac{i}{\sqrt{2}}
    \phi_R
    \left[
        \overline{\psi}_{e_i}
        P_R
        \slashed{\partial}
        P_L
        \psi_{e_i}
        -
        \partial_\mu
        \overline{\psi}_{e_i}
        P_R
        \gamma^\mu
        P_L
        \psi_{e_i}
    \right]
\end{align*}
Hence, the charged leptons have the full interaction piece involving the Dirac fields 
\begin{align}
    \mathcal{L}_{\phi_R e_i e_i}
    &=
    \frac{\lambda_{L_i^* L_i}}{M_*} 
    \frac{i}{\sqrt{2}}
    \phi_R
    \left[
        \overline{e}_{i}
        P_R
        \slashed{\partial}
        P_L 
        e_i
        -
        \partial_\mu
        \overline{e}_{i}
        P_R
        \gamma^\mu
        P_L
        e_i
    \right]
    +
    \frac{\lambda_{E_i^* E_i}}{M_*} 
    \frac{i}{\sqrt{2}}
    \phi_R
    \left[
        \overline{e}_{i}
        P_L
        \slashed{\partial}
        P_R 
        e_i
        -
        \partial_\mu
        \overline{e}_{i}
        P_L
        \gamma^\mu
        P_R
        e_i
    \right]
\end{align}
The associated matrix element can then be written down from Eq. \ref{matrixElementFermionCombo2}:
\begin{align}
    i \mathcal{M}_{ \phi_R \rightarrow l^+ l^- }
    &=
    \overline{u}_e^s(k)
    \left(
        i 
        \frac{\lambda_{L_i^* L_i} + \lambda_{E_i^* E_i}}{2 \sqrt{2} M_*}
        \gamma^\mu 
        \left(
            k_\mu 
            -
            k'_\mu
        \right)
        -
        i 
        \frac{\lambda_{L_i^* L_i} - \lambda_{E_i^* E_i}}{2 \sqrt{2} M_*}
        \gamma^\mu 
        \gamma_5
        \left(
            k_\mu 
            -
            k'_\mu
        \right)
    \right) 
    v^{s'}_e(k') 
\end{align}
Assuming no other contributions to the $\phi \rightarrow l^+ l^-$ decay channel, we can write down the squared matrix element from Eq. \ref{matrixElementSqrFermionCombo2}:
\begin{align}
    | \mathcal{M}_{ \phi_R \rightarrow l^+ l^- } |^2
    &=
    \left(
        \frac{
            \lambda_{L_i^* L_i}^2
            +
            \lambda_{E_i^* E_i}^2
        }{M_*^2}
    \right)
    m_{l}^2
    m_\phi^2
    \left(
        1
        -
        4
        \frac{m_{l}^2}{m_\phi^2}
    \right)
    +
    \left(
        \frac{
            2
            \lambda_{L_i^* L_i}
            \lambda_{E_i^* E_i}
        }{M_*^2}
    \right)
    m_{l}^2
    m_\phi^2
\end{align}
The decay width to charged lepton pairs is then given by 
\begin{align}
    \Gamma_{\phi_R \rightarrow l^+ l^-}
    &=
    \frac{
        \left(
            \left(
                \lambda_{L_i^* L_i}^2
                +
                \lambda_{E_i^* E_i}^2
            \right)
            \left(
                1
                -
                4
                \frac{m_{l}^2}{m_\phi^2}
            \right)
            +
            2
            \lambda_{L_i^* L_i}
            \lambda_{E_i^* E_i}
        \right)
    }{
        16
        \pi 
    } 
    \frac{
        m_{l}^2
        m_\phi
    }{
        M_*^2
    }
    \lambda^{1/2}
    \left(
        1,
        \frac{
            m_l^2
        }{ 
            m_\phi^2
        },
        \frac{
            m_l^2
        }{ 
            m_\phi^2
        }
    \right)
\end{align}

The neutrinos will have the interaction:
\begin{align}
    \mathcal{L}_{\phi_R \nu_i \nu_i}
    &=
    \frac{\lambda_{L_i^* L_i}}{M_*} 
    \frac{i}{\sqrt{2}}
    \phi_R
    \left[
        \overline{\nu}_{i}
        P_R
        \slashed{\partial}
        P_L 
        \nu_i
        -
        \partial_\mu
        \overline{\nu}_{i}
        P_R
        \gamma^\mu
        P_L
        \nu_i
    \right]
\end{align}
which has the associated matrix element from Eq. \ref{matrixElementFermionCombo1}
\begin{align}
    i \mathcal{M}_{ \phi_R \rightarrow \overline{\nu}_l \nu_l }
    &=
    \overline{u}_{\nu}^s(k)
    \left(
        i 
        \frac{
            \lambda_{L_i^* L_i}
        }{2 \sqrt{2} M_*} 
        \gamma^\mu 
        (1 - \gamma_5) 
        \left(
            k_\mu 
            -
            k'_\mu
        \right)
    \right) 
    v^{s'}_{\nu}(k') 
\end{align}
Again, assuming no additional contributions to $\phi \rightarrow \overline{\nu}_i \nu_i$, the squared matrix element can be retrieved from Eq. \ref{matrixElementSqrFermionCombo1}:
\begin{align}
    |\mathcal{M}_{ \phi_R \rightarrow \overline{\nu}_l \nu_l }|^2
    &=
    \left(
        \frac{
            \lambda_{L_i^* L_i}^2
        }{M_*^2}
    \right)
    m_{\nu_i}^2
    m_\phi^2
    \left(
        1
        -
        4 
        \frac{m_{\nu_i}^2}{m_\phi^2}
    \right)
\end{align}
Note that, assuming the neutrinos are massless, this matrix element vanishes.

\subsection{Modulus decay to gravitino pairs}
\label{app:gravitino}

Nakamura and Yamaguchi (NY) Ref. ~\cite{Nakamura:2006uc} consider the case
of particular models which lead to helicity-unsuppressed modulus 
decay to gravitinos (case {\bf 1}):
\be
\Gamma (\phi_R\to\psi_\mu \psi_\mu )=\frac{1}{288\pi}
d_{3/2}^2\frac{m_{\phi}^3}{m_P^2}
\left(1-4m_\psi^2/m_\phi^2\right)^{1/2}
\ee
where $d_{3/2}$ is defined in terms of the K\"ahler function as 
$\langle G_{\phi\phi^*}\rangle^{-1/2}\langle e^{G/2}G_\phi\rangle\equiv d_{3/2}\frac{m_{3/2}^2}{m_\phi}$. The dimensionless constant $d_{3/2}$ is 
then model-dependent depending on the form of the K\"ahler function but is 
expected to be of order unity. 
We have appended a phase space factor to the formula of NY. While the NY formula 
is technically valid in the high energy limit where gravitino decay is dominantly to goldstino components, we adopt this form even in the lower energy limit.
This form for modulus decay to gravitino pairs was also obtained by Endo {\it et al.} 
Ref.~~\cite{Endo:2006zj}.  

Another possibility is that for other forms of the K\"ahler function~\cite{Dine:2006ii}, 
the modulus decay to gravitinos is helicity suppressed (case {\bf 2}).
In that case, we use the above formula but with the replacement $m_\phi^3\to m_{3/2}^2 m_\phi$.

%%%%%%%%%%%%%%%%%%%%%%%%%%%%%%%%%%%%%%%%%%%%%%%%%%%%%%

%\section*{References}
\bibliography{mod1}
\bibliographystyle{elsarticle-num}

\end{document}